\documentclass[11pt,a4paper,aps,nofootinbib]{article}
\usepackage{jcappub}
\usepackage[labelfont=bf]{caption} 
\usepackage[utf8]{inputenc}
\usepackage{color}
\usepackage{amsmath, bm, physics}
\usepackage{amssymb, mathtools}
\usepackage{subcaption}
\usepackage{hyperref}
\usepackage{multirow}
\date{\today}

\title{Axion inflation in the regime of homogeneous backreaction}

\author[a,b]{Matteo Barbon,}
\author[a,c]{Nadir Ijaz,
}
\author[a,c]{and Marco Peloso}

\affiliation[a]{Dipartimento di Fisica e Astronomia ``G. Galilei", Universit\`a degli Studi di Padova, via Marzolo 8, I-35131 Padova, Italy}
\affiliation[b]{Scuola Galileiana di Studi Superiori, Università degli Studi di Padova, via Venezia 20, I-35131 Padova, Italy}
\affiliation[c]{INFN, Sezione di Padova, via Marzolo 8, I-35131 Padova, Italy}

\emailAdd{matteo.barbon.1@studenti.unipd.it}
\emailAdd{nadir.ijaz@phd.unipd.it}
\emailAdd{marco.peloso@pd.infn.it}

\abstract{
We investigate the Stochastic Gravitational Wave Background (SGWB) produced in models of axion inflation coupled to gauge fields. Achieving a detectable signal at Pulsar Timing Array, astrometry, or interferometer frequencies requires a sufficiently strong amplification of the gauge fields, at a level that induces significant backreaction on the inflaton background dynamics. Numerical studies based on the approximation of homogeneous backreaction (i.e., neglecting inhomogeneities of the inflaton field) exhibit oscillations in the inflaton velocity, with corresponding peaks in the SGWB spectrum. The most recent lattice simulations have questioned the validity of this regime, showing examples characterized by a rapid increase in the inflaton gradient energy and the breakdown of homogeneous backreaction. We compute this energy density perturbatively within the assumption of homogeneous backreaction, obtaining examples with a detectable SGWB and with an inflaton gradient energy that remains subdominant to the inflaton zero-mode kinetic energy throughout inflation.
}

\begin{document}

\maketitle

\section{Introduction} 
\label{sec:intro}

Large-scale measurements~\cite{Planck:2018jri,BICEP:2021xfz} strongly support early-universe cosmology with initial conditions provided by cosmological inflation~\cite{Guth:1980zm,Linde:1981mu,Albrecht:1982wi}. However, particle-physics realizations of inflation encounter significant theoretical challenges, especially in the so-called large-field models, which are widely studied because they can potentially produce an observable tensor signal~\cite{Lyth:1996im}. In these models, the inflaton traverses a field range $\Delta \phi$ of order the Planck scale, $M_p \equiv 1/\sqrt{8\pi G_N}$ (with $G_N$ the Newton constant). As a result, higher-dimensional operators of the form $\Delta V \propto \phi^{4+n}/M^n$ can easily spoil the required flatness of the potential unless suppressed by a scale $M > M_p$, which appears unnatural in the context of effective field theory.

A possible solution to this problem is to postulate~\cite{Freese:1990rb,Adams:1992bn} that the theory possesses an approximate invariance under a shift symmetry $\phi \to \phi + {\rm constant}$. This is, for instance, the case for axion-like particles (ALPs), where the continuous shift symmetry can be broken by instantons or other non-perturbative string effects down to a discrete subgroup, $\phi \to \phi + 2\pi f$, with $f$ a mass-dimension-one parameter known as the axion decay constant. This breaking permits the generation of a periodic potential $V_0$ for the inflaton–axion, which is protected from large radiative corrections (technical naturalness). In fact, any coupling that respects the shift symmetry can contribute to the potential only in combination with terms that break the symmetry, so that the smallness of $V_0$ also suppresses any radiative corrections it receives. Despite the simplest potential $V = \Lambda^4 \left[ 1 - \cos \frac{\phi}{f} \right]$ considered in~\cite{Freese:1990rb} being now ruled out by Cosmic Microwave Background (CMB) measurements~\cite{Planck:2018jri}, the protection mechanism introduced in~\cite{Freese:1990rb} has led to a widely investigated class of inflationary models referred to as axion inflation~\cite{Pajer:2013fsa}.~\footnote{A relatively simple realization that is still allowed by data is two-field aligned axion inflation~\cite{Kim:2004rp}, which admits metastable inflationary trajectories~\cite{Peloso:2015dsa} leading to a phenomenology that might be a target in the near future of Cosmic Microwave Background (CMB) polarization measurements such as CMB-S4~\cite{CMB-S4:2020lpa} and LiteBIRD~\cite{LiteBIRD:2023iei}. The inflationary evolution for this model has recently been solved in~\cite{Greco:2024ngr}, which obtained relatively simple analytic expressions for the trajectory and the associated phenomenology in terms of the model parameters.}

The decay of the axion inflaton and reheating are typically due to the shift-symmetric operator $\phi F {\tilde F}$, where $F$ is the field strength of a gauge field and ${\tilde F}$ is its dual.~\footnote{The decay into fermions from a shift-symmetric coupling is instead helicity suppressed, and generally subdominant~\cite{Pajer:2013fsa}.} Interestingly for this work, this coupling can have far-reaching effects already during inflation. The first application considered in the literature was the generation of primordial magnetic fields~\cite{Turner:1987bw,Garretson:1992vt,Anber:2006xt}. Being a shift-symmetric coupling, this production is proportional to the inflaton derivative, and therefore it occurs at the expense of its kinetic energy. This induces an additional effective friction (besides the usual Hubble friction term) on the motion of the inflaton, which, as noted by Anber and Sorbo~\cite{Anber:2009ua} (AS), can in some cases significantly backreact on the inflaton dynamics.~\footnote{In this work we focus on the coupling of an axion to a U(1) gauge field. Refs.~\cite{Adshead:2012kp,Dimastrogiovanni:2016fuu} proposed models with, respectively, a rolling axion inflaton and an axion spectator field coupled to an SU(2) gauge multiplet. Also in this context backreaction can be highly relevant, as studied in Refs.~\cite{Maleknejad:2018nxz,Ishiwata:2021yne,Iarygina:2023mtj,Dimastrogiovanni:2025snj}.} Shortly afterwards, Ref.~\cite{Barnaby:2010vf} computed the primordial scalar and tensor perturbations sourced by the produced gauge fields in the regime of weak backreaction. The sourced scalars were found to be highly non-Gaussian, and therefore must remain subdominant to the standard vacuum modes (induced by the inflationary expansion). This limits the gauge field amplification to a level that provides negligible backreaction on the background dynamics~\cite{Barnaby:2010vf,Planck:2015zfm,Jamieson:2025ngu}.

It is important to stress that this constraint only applies to the moments during which the CMB modes were produced. In fact, the amplification of the gauge field is exponentially sensitive to the inflaton velocity, which generically increases during inflation. In the presence of this pseudo-scalar interaction, one often observes a transition between the weak and strong backreaction regimes during the inflationary stages that follow the production of the CMB~\cite{Barnaby:2011qe}. This can have very interesting consequences for the generation of a potentially detectable stochastic gravitational wave background (SGWB). While Ref.~\cite{Barnaby:2010vf} showed that the stringent limits from scalar non-Gaussianity force the sourced tensor signal to be highly subdominant to the vacuum one, the increased amplification at later stages of inflation can lead to a significantly stronger SGWB at smaller scales, potentially detectable by gravitational wave (GW) interferometers~\cite{Cook:2011hg,Barnaby:2011qe,Domcke:2016bkh}. This is particularly interesting since, as noted in~\cite{Sorbo:2011rz}, the sourced SGWB is mostly composed of a single chiral polarization. A number of works have studied the prospects for detecting a chiral SGWB from CMB~\cite{Gluscevic:2010vv} and Pulsar Timing Array (PTA)~\cite{Cruz:2024esk} measurements, as well as from space-borne~\cite{Seto:2006hf,Seto:2006dz,Domcke:2019zls,Orlando:2020oko} and ground-based~\cite{Seto:2007tn,Seto:2008sr,Crowder:2012ik,Mentasti:2023gmg,Abac:2025saz} interferometers.~\footnote{Quite generally, the production of an observable SGWB at CMB scales needs to be confronted with limits from scalar non-Gaussianity~\cite{Barnaby:2010vf,Barnaby:2012xt,Namba:2015gja}. Correspondingly, the generation of an observable SGWB at small scales through the $\phi F \tilde{F}$ mechanism can be considered viable only if it can be demonstrated that it does not lead to an excessive production of scalar perturbations, which at such scales typically results in the overabundance of Primordial Black Holes (PBHs)~\cite{Linde:2012bt,Bugaev:2013fya,Garcia-Bellido:2016dkw,Garcia-Bellido:2017aan,Ozsoy:2023ryl}. Besides the uncertainty on the amplitude of the scalar perturbations generated in the case of strong backreaction, which we address in this work, the PBH abundance is strongly affected by the statistics of the scalar modes, which are still uncertain~\cite{Caravano:2022epk}.}

The original AS solution assumed a constant Hubble rate $H$ and inflaton speed $\dot{\phi}$, leading to the approximate solution~\eqref{A-approx} for the gauge mode functions~\cite{Anber:2009ua}. Neither $H$ nor $\dot{\phi}$ are perfectly constant during inflation, but they evolve adiabatically in proportion to the slow-roll parameter $\epsilon$. Given that the inflaton speed typically increases during inflation, and that the amplitude of the gauge modes is exponentially sensitive to this speed~\cite{Anber:2009ua}, one can expect the system to enter the strong backreaction regime at relatively late stages of inflation, after the CMB modes are produced~\cite{Barnaby:2011qe}. This, as we already mentioned, allows for interesting signatures at small scales (from modes produced in these later stages of inflation)~\cite{Cook:2011hg,Barnaby:2011qe,Domcke:2016bkh}, while still respecting the strong limits from non-Gaussianity at CMB scales~\cite{Barnaby:2010vf,Planck:2015zfm,Jamieson:2025ngu}.

The above-cited studies of backreaction are based on the natural assumption that, due to the adiabatic evolution of the system, the solution~\eqref{A-approx} remains accurate even if $H$ and $\dot{\phi}$ are not perfectly constant. In particular, this implies that, at any given instant, the backreaction term in which the solution~\eqref{A-approx} is employed is only sensitive to the inflaton speed at that precise moment. This gives rise to a steady-state evolution of the system, in which the “driving force” acting on the inflaton (the derivative of the potential) is perfectly balanced by the friction from dissipation evaluated at that same time. We denote this as {\it local backreaction}.

The numerical evolutions of the system showed that this is not the case.~\footnote{In fact, obtaining concrete realizations of local backreaction is far from trivial. See Ref.~\cite{Creminelli:2023aly} for an explicit construction.} The earlier numerical studies assumed a grid of momenta for the gauge modes coupled to a homogeneous inflaton. Following~\cite{Figueroa:2023oxc}, we denote this as {\it homogeneous backreaction}. The assumption of a homogeneous inflaton greatly simplifies the evolution equations of the system, eliminating couplings between gauge modes. Thanks to this, the system can be simulated with a one-dimensional grid~\cite{Cheng:2015oqa,Notari:2016npn,DallAgata:2019yrr}.

Refs.~\cite{Cheng:2015oqa,Notari:2016npn,DallAgata:2019yrr} numerically integrated the evolution for the case of a homogeneous inflaton coupled to a large set of gauge-field modes $A(k)$. These works showed that homogeneous backreaction in this class of models does not proceed in a steady-state fashion, but that the inflaton speed oscillates in amplitude with a period of about $\sim 4$–$5$ e-folds, with the inflaton kinetic energy remaining subdominant to the potential (thus, without causing a breakdown of inflation). A convincing explanation for this oscillatory behavior was put forward in Ref.~\cite{Domcke:2020zez}, which solved the system numerically by adopting a recursive approach.~\footnote{This iterative method consists of numerically computing the motion of the inflaton and scale factor using the gauge mode solution obtained in the previous step as external input; these solutions for the inflaton and the scale factor are then employed as external functions to improve the gauge mode solutions; this cycle is repeated until convergence.} The oscillations in the inflaton speed arise because the contribution of each gauge-field mode to backreaction at a given time $t$ depends on the inflaton speed at earlier moments $t' < t$, when that mode was being amplified. This dependence creates a memory effect, with backreaction responding to past dynamics rather than being determined solely by the immediate balance between the potential slope and dissipation. Consequently, the axion velocity oscillates around the AS solution: brief intervals of accelerated axion motion lead to enhanced gauge-field amplification and stronger backreaction, which in turn slows the axion down. As the axion decelerates, gauge-field production diminishes and backreaction weakens, paving the way for another phase of rapid axion motion and renewed gauge-field growth. The emergence of this oscillatory pattern was proven analytically in~\cite{Peloso:2022ovc}, which studied the system of perturbations around the AS solution (still in the approximation of a homogeneous inflaton) and obtained the explicit expression of the kernel that accounts for this memory effect.

An alternative numerical method for the study of homogeneous backreaction~\cite{Gorbar:2021rlt,Durrer:2023rhc,vonEckardstein:2023gwk} is based on the so-called {\it gradient expansion formalism} (GEF)~\cite{Sobol:2019xls}. In this formalism, rather than employing the set of gauge modes ${\bf A}_k$, one can rewrite the equations governing homogeneous backreaction as a tower of equations for two-point correlation functions of “electric” and “magnetic” combinations, such as $\left\langle {\bf E} \cdot {\rm rot}^n {\bf B} \right\rangle$, with an increasing number of spatial derivatives. This tower of equations is then truncated at some high order $n$ and integrated numerically. This formalism has been recently extended in Refs.~\cite{Domcke:2023tnn} and~\cite{Durrer:2024ibi} to include scalar perturbations.

If present in the full system (namely, that in which backreaction is accounted for without any approximation), these oscillations can lead to a very distinctive phenomenology, with, for example, peaks in the GW power spectrum corresponding to modes produced while the inflaton speed is at a maximum in its oscillatory pattern. The peaks would have a characteristic separation in frequency, related to the above-mentioned period of $\sim 4$–$5$ e-folds for the oscillations of $\dot{\phi}$, which would constitute a distinctive signature of this mechanism, producing peaked signals potentially across different GW observatories, as proposed in~\cite{Garcia-Bellido:2023ser}, and further analyzed in~\cite{Ozsoy:2024apn,vonEckardstein:2025oic}.~\footnote{GW production from gauge fields amplified by the $\phi F {\tilde F}$ mechanism has been the subject of several investigations. Besides the above-cited works, and the related analyses of~\cite{Caprini:2019pxz,LISACosmologyWorkingGroup:2022jok,LISACosmologyWorkingGroup:2024hsc} with a focus on the LISA mission, interesting GW production, with possibly multiple peaks in the spectrum, can occur in models in which $\phi$ is not the inflaton, but a spectator field (or a set of spectators)~\cite{Barnaby:2012xt,Namba:2015gja,Dimastrogiovanni:2016fuu,Thorne:2017jft,Talebian:2022cwk,Dimastrogiovanni:2023juq,Putti:2024uyr,Caravano:2024xsb,He:2025ieo,Zhang:2025cyd}. Ample production can also take place at the end of inflation and reheating~\cite{Adshead:2018doq,Adshead:2019lbr,Adshead:2019igv,Bastero-Gil:2022fme}. Refs.~\cite{Michelotti:2024bbc} and~\cite{Kume:2025lvz} computed instead the production of GW and scalar modes in the presence of a non-standard axion kinetic term and small sound speed.} 

All the above methods for dealing with backreaction are based on approximations of the equations governing the dynamics of the system.~\footnote{For a yet different approach see Ref.~\cite{Galanti:2024jhw}, in which backreaction is computed via the mean-field Hartree approximation.} Spatial gradients in the axion field give rise to non-linear scatterings among gauge quanta, as well as between gauge and axion quanta, enabling energy exchange across different wavenumbers. This alters the spectrum of the fluctuations in a way that cannot be captured by homogeneous backreaction, where the gauge quanta continue to satisfy a linearized equation and therefore do not interact with one another. Only very recently has the full set of equations been solved via lattice simulations~\cite{Caravano:2021bfn,Caravano:2022epk,Figueroa:2023oxc,Caravano:2024xsb,Sharma:2024nfu,Figueroa:2024rkr,Lizarraga:2025aiw,Jamieson:2025ngu}. We denote this as {\it full backreaction}. Lattice simulations during inflation are a notoriously challenging task, and only a limited number of e-folds of expansion can be covered due to the large number of modes required to span the relevant dynamical range from the beginning to the end of the simulation.~\footnote{These simulations build on earlier works that studied the effect of the axion-gauge coupling during reheating. For a set of works on this topic, see Refs.~\cite{Adshead:2015pva,Adshead:2018doq,Cuissa:2018oiw,Adshead:2019lbr,Adshead:2019igv,Adshead:2023mvt}.} The first of these works~\cite{Caravano:2021bfn,Caravano:2022epk} confirmed the accuracy of the analytical results of~\cite{Barnaby:2010vf,Barnaby:2011vw} in the weak backreaction regime, and reproduced the first oscillation of $\dot{\phi}$ at the onset of the strong backreaction regime that had emerged from earlier studies based on the homogeneous inflaton approximation. The following works, starting from Ref.~\cite{Figueroa:2023oxc}, had an improved dynamical range, allowing the simulation of the evolution beyond the first oscillation. From these studies, a much more chaotic behavior emerged. In the cases studied in~\cite{Figueroa:2023oxc,Figueroa:2024rkr}, the gradient energy of the inflaton rapidly grows during the first oscillation of $\dot{\phi}$, becoming comparable with the kinetic energy of the homogeneous backreaction solution toward the completion of this first oscillation. This signals a departure of the full solution from the one obtained with the approximation of a homogeneous inflaton. The subsequent integration of the full system shows that the following oscillations of the homogeneous solution are either strongly reduced or absent, depending on the values assumed by the model parameters. The results shown in~\cite{Figueroa:2023oxc,Figueroa:2024rkr} are in fact extremely sensitive to the input parameters. These simulations cover the last $\sim 7$ e-folds of inflation, which is a remarkable achievement given the extreme difficulty of the task.

While these collective results have significantly increased our knowledge of the system, we are still far from a complete understanding of it. The strong sensitivity of the results to the parameters, and the very high cost of lattice simulations, leave, in our opinion, some open questions. Here, we formulate two of them: 1) Is there a range of parameters that leads to interesting phenomenology, but for which backreaction is not strong enough to trigger significant inflaton perturbations, so that homogeneous backreaction remains valid throughout the evolution? 2) Even in cases where homogeneous backreaction ceases to be valid, what is the subsequent fate of the gradient energy? Typically, gradients are rapidly diluted away by inflation. Could we expect that the inflationary expansion dilutes the gradients to a level at which homogeneous backreaction becomes accurate again? This might lead to a qualitatively new pattern of oscillations,~\footnote{We thank Matias Zaldarriaga for an insightful discussion on this topic.} possibly with an increased period (as required for the depletion of the gradient energy). While, unfortunately, the latter question will need to await more extended lattice simulations (and not focused solely on the end of inflation, as a sufficiently long phase might be needed for the depletion of gradients), we hope that the first question can be addressed along the lines opened in Refs.~\cite{Domcke:2023tnn,Durrer:2024ibi}, where inflaton inhomogeneities are computed perturbatively.

Differently from these papers, which added inflaton perturbations to the GEF equations, in the present work we follow the scheme of Ref.~\cite{Garcia-Bellido:2023ser}, where homogeneous backreaction is studied with a one-dimensional lattice of gauge modes, which we supplemented by the perturbative computation of inflaton perturbations. By monitoring the various contributions to the energy density, and in particular the gradient energy density of the inflaton, we can obtain an indication of the accuracy of the homogeneous backreaction. In particular, we postulate a criterion for the validity of this approximate regime by comparing this contribution against the kinetic energy of the inflaton zero mode (which, in turn, is always much smaller than its potential energy during inflation). We regard as reliable the phenomenological results for modes that exit the horizon while the inflaton gradient energy remains sufficiently subdominant. As with any approximation of such an extremely complex system, we do not expect the results to be fully accurate once the gradient energy becomes comparable to the kinetic energy. In fact, the comparison on the background dynamics with lattice simulations presented in~\cite{Domcke:2023tnn} show that some departures take place when the ratio between the spatial gradient and the kinetic energies reaches the $\sim 5\%$ level. It seems reasonable to expect that, for ratios of this amount this approximation should at least capture the correct order of magnitude and the approximate spectral shape of the sourced spectra (both inflaton and tensor perturbations). Conversely, we consider as unreliable the results obtained in the homogeneous approximation once the gradient energy of the perturbations approximates or even exceeds the kinetic energy. While this criterion appears reasonable, we admittedly lack a rigorous mathematical proof on the precise needed threshold, and we believe that only full lattice simulations can allow its verification.~\footnote{We thank Valerie Domcke for an insightful discussion on this topic.}

This work is structured as follows. In Section~\ref{sec:background} we introduce the model and the equations governing homogeneous backreaction. In Section~\ref{sec:perturbations} we provide the formalism to compute the sourced tensor and inflaton perturbations in this regime. In Section~\ref{sec:SGWB} we list possible channels of detection for the SGWB sourced by this mechanism, as well as constraints that need to be satisfied by it and by the scalar perturbations that are simultaneously produced. In Section~\ref{sec:example} we examine these potential detections and constraints within a specific example model previously studied in~\cite{Garcia-Bellido:2023ser}, now with an improved awareness of the need to assess when predictions based on homogeneous backreaction can be considered reliable. In Section~\ref{sec:peaks} we study how changes in relevant model parameters (specifically, the slope of the inflaton potential and the coupling strength to gauge fields) impact the number of peaks in the SGWB spectrum obtained in the regime of homogeneous backreaction. In Section~\ref{sec:conclusions} we present our conclusions. The present work is concluded with five appendices, which present some more technical computational details. In particular, in Appendix~\ref{app:constant-xi} we review results from the literature for the case of constant inflaton speed and Hubble parameter, and verify that our more general expressions agree with them in this limit. In Appendix~\ref{app:code} we provide the explicit expressions for the correlators evaluated by our code. In Appendix~\ref{app:rho} we show how these correlators are employed to provide the various energy density contributions. In Appendix~\ref{app:f-Omega} we provide the explicit relations for the present frequency and energy density of the SGWB. Finally, in Appendix~\ref{app:parameters} we provide details of the model studied in Section~\ref{sec:example}.

\section{The model and the background evolution in the regime of homogeneous backreaction}
\label{sec:background}

We consider the action
\begin{equation}
S = \int d^4 x \sqrt{-g} \left[ \frac{M_p^2}{2} R - \frac{1}{2} \left( \partial \phi \right)^2 - V \left( \phi \right) - \frac{1}{4} F^2 - \frac{1}{8 \sqrt{-g}} \frac{\phi}{f} \epsilon^{\mu \nu \alpha \beta} F_{\mu \nu} F_{\alpha \beta} \right] \;,
\label{model}
\end{equation}
with $\epsilon^{\mu \nu \alpha \beta}$ totally antisymmetric and $\epsilon^{0123} \equiv +1$. We consider the line element
\begin{equation}
d s^2 = a^2 \left( \tau \right) \left[ - d \tau^2 + \left( \delta_{ij} + h_{ij} \right) d x^i d x^j \right] \;,
\label{line}
\end{equation}
where $h_{ij}$ are the transverse and traceless GW modes, that will be considered in the next section. In this section we set them to zero, and consider a homogeneous inflaton-axion 
\begin{equation}
\phi \left( \tau ,\, \vec{x} \right) = \varphi \left( \tau \right) \;. 
\label{infla-back}
\end{equation}

The inflaton has a Chern-Simons coupling (the last term in eq.~\eqref{model}) with a U(1) gauge field, expressed in terms of the gauge field strength $F_{\mu \nu} \equiv \partial_\mu A_\nu - \partial_\nu A_\mu$ and of the mass-dimension one axion decay constant $f$. For a homogeneous inflaton we can work in the gauge $\vec{A}_0 = \vec{\nabla} \cdot \vec{A} = 0$, and decompose the vector components as
\begin{eqnarray}
A_i &=& \int \frac{d^3 k}{\left( 2 \pi \right)^{3/2}} \, {\rm e}^{i \vec{k} \cdot \vec{x}} \sum_{\lambda = \pm}
\epsilon_{i,\lambda} \left( {\hat k} \right) {\hat A}_\lambda \left( \tau ,\, \vec{k} \right)  \;, 
\label{A-deco1}
\end{eqnarray}
where the two helicity operators satisfy $\vec{k} \cdot \vec{\epsilon}_\lambda \left( {\hat k} \right) = 0$, $\vec{k} \times \vec{\epsilon}_\lambda \left( {\hat k} \right) = - i \lambda k \vec{\epsilon}_\lambda \left( {\hat k} \right)$, $\vec{\epsilon}_\lambda \left( - {\hat k} \right) =  \vec{\epsilon}_{-\lambda} \left( {\hat k} \right) =  \vec{\epsilon}_\lambda^{\;*} \left( {\hat k} \right)$, and $\vec{\epsilon}_\lambda^{\;*} \left({\hat k} \right) \vec{\epsilon}_{\lambda'} \left( {\hat k} \right)  = \delta_{\lambda \lambda'}$. We quantize 
\begin{eqnarray}
{\hat A}_\lambda \left( \tau ,\, \vec{k} \right)  &=& A_\lambda \left( \tau ,\, k \right) \, {\hat a}_\lambda \left( \vec{k} \right) + A_\lambda^* \left( \tau ,\, k \right) \, {\hat a}_\lambda^\dagger \left( -\vec{k} \right)  \;, 
\label{A-deco2}
\end{eqnarray}
where annihilation/creation operator satisfy 
\begin{equation}
\left[ {\hat a}_\lambda \left( \vec{k} \right) ,\, {\hat a}_{\lambda'}^\dagger \left( \vec{k}' \right) \right] = \delta_{\lambda \lambda'} \, \delta^{(3)} \left (\vec{k} - \vec{k}' \right) \;, 
\end{equation}
while the mode function satisfies
\begin{equation}
A_\lambda'' + \left( k^2 - 2 \lambda \, \xi \; a H \, k \,  \right) A_\lambda = 0 
\;\;\; , \;\;\; \xi \equiv \frac{\dot{\varphi}}{2 H f} \;, 
\label{A-eom}
\end{equation}
where $H \equiv \frac{\dot{a}}{a}$ is the Hubble rate, and where prime and dot denote, respectively differentiation with respect to conformal time $\tau$, and to cosmological time $t$ (defined through $dt = a \, d \tau$). Without loss of generality, we assume that $\dot{\varphi} > 0$, so that the chirality $A_+$ is tachyonically amplified close to horizon crossing, while the mode $A_-$ essentially remains in its vacuum state, and can be disregarded for the remainder of this work. 

Although we are not necessarily assuming that the gauge field is the Standard Model electromagnetic field (nor the hypercharge) it is convenient to adopt electromagnetic notation and define the ``electric'' and ``magnetic'' configurations
\begin{eqnarray}
E_i &\equiv& - \frac{1}{a^2} A_i' = - \frac{1}{a^2} \int \frac{d^3 k}{\left( 2 \pi \right)^{3/2}} {\rm e}^{i \vec{k} \cdot \vec{x}} \; \epsilon_{i,+} \left( {\hat k} \right) \, {\hat A}_+' \left( \tau ,\, \vec{k} \right) \;, 
\nonumber\\
B_i &\equiv&  \frac{1}{a^2} \epsilon_{ijk} \partial_j A_k = \frac{1}{a^2} \int \frac{d^3 k}{\left( 2 \pi \right)^{3/2}} {\rm e}^{i \vec{k} \cdot \vec{x}} \; k \, \epsilon_{i,+} \left( {\hat k} \right) {\hat A}_+ \left( \tau ,\, \vec{k} \right) \;, 
\label{EB}
\end{eqnarray}
in terms of which the background equations following from \eqref{model} acquire the form 
\begin{eqnarray}
&& \varphi'' + 2 \, {\cal H} \, \varphi' + a^2 \, \frac{d V}{d \varphi} = \frac{a^2}{f} \, \left\langle \vec{E} \cdot \vec{B} \right\rangle \;, \nonumber\\
&& {\cal H}^2 = \frac{1}{3 M_p^2} \left[ \frac{\varphi^{'2}}{2} + a^2 \, V + \frac{a^2}{2} \left\langle \vec{E}^2+\vec{B}^2 \right\rangle
\right] \;, 
\label{bck-eom}
\end{eqnarray}
where ${\cal H} \equiv \frac{a'}{a} = a \, H$ and where $\left\langle \cdots \right\rangle$ denotes spatial averaging. The system is said to be in a regime of strong backreaction when the amplified gauge field provides a non-negligible contribution to these equations, thus significantly modifying the background dynamics. As long as the inflaton perturbations remain ``small'' (we attempt to quantify this condition in Section \ref{sec:example}), the solutions of these equations, with the gauge field given by eq.~\eqref{A-eom}, provide a good description of the background dynamics. Following~\cite{Figueroa:2024rkr}, we denote this regime as that of homogeneous backreaction. When this is no longer the case, the inflaton acquires strong spatial gradients that modify the vector field amplification and the precise way in which they backreact on the background dynamics. When this happens, the analysis performed in this work loses its validity, and one needs to resort to lattice simulations such as those developed in~\cite{Caravano:2022epk,Figueroa:2024rkr}. 

\section{Tensor and scalar perturbations}
\label{sec:perturbations}

The amplified gauge modes source primordial tensor and scalar perturbations. This computation was first performed analytically in~\cite{Barnaby:2010vf,Barnaby:2011vw} in the regime of negligible backreaction, and adiabatically evolving $\xi$ and $H$. Ref.~\cite{Garcia-Bellido:2023ser} studied the tensor production in the regime of homogeneous backreation. As shown there, the tensor field in the line element~\eqref{line} is decomposed into helicities as 
\begin{equation} 
{\hat h_{ij}} \left( \tau ,\, \vec{x} \right) = \frac{2}{M_p \, a \left( \tau \right)} \int \frac{d^3 k}{\left( 2 \pi \right)^{3/2}} \, {\rm e}^{i \vec{k} \cdot \vec{x}} \, \sum_{\lambda = \pm} \Pi_{ij,\lambda}^* \left( {\hat k} \right) {\hat Q}_\lambda \left( \tau ,\, \vec{k} \right) \;, 
\label{h-deco} 
\end{equation} 
where the polarization operators are constructed from the vector ones as $\Pi_{ij,\lambda}^* \left( {\hat k} \right) \equiv \epsilon_{i,\lambda}\left( {\hat k} \right) \epsilon_{j,\lambda}\left( {\hat k} \right)$, and where the canonical variables ${\hat Q}_\lambda$ satisfy
\begin{equation} 
\left[ \frac{\partial^2}{\partial \tau^2} + k^2 - \frac{a''}{a} \right] {\hat Q}_\lambda \left( \tau ,\, \vec{k} \right) = - \frac{a^3}{M_p} \Pi_{ij,\lambda} \left( {\hat k} \right) \int \frac{d^3 x}{\left( 2 \pi \right)^{3/2}} {\rm e}^{-i \vec{k} \cdot \vec{x}} \left( {\hat E_i} \, {\hat E_j} + {\hat B_i} \, {\hat B_j} \right) \;. 
\label{eq-lambda} 
\end{equation} 

The canonical variable corresponding to the inflaton perturbation is instead defined through 
\begin{equation}
\phi \left( \tau ,\, \vec{x} \right) = \varphi \left( \tau \right) +   \hat {\delta \phi} \left( \tau ,\, \vec{x} \right) \;\;,\;\; \hat {\delta \phi} \left( \tau ,\, \vec{x} \right) = \frac{1}{a \left( \tau \right)} \int \frac{d^3 k}{\left( 2 \pi \right)^{3/2}} \, {\rm e}^{i \vec{k} \cdot \vec{x}} \, {\hat Q}_\phi \left( \tau ,\, \vec{k} \right) \;, 
\end{equation} 
and it satisfies (see e.g.~\cite{Namba:2015gja})
\begin{equation}
\left[ \frac{\partial^2}{\partial \tau^2} + k^2 - \frac{a''}{a} + a^2 \, \frac{\partial^2 V}{\partial \varphi^2} \right] {\hat  Q}_\phi \left( \tau ,\, \vec{k} \right) = \frac{a^3}{f} \int \frac{d^3 x}{\left( 2 \pi \right)^{3/2}} {\rm e}^{-i \vec{k} \cdot \vec{x}} \, {\hat E_i} \, {\hat B_i} \;. 
\label{eq-phi}
\end{equation}
This equation disregards metric perturbations that lead to interactions of gravitational strength, and are therefore subdominant to the direct coupling with the vector field in the regime in which $
\delta \phi$ is amplified at significant level~\cite{Barnaby:2011vw,Durrer:2024ibi}.~\footnote{Including these terms would also provide slow-roll corrections to ${\hat O}_\phi$, and thus lead to minor modifications of the spectral tilt of the vacuum scalar modes and to the Green function of this operator. These effects play no relevant role in our investigation, and, therefore, for consistency, we disregard all the effects of scalar metric perturbations in this work.}

The two equations~\eqref{eq-lambda} and~\eqref{eq-phi} have the identical structure
\begin{align}
& {\hat O}_X \; {\hat Q}_X \left( \tau ,\, \vec{k} \right) = \hat{\cal S}_X \left( \tau ,\, \vec{k} \right) \;, \nonumber\\
& {\hat O}_X \equiv \frac{\partial^2}{\partial \tau^2} + k^2 - \frac{a''}{a} + \delta_{X \phi} \; a^2 \, \frac{\partial^2 V}{\partial \varphi^2} \;\;\;,\;\;\; X = \left\{ \lambda ,\, \phi \right\} \;, 
\label{system-X}
\end{align}
in terms of the sources
\begin{align}
& \hat{\cal S}_\lambda \left( \tau ,\, \vec{k} \right) \equiv - \frac{a^3}{M_p} \Pi_{ij,\lambda} \left( {\hat k} \right) \int \frac{d^3 x}{\left( 2 \pi \right)^{3/2}} {\rm e}^{-i \vec{k} \cdot \vec{x}} \left( {\hat E_i} \, {\hat E_j} + {\hat B_i} \, {\hat B_j} \right) \;, \nonumber\\
& \hat{\cal S}_\phi \left( \tau ,\, \vec{k} \right) \equiv \frac{a^3}{f} \int \frac{d^3 x}{\left( 2 \pi \right)^{3/2}} {\rm e}^{-i \vec{k} \cdot \vec{x}} \, {\hat E_i} \, {\hat B_i} \;, 
\label{surces-X}
\end{align}
admitting the formal solutions  
\begin{equation}
{\hat Q}_X \left( \tau ,\, \vec{k} \right) = {\hat Q}_{X,v} \left( \tau ,\, \vec{k} \right) + {\hat Q}_{X,s} \left( \tau ,\, \vec{k} \right) \;\;,\;\; X = \lambda ,\, \phi \;, 
\end{equation}
where ${\hat Q}_{X,v} \left( \tau ,\, \vec{k} \right)$ is the vacuum solution, while ${\hat Q}_{X,s} \left( \tau ,\, \vec{k} \right)$ is sourced by the amplified vector field and it is formally given by 
\begin{equation}
{\hat Q}_{X,s} \left( \tau ,\, \vec{k} \right) = \int_{-\infty}^\tau \, d \tau' \, {\tilde G}_{X,k} \left( \tau ,\, \tau' \right) \, \hat{\cal S}_X \left( \tau' ,\, \vec{k} \right) \;\;,\;\; X = \lambda ,\, \phi \,, 
\label{formal-sol}
\end{equation}
where ${\tilde G}_{X,k}$ are the non-distributional part of the retarded Green functions of the operators ${\hat O}_X$. They are given by
\begin{equation}
{\tilde G}_{X,k} \left( \tau ,\, \tau' \right) = 2 \, {\rm Im } \left[ F_X^* \left( \tau ,\, k \right) \, F_X \left( \tau' ,\, k \right) \right] \;, 
\end{equation}
where
\begin{equation}
{\hat O}_X F_{X} \left( \tau ,\, k \right) = 0 \;\;,\;\; \lim_{\tau = - \infty} F_X \left( \tau ,\, k \right) = \frac{{\rm e}^{-i k \tau}}{\sqrt{2 k}} \,,  
\end{equation}
see~\cite{Garcia-Bellido:2023ser} for a detailed derivation.

As we shall see, physical quantities of our interest (in particular, energy densities of the sourced modes) are proportional to the correlators ${\cal C}_X^{(m)} \left( \tau ,\, k \right)$, defined through
\begin{equation}
\!\!\!\!\!\!\!\! \!\!\!\!\!\!\!\! 
\left\langle \frac{\partial^m}{\partial \tau^m} \left( \frac{{\hat Q}_{X,s} \left( \tau ,\, \vec{k} \right)}{a \left( \tau \right)} \right) \frac{\partial^m}{\partial \tau^m} \left( \frac{{\hat Q}_{X',s} \left( \tau ,\, \vec{k}' \right)}{a \left( \tau \right)} \right)  \right\rangle' \equiv \frac{2 \pi^2 M_p^2}{k^{3-2m}} \, {\cal C}_X^{(m)} \left( \tau ,\, k \right) \, \delta_{X,X'} \;, 
\label{C-X-def}
\end{equation}
for $X = \lambda ,\, \phi$ and $m=0,1$. This gives~\footnote{We note that the contribution with the time derivative acting on the extremum of integration vanishes, since ${\tilde G}_{X,k} \left( \tau ,\, \tau \right) = 0$.}
\begin{align}
\!\!\!\!\!\!\!\! 
{\cal C}_X^{(m)} \left( \tau ,\, k \right) &= \frac{k^{3-2m}}{2 \pi^2 M_p^2} \nonumber\\
& \!\!\!\!\!\!\!\!  \!\!\!\!\!\!\!\!  \times \int_{-\infty}^\tau \, d \tau' \, \frac{\partial^m}{\partial \tau^m} \left( \frac{{\tilde G}_{X,k} \left( \tau ,\, \tau' \right)}{a \left( \tau \right)} \right) \int_{-\infty}^\tau \, d \tau'' \, \frac{\partial^m}{\partial \tau^m} \left( \frac{{\tilde G}_{X,k} \left( \tau ,\, \tau'' \right)}{a \left( \tau \right)} \right) \,  \left\langle \hat{\cal S}_X \left( \tau' ,\, \vec{k} \right) \, \hat{\cal S}_{X} \left( \tau'' ,\, \vec{k}' \right) \right\rangle\Large'  \;.  
\label{C-X}
\end{align}

In the above expressions, the prime after the expectation value denotes the expectation value at the net of an overall Dirac delta-function $\delta^{(3)} \left( \vec{k} + \vec{k}' \right)$. The chosen rescaling in front of the correlators ensures that they are dimensionless, so that they are convenient intermediate quantities for numerical evaluations. Moreover, for the $m=0$ case, they have the same scaling dependence~\footnote{In fact they have also the same time dependence in the GW case, where the spectra coincide up to a numerical factor, and in the scalar case in the limit of constant $H$ and $\dot{\varphi}$.} of the power spectra of the sourced scalar curvature ${\hat \zeta}_s \left( \tau ,\, \vec{k} \right) \equiv - \frac{H}{\dot{\varphi}} \frac{{\hat Q}_{\phi,s} \left( \tau ,\, \vec{k} \right)}{a}$ (in spatially flat gauge) and of the sourced GW modes ${\hat h}_{\lambda,s} \left( \tau ,\, \vec{k} \right) \equiv \frac{2}{M_p} \, \frac{{\hat Q}_{\lambda,s} \left( \tau ,\, \vec{k} \right)}{a}$. In fact
\begin{align}
P_{\zeta,s} \left( \tau ,\, k \right) &\equiv \frac{k^3}{2 \pi^2} \, \left\langle {\hat \zeta}_s \left( \tau ,\, \vec{k} \right) {\hat \zeta}_s \left( \tau ,\, \vec{k}' \right) \right\rangle' = \frac{H^2 \, M_p^2}{\dot{\varphi}^2} \, {\cal C}_\phi^{(0)} \left( \tau ,\, k \right) \;, \nonumber\\
P_{\lambda,s} \left( k \right) &\equiv \frac{k^3}{2 \pi^2} \, \left\langle {\hat h}_{\lambda,s} \left( \tau ,\, \vec{k} \right) {\hat h}_{\lambda,s} \left( \tau ,\, \vec{k}' \right) \right\rangle' = 4 \, {\cal C}_\lambda^{(0)} \left( \tau ,\, k \right) \;. 
\label{PS-sourced}
\end{align} 

The computation of the two-point function of the GW source is performed in~\cite{Garcia-Bellido:2023ser}, leading to their expression (2.24). We compute the two-point function of the scalar field source analogously (see also the detailed derivation provided in~\cite{Barnaby:2011vw} for the case of constant $H$ and $\xi$). Inserting these correlators into 
~\eqref{C-X}, we obtain 
\begin{align}
{\cal C}_\lambda^{(m)} =& \frac{k^{3-2m}}{16 \pi^2 M_p^4} \int \frac{d^3 p}{\left( 2 \pi \right)^3} \left( 1 + \lambda \, {\hat k} \cdot {\hat p} \right)^2 \left( 1 + \lambda \, {\hat k} \cdot {\hat q} \right)^2 \times \nonumber\\
& \left\vert \int_{-\infty}^\tau d \tau' \, \frac{1}{a \left( \tau' \right)} \, \frac{\partial^m}{\partial \tau^m} \left( \frac{{\tilde G}_{\lambda,k} \left( \tau ,\, \tau' \right)}{a \left( \tau \right)} \right) \; \left[ A_+' \left( \tau' ,\, p \right) A_+' \left( \tau' ,\, q \right) + p \, q \, A_+ \left( \tau' ,\, p \right) A_+ \left( \tau' ,\, q \right) \right] \right\vert^2_{\vec{q} = \vec{k} - \vec{p}} \;,
\label{C-lambda}
\end{align}
for the GW correlator, and
\begin{align}
{\cal C}_\phi^{(m)} =& \frac{k^{3-2m}}{16 \pi^2 f^2 M_p^2} \int \frac{d^3 p}{\left( 2 \pi \right)^3} \left( 1 - {\hat p} \cdot {\hat q} \right)^2 \times 
\nonumber\\
& \left\vert \int_{-\infty}^\tau d \tau' \, \frac{1}{a \left( \tau' \right)} \, \frac{\partial^m}{\partial \tau^m} \left( \frac{{\tilde G}_{\phi,k} \left( \tau ,\, \tau' \right)}{a \left( \tau \right)} \right) \; \left[ A_+' \left( \tau' ,\, p \right) \, q \, A_+ \left( \tau' ,\, q \right) + A_+' \left( \tau' ,\, q \right) \, p \, A_+ \left( \tau' ,\, p \right)  \right] \right\vert^2_{\vec{q} = \vec{k} - \vec{p}} \;,
\label{C-phi}
\end{align}
for the correlator of the inflaton perturbations (we recall that $m$ is taken to be either $0$ or $1$ in these expressions). 

\section{SGWB observables and constraints}
\label{sec:SGWB}

In this work we are interested in the sourced GW energy density during inflation, that we compute, through eqs.~\eqref{rho-GW} and~\eqref{rho-code}. We are also interested in their fractional energy density today
\begin{equation}
\Omega_{\rm GW} \left( f \right) \equiv \frac{1}{\rho_{\rm crit,0}} \, \frac{d \rho_{\rm GW,0}}{d \ln f} \;,  
\label{OmGW}
\end{equation}
where $\rho_{\rm crit}$ is the critical energy density of a flat universe and the suffix $0$ denotes the present time. The frequency $f$ and the fractional energy density $\Omega_{\rm GW}$ are related to the wavenumber $k$ and the primordial power spectrum $P_\lambda$ by, respectively, eq.~\eqref{f-kt} and~\eqref{OmGW-Pla}. 

As in~\cite{Garcia-Bellido:2023ser}, we consider the limits in the $f-\Omega_{\rm GW} \, h^2$ plane (where $h$ is the rescaled current Hubble rate from the standard relation $H_0 \equiv 100 h {\rm km/s/Mpc}$) computed in Ref.~\cite{Schmitz:2020syl}, plus the forecasted sensitivity curve for THEIA~\cite{Garcia-Bellido:2021zgu}. We also include some constraints not directly studied in~\cite{Garcia-Bellido:2023ser}. 

First of all, we consider the Big-Bang Nucleosynthesis (BBN) bound on the energy density of the SGWB modes that were inside the horizon at the BBN time (corresponding to present frequencies $f \geq f_{\rm BBN,0} \simeq 1.5 \cdot 10^{-12} \, {\rm Hz}$~\cite{Caprini:2018mtu}). Translated to the current energy density this gives~\cite{Caprini:2018mtu} 
\begin{equation}
\int_{f_{\rm BBN,0}} \frac{d f}{f} \, \Omega_{\rm GW} \left( f \right) h^2 \lesssim 5.6 \cdot 10^{-6} \,  \Delta N_{\rm eff} \simeq 1.8 \cdot 10^{-6} \;, 
\label{Delta-N-eff}
\end{equation} 
where we have employed the $2 \sigma$ limit $\Delta N_{eff} \lesssim 0.33$ of Ref.~\cite{Planck:2018vyg} on the number of additional effective neutrino species. 

We do not consider limits on the integrated SGWB energy density as they depend on its statistics~\cite{Caprini:2018mtu}, which is unclear for the sourced background produced in the regime of strong backreaction. Concerning the CMB limits, we impose that the sourced signal is extremely small at the largest scales ($\xi \lesssim 2.5$), so to respect the stringent non-Gaussianity constraint~\cite{Barnaby:2010vf,Planck:2015zfm}  on the sourced scalars. As a consequence, the CMB limit $r \lesssim 0.03$~\cite{Tristram:2021tvh,Galloni:2022mok} on the tensor-to-scalar ratio is imposed on the vacuum modes. We also verified that the scalar power spectra obtained at relatively smaller CMB scales (up to the present wavenumber $k \sim 5 \, {\rm Mpc}^{-1}$) in all the examples studied in the next section are well below the upper bound reported in
Figure 20 (bottom panel) of Ref.~\cite{Planck:2018jri}.

Finally, we consider limits on sourced scalar modes from CMB spectral distortions. A guaranteed contribution to these distortions comes from the energy released due to Silk damping~\cite{Silk:1967kq} of primordial small-scale perturbations after horizon re-entry~\cite{Sunyaev:1970plh,Daly:1991uob,1991MNRAS.248...52B,Hu:1994bz}. These deviations from a pure black-body spectrum are typically parametrized in terms of $\mu-$ and $y-$distortions, related to the power spectrum of primordial modes by window functions~\footnote{We do not consider the distortions caused by the tensor modes, as they are typically subdominant~\cite{Chluba:2014qia,Putti:2024uyr}.}
\begin{equation}
\mu = \int_{k_{\rm min}}^\infty d \ln k \, P_\zeta \left( k \right) \, W_\mu \left( k \right) \;\;,\;\; 
y = \int_{k_{\rm min}}^\infty d \ln k \, P_\zeta \left( k \right) \, W_y \left( k \right) \;,
\label{mu-y}
\end{equation} 
with $k_{\rm min} \simeq 1 \, {\rm Mpc}^{-1}$, and where we use the analytic approximation for the window functions developed in Refs~\cite{Chluba:2012gq,Chluba:2012we,Chluba:2015bqa,Lucca:2019rxf} and summarized in~\cite{Putti:2024uyr} (see their Figure 1), where we also refer the reader for the precise definition of the $\mu$ and $y$ parameters, as well as for a more extended discussion. To date, the most precise measurement of the CMB spectrum is provided by COBE/FIRAS, which constrained $\mu \leq 9 \cdot 10^{-5}$ and $y \leq 1.5 \cdot 10^{-5}$ at a $95\%$ confidence level (CL)~\cite{Fixsen:1996nj}. A later analysis by~\cite{Bianchini:2022dqh} improved the $\mu-$distortion limit to $\mu \leq 4.7 \cdot 10^{-5}$ at a $95\%$ CL, thanks to an improved component separation and Bayesian analysis approach. In the last decade there has been an intense ongoing discussion about potential future missions (PIXIE~\cite{Kogut:2010xfw}, PRISM~\cite{PRISM:2013fvg}, COSMO~\cite{Masi:2021azs}, BISOU~\cite{Maffei:2021xur}) that might detect $\mu-$ and $y-$distortions down to the $\left[ 10^{-9} - 10^{-8} \right]$ range~\cite{Chluba:2019kpb,Chluba:2019nxa}.

The values of $\Delta N_{\rm eff} ,\, \mu ,$ and $y$ associated with the evolution studied in Subsection~\ref{subsec:baseline-parameters} are reported in that subsection. Those associated with the evolusions shown in Subsection~\ref{subsec:vary-parameters} are reported in Table~\ref{tab:param-res} of Appendix~\ref{app:parameters}.

\section{An example from the literature}
\label{sec:example}

In this section we study the potential detection and constraints on the SGWB produced by a specific example of axion inflation with coupling to gauge fields. We consider the potential employed in~\cite{Garcia-Bellido:2023ser} and given in Appendix~\ref{app:parameters}. The potential is characterized by three `main' regions (first, third, and fifth lines of eq.~\eqref{potential}, respectively). The CMB modes are produced while the inflaton is in the first `main' region, which has the shape of a quartic hill-top potential, arranged to be sufficiently flat so to give a slow motion of the inflaton and negligible gauge fields amplification. Therefore, at CMB scales there is no appreciable backreaction of the gauge fields on the inflaton dynamics, and the tensor and scalar perturbations coincide with the vacuum ones. In the second and third ``main'' regions the potential is linear, with slope arranged so that the gauge field production is relevant in the second `main' region, and negligible~\footnote{This avoids excessive production of PBHs~\cite{Linde:2012bt,Garcia-Bellido:2016dkw} and of high frequency GWs~\cite{Adshead:2019lbr,Adshead:2019igv}, and it respects the existing ground‑based interferometer bounds \cite{KAGRA:2021kbb}.} in the third one. Each `main' region is joined with the adjacent one(s) by a section in which the potential is quadratic, with parameters chosen so that the potential and its first derivative are continuous everywhere. For definiteness, we impose that inflation ends $60$ e-folds after the starting value of the inflaton in the simulation. This in general requires a further modification of the potential beyond what we provide (for instance, a final quadratic part with a sufficiently small curvature so to avoid overproduction of GW at this stage~\cite{vonEckardstein:2025oic}), or the addition of new degrees of freedom, that become relevant at this point, causing inflation to end. We do not address this point specifically, as this is not the focus of our study. The main purpose of the potential~\eqref{potential} is to provide an intermediate and controlled region, so that in this region, and only there, gauge field amplification is relevant~\cite{Garcia-Bellido:2023ser}.~\footnote{As the potential and its dreivative are continuous, this ``intermediate region'' consists in the second ``main'' region described above as well as portions of the two regions that surround it.} We divide our study in two parts. In Subsection~\ref{subsec:baseline-parameters} we adopt the same choice of parameters as in Ref.~\cite{Garcia-Bellido:2023ser} and investigate whether their results are compatible with the approximation of homogeneous backreaction. In Subsecton~\ref{subsec:vary-parameters} we then discuss the effect of varying the key model parameters. 

\subsection{`Baseline' parameters}
\label{subsec:baseline-parameters}

Let us consider the specific example studied in~\cite{Garcia-Bellido:2023ser}, characterized by the axion decay constant $f=M_p/57$ and by the potential parameters specified in the caption of Figure~\ref{fig:potential}. The background fields evolution for this model and choice of parameters is presented in Section~4 of~\cite{Garcia-Bellido:2023ser}, and it is characterized by an oscillatory behavior of the inflaton speed during the stage of strong homogeneous backreaction. This gives rise to peaks in the GW spectrum (for modes that were produced while the inflaton speed was at a maximum), as shown in their Figure 4.

However, the work in~\cite{Garcia-Bellido:2023ser} lacked a study of the validity of the homogeneous backreaction approximation (employed in the background simulation), which we address now. As we already wrote, we propose a criterion for the validity of this regime by comparing the spatial gradient contribution of the inflaton inhomogeneities energy density (computed at first order in perturbation theory) with the kinetic energy of the inflaton zero mode.
We consider the phenomenological results to be unreliable for modes that cross the horizon after the spatial gradient energy of the inflaton becomes a significant fraction of the kinetic energy of the inflaton zero mode. As discussed in the Introduction, we estimate that “significant” corresponds to a few percent.

\begin{figure}[h!]
\centering
\includegraphics[width=0.75\linewidth]{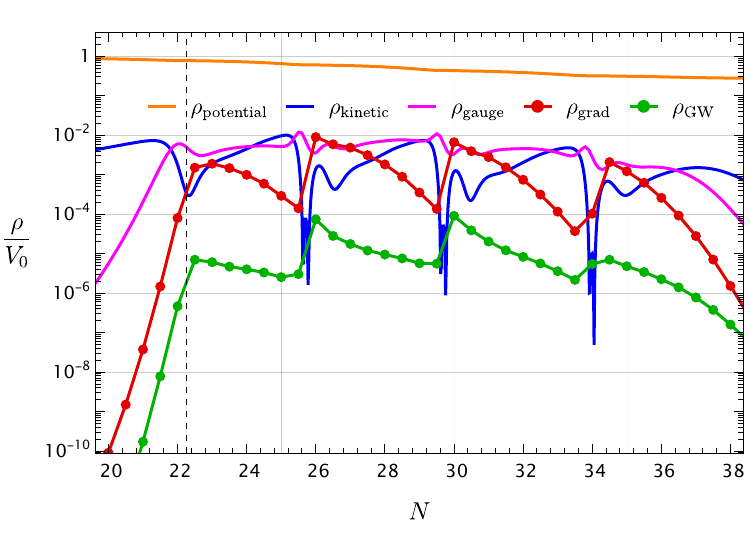}
\caption{Evolution as a function of the number of e-folds of various contributions to the energy density. From top to bottom (referring to the left portion of the figure): the dominant (orange) curve is the potential energy of the inflaton; the second (blue) curve is the kinetic energy of the inflaton zero mode; the third (magenta) curve is the energy density in gauge fields; the fourth (red) curve is the gradient energy density of the inflaton perturbations; the fifth (green) curve is the GW energy density. The (black) vertical dashed line indicates the moment at which the gradient energy of the inflaton perturbations becomes equal to the kinetic energy of the zero mode.}
\label{fig:EnergyDensities_57_495}
\end{figure}

To this purpose, we evaluated (using the relations in Appendix~\ref{app:rho}) the various components of the energy densities of the fields in the model, and show them in Figure~\ref{fig:EnergyDensities_57_495}. As we can see in the figure, the potential energy density dominates throughout the evolution, confirming the validity of the slow-roll regime during the entire inflationary phase; the kinetic energy, by contrast, exhibits an oscillatory behavior in agreement with that of $\xi$ in Figure 2 of~\cite{Garcia-Bellido:2023ser}. The sourced fields $h_\pm$ and  $\delta\phi$ are set to zero at the beginning of the simulation, and, therefore, their energy densities start to grow only when of the gauge field grows, which in turn happens when the inflaton potential becomes sufficiently steep. Their energy density exhibits an exponential growth starting from $N \gtrsim 20$, which, after a few e-folds, leads to an oscillatory behaviour. The oscillations are synchronized (with some delay) with those of the inflaton velocity: since the gauge fields are amplified by the motion of the inflaton, the increase of $\rho_{\rm kinetic}$ is followed by an increase of $\rho_{\rm gauge}$; the backreaction of the gauge fields causes in turn the decrease of $\rho_{\rm kinetic}$; at the beginning of this decrease, the gauge fields are still increasing due to a memory effect
\cite{Domcke:2020zez,Peloso:2022ovc} (as discussed in the Introduction); eventually the prolonged decrease of the inflaton velocity shuts down the production of gauge fields, and the expansion of the universe dilutes their energy density; this decreases the backreaction, and the inflaton speeds up again. This gives rise to the pattern of synchronized oscillations. As the gauge field source inflaton perturbations and GW, also the latter exhibit an oscillatory behaviour, synchronized with that of $\rho_{\rm kinetic}$ and $\rho_{\rm gauge}$, with a further small delay. 

\begin{figure}[h!]
\centering
\includegraphics[width=0.7\linewidth]{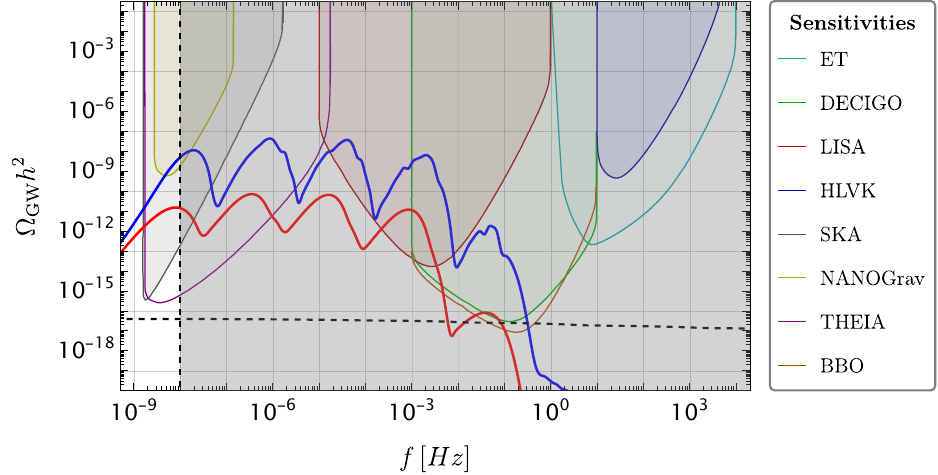}
\caption{The figure shows the fractional energy density of the vacuum (nearly horizontal black dashed curve) and sourced (dominant blue and subdominant red curve for the left and red helicities, respectively) SGWB, computed under the assumption of homogeneous backreaction, compared to the sensitivity of various  observatories.  The shaded region beyond the vertical dashed line represents the frequency spectrum produced after the breakdown of the homogeneous backreaction approximation.}
\label{fig:GW_57_495}
\end{figure}

As we already remarked, the potential is designed so to have negligible sourced perturbations at CMB scales and agree with CMB phenomenology (see Appendix~\ref{app:parameters}). Concerning the other limits discussed in Section~\ref{sec:SGWB}, the total energy density associated with the sourced GW shown in Figure~\ref{fig:GW_57_495} results in $\Delta N_{\rm eff} \simeq 2 \cdot 10^{-2}$ in Eq.~\eqref{Delta-N-eff}, well below the current bounds. Concerning instead spectral distoritions, the scalar perturbations evaluated from this example give $\mu = 10^{-9}$ and $y= 8 \cdot 10^{-10}$ in eq.~\eqref{mu-y}, namely at a level slightly below the target range of the potential future missions mentioned in the previous section.

We stress however that these results are obtained under the assumptions that the produced GWs and inflaton perturbations do not affect the dynamics of the inflaton zero mode and gauge field. While our results show that this is true for the sourced GWs, we cannot safely assume that this is the case also for the inflaton perturbations, whose gradient energy density becomes comparable to $\rho_{\rm kinetic}$ at $N \simeq 22$. This presumably invalidates the results from this moment on, which are based on the homogeneous backreaction approximation, in an analogous way to what is observed in the examples obtained from the full lattice simulations starting from Ref.~\cite{Figueroa:2023oxc,Figueroa:2024rkr}. This has a dramatic effect for the GW phenomenology obtained in this example. Figure~\ref{fig:GW_57_495} shows the final GW spectrum produced in the model, under the assumption of homogeneous backreaction. The curves reproduce with excellent agreement Figure~4 of~\cite{Garcia-Bellido:2023ser}. However, we now question the validity of the spectrum for modes produced after $N \sim 22$. This corresponds to frequencies $f \gtrsim 6 \cdot 10^{-9}$ Hz, which cover the vast majority of the figure. 

We thus conclude that we cannot rely on the validity of the results in the specific example studied in Ref.~\cite{Garcia-Bellido:2023ser}, apart from the rise of the first peak in the outmost left portion of the figure. In the following subsection we study how this changes for different choices of the parameters of the model. 

\subsection{Changing the model parameters}
\label{subsec:vary-parameters}

In the model that we are considering, significant inflationary amplification of gauge fields only occurs while the inflaton is in the intermediate region of the potential. This choice, and the simple form of the potential in this region, allow us to immediately characterize the  relevant characteristics of the model and to study how they impact the phenomenological results. The key parameters are:

\begin{itemize}
\item the axion decay constant $f$; 
\item the slope of the intermediate region;
\item the width of the intermediate region $\Delta \phi_{2,3} \equiv \phi_3 - \phi_2$;
\item the position of the intermediate region within the full potential. 
\end{itemize}

\begin{figure}[h]
\hspace*{-0.8cm}
\includegraphics[width=1.1\linewidth,height=0.25\linewidth]{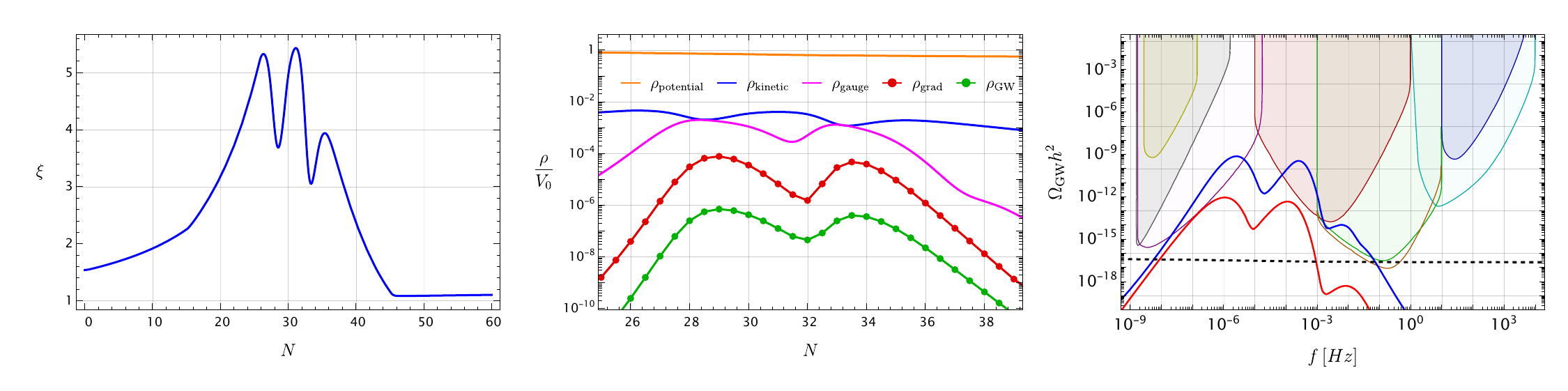}%
\caption{Results for a flatter inflaton potential in the intermediate region (slope $v'=-0.1516$ rather than $-0.2822$), while all the other parameters coincide with those of the `baseline' model that was used in the two previous figures. Left panel: evolution of the parameter $\xi$ controlling the gauge‑field amplification. Central panel: evolution of various energy density components. Right panel: present GW fractional energy density. The main difference with the results in Figure~\ref{fig:EnergyDensities_57_495} is that now the inflaton spatial gradient energy remains significantly smaller than the kinetic energy for the whole evolution.}
\label{fig:Results_57_27}
\end{figure}

Let us begin by discussing the impact of a variation in the parameter $v' \equiv \frac{M_p}{V_0} \, \frac{\partial V}{\partial \varphi}$ (where $V_0$ is the initial value of the potential) controlling the slope of the potential in the intermediate region, characterized by significant gauge field amplification. We expect that decreasing the absolute value of the slope reduces the inflaton speed, therefore decreasing the gauge field amplification and the sourced primordial perturbations. Starting from the 
`baseline' parameters studied in the previous subsection, a set of simulations in which we varied only~\footnote{When in this subsection we specify which parameters of the potential we vary or leave fixed, we do not refer to the parameters $p_1 - p_8$, which are varied or kept fixed so as to guarantee that the potential and its derivative are continuous everywhere.} $v'$ shows indeed a decrease of the production as the steepness of that potential region is lowered. This shifts to later times (greater values of $N$) the moment in which the gradient energy of the inflaton perturbations reaches or becomes a significant fraction of the kinetic energy of the inflaton zero mode, namely the moment of breakdown of the  homogeneous backreaction.~\footnote{More properly, our estimate for this moment. We stress once again that we rely on the assumption that the comparison of the total energy in the perturbations vs. the kinetic energy of the zero mode is a valid indicator of the breakdown of the homogeneous backreaction. As we already wrote, this needs to be checked with full lattice simulations. We do not repeat this caveat further, but it applies to all the following discussion.} In Figure~\ref{fig:Results_57_27} we show one example in which the decrease of $\left\vert v' \right\vert$ leads to an evolution for which the ratio between the axion gradient and (zero mode) kinetic energy never exceeds $\simeq 4 \%$.~\footnote{The maximum ratio between the axion gradient and kinetic energy attained in any of the simulations discussed in this section is reported in Table~\ref{tab:param-res} of Appendix~\ref{app:parameters}.} The SGWB spectrum exhibits two dominant and one smaller potentially detectable peaks, all arising within a regime of reliable homogeneous backreaction.

Let us now turn our attention to the axion decay constant $f$,~\footnote{The decay constant is limited from below to $f \gtrsim 10^{-2} \, M_p$ to avoid overproduction of sourced scalar perturbations at CMB scale (the precise limit being dependent on the inflaton potential), and from above to $f \lesssim M_p$ by theoretical considerations.} that is inversely proportional to the strength of the interaction between the axion and the gauge field. Similarly to the decrease of the slope of the potential, an increase of $f$ also reduces the gauge field amplification (we note that both choices reduce the parameter $\xi \equiv \frac{\dot{\phi}}{2fH}$). 

\begin{figure}[h]
\hspace*{-0.8cm}
\includegraphics[width=1.1\linewidth,height=0.25\linewidth]{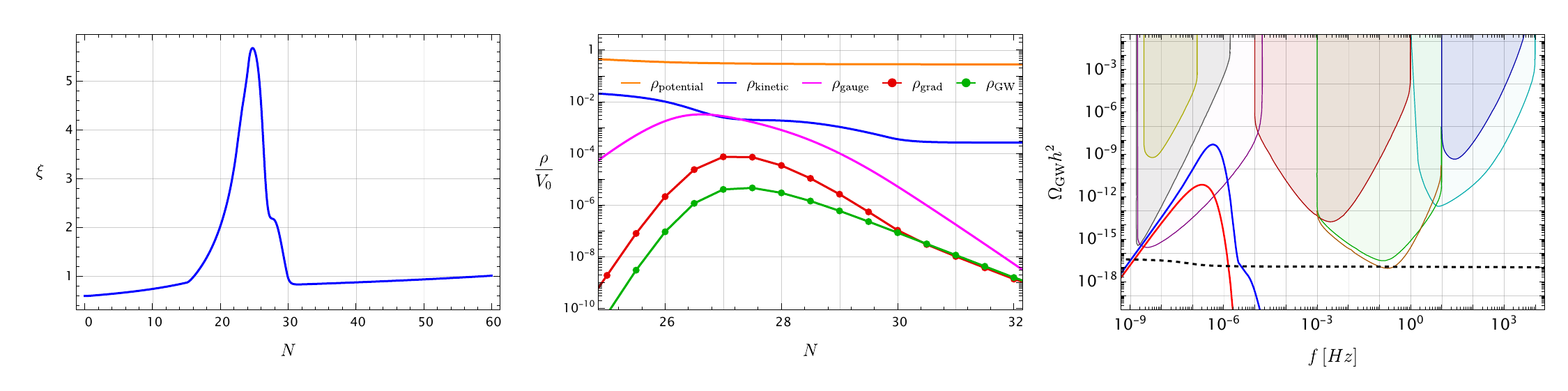}%
\caption{Results for a decreased inflaton-gauge interaction with respect to the `baseline' model (${\tilde f}^{-1} = 22$ rather than $57$), while the potential is unchanged. The panels are organized as in the previous figure.}
\label{fig:Results_c495}
\end{figure}

Starting from the `baseline' parameters, we decreased the inverse of the decay constant without changing the inflaton potential,~\footnote{This requires changing the parameters in the rescaled potential integrated numerically according to eq.~\eqref{kappa}.} noticing indeed a decrease of the gradient energy. Figure~\ref{fig:Results_c495} shows results for the specific choice ${\tilde f}^{-1} \equiv \frac{M_p}{f} = 22$, where we see that the gradient energy indeed remains significantly smaller than the kinetic one throughout the inflationary evolution. 

\begin{figure}[h]
\hspace*{-0.8cm}
\includegraphics[width=1.1\linewidth,height=0.25\linewidth]{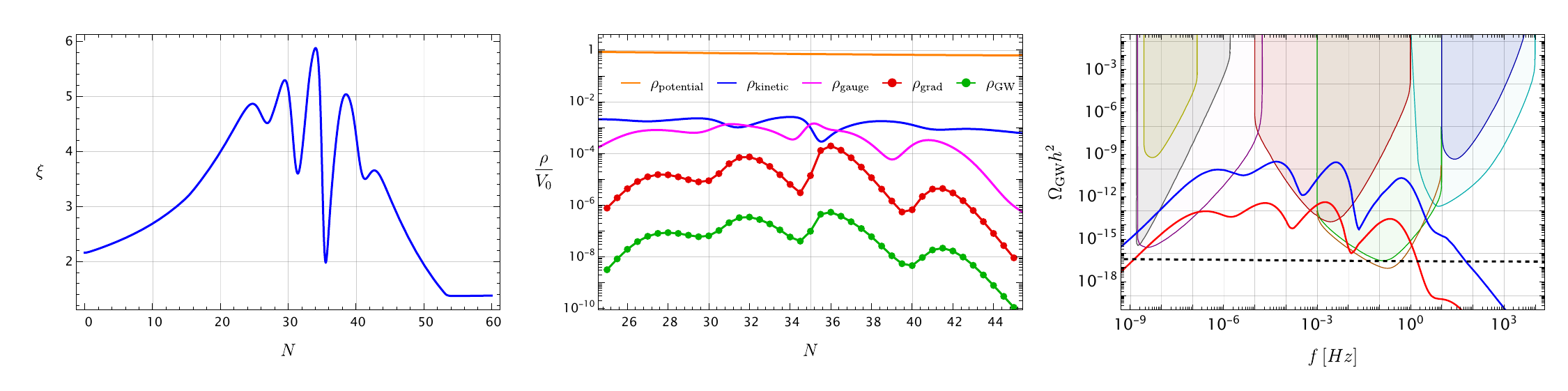}\\%
\hspace*{-0.8cm}
\includegraphics[width=1.1\linewidth,height=0.25\linewidth]{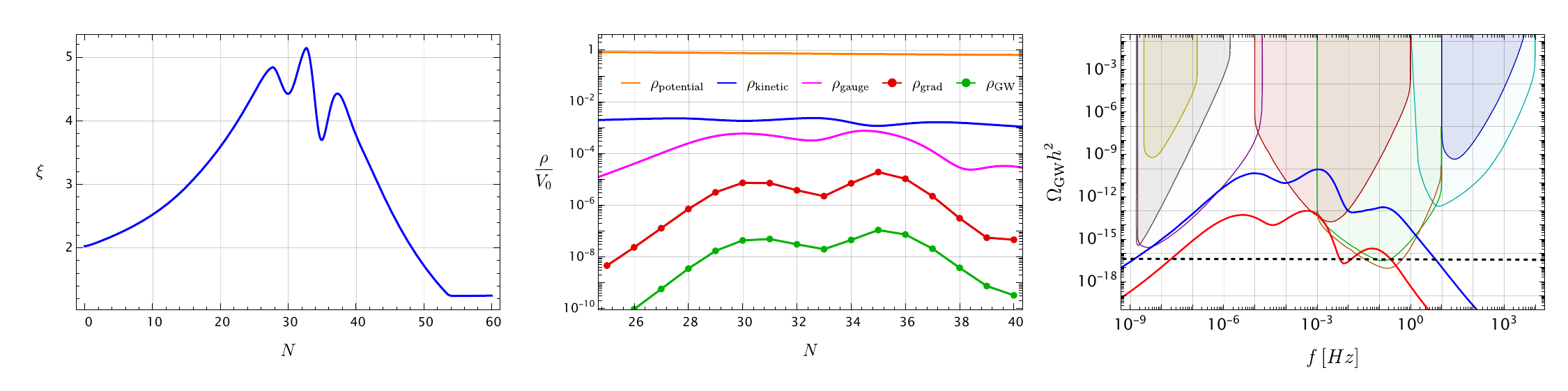}%
\caption{Results for an increased inflaton-gauge interaction (${\tilde f}^{-1} = 80$ for the top row and $75$ for the bottom row, in contrast to $57$ in the ``baseline'' model) and flatter potential in the intermediate region (slope $v' = -0.1237$ for the top row and $-0.1129$ for the bottom row, in contrast to $-0.2822$ in the ``baseline'' model). The panels are organized as in the two previous figures.
}
\label{fig:Results_f_c5}
\end{figure}

A main qualitative difference between this and the two previous cases is that Figure \ref{fig:Results_c495} is characterized by only one marked oscillation of $\xi$, and, correspondingly, only one marked peak in the SGWB spectrum. This is due to the fact that, since the slope of the potential is unchanged, while the strength $f^{-1}$ of the inflaton-gauge coupling is decreased, the gauge field production is now reduced. Corespondingly, backreaction is reduced, and the inflaton spends less time in the intermediate region of the potential. Once it exits this region, gauge field amplification is effectively shut off, and no further oscillations can develop. To contrast this, we show in Figure~\ref{fig:Results_f_c5} the results of two simulations where we varied both the coupling strength $f^{-1}$ and the slope of the potential $v'$ in the intermediate region.~\footnote{All the other parameters of the rescaled potential are changed according to eq.~\eqref{kappa}, corresponding to keeping them invariant in the model potential.} The evolution of the parameter $\xi$ and the SGWB spectrum now indeed show several peaks, all produced while the gradient axion energy is smaller than the kinetic one. 

More specifically, the two rows of the figure show the results of two different evolutions, to provide a measure of the sensitivity to the model parameters. Parameters in the top row have been chosen so that the spatial gradient energy is only slightly smaller than the kinetic one; on the contrary, in the evolution shown in the second row the gradient energy remains about two order of magnitude smaller than the kinetic energy. The SGWB background produced in the second evolution is only slightly smaller than that produced in the first evolution, and above detection. 

\begin{figure}[h]
\hspace*{-0.8cm}
\includegraphics[width=1.1\linewidth,height=0.25\linewidth]{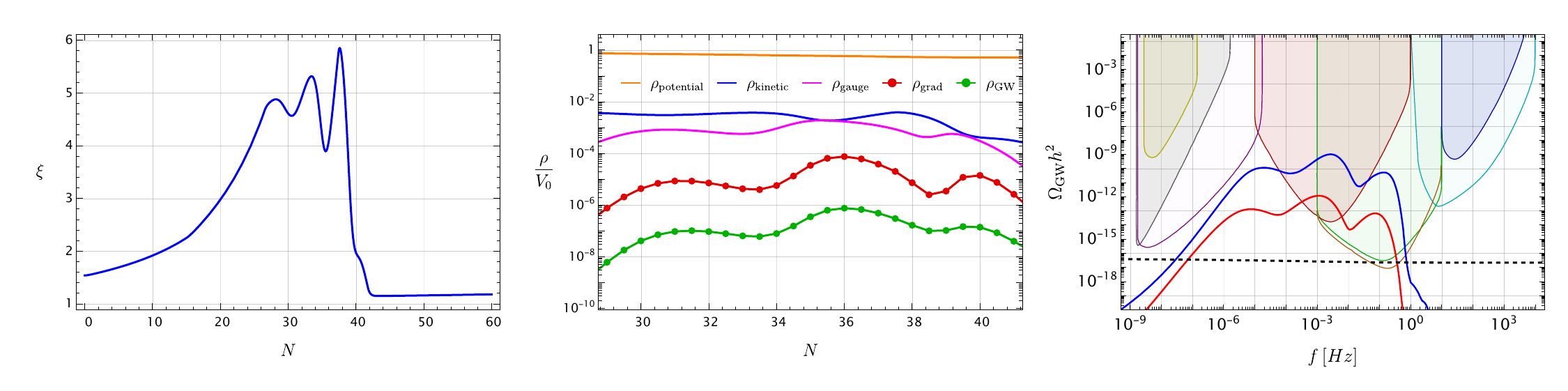}%
\caption{Results for an increased width ($\Delta {\tilde \phi}_{2,3} = 104$, rather than $35$), and for a flatter potential (slope $v' = -0.1351$ rather than $-0.2822$) in the intermediate region with respect to the `baseline' model. The panels are organized as in the three previous figures.}
\label{fig:Results_Deltaphi_c5}
\end{figure}

Another way to increase the number of oscillations is to increase the width of the intermediate region, so that the inflaton spends more time in the region of the potential associated with significant gauge field amplification. We do this in the simulation shown in Figure \ref{fig:Results_Deltaphi_c5}, where we increased the width $\Delta \phi_{2,3} \equiv \phi_3 - \phi_2$ by setting ${\tilde \phi}_2 \equiv \frac{\phi_2}{f} = 186$ (as in the `baseline' model) and ${\tilde \phi}_3 = 290$ (rather than the `baseline' model value $221$). Increasing $\Delta \phi_{2,3}$ also increases the overall gauge field production. We compensated for this by decreasing the slope of the potential of the intermediate region, while the axion decay constant $f$ and the other model parameters are kept as in the `baseline' model.

\section{Peaks in $\Omega_{\rm GW} \left( f \right)$}
\label{sec:peaks}

\begin{figure}[h]
\centering
\includegraphics[width=0.7\linewidth]{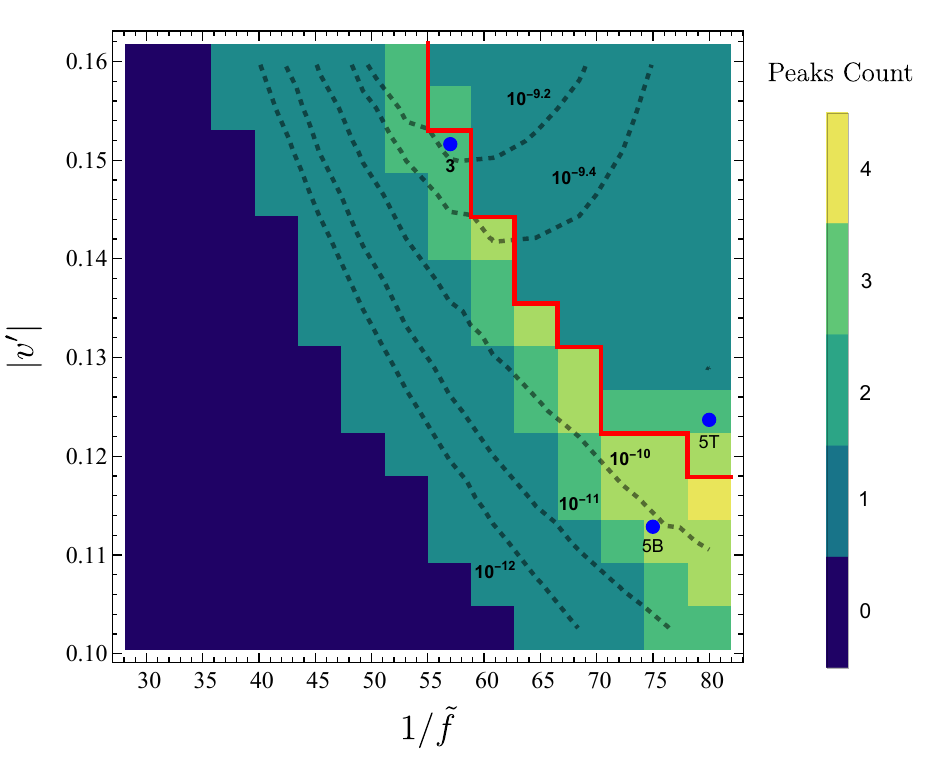}
\caption{The colours indicate the number of peaks in $\Omega_{\rm GW} h^2$ greater than $10^{-14}$, produced in the regime of homogeneous backreaction (see the text for details). The region below the (red) solid line is characterized by parameters for which the homogeneous backreaction assumption holds all throughout inflation. The five superimposed (black) dashed contour lines indicate the value of $\Omega_{\rm GW} h^2$ in the highest peak. The three marked points refer to the evolutions presented in Figures~\ref{fig:Results_57_27} and~\ref{fig:Results_f_c5}. 
}
\label{fig:peaks}
\end{figure}

As discussed, local maxima in the inflaton velocity result in localized peaks in the SGWB spectrum~\cite{Garcia-Bellido:2023ser}. We are interested in exploring how changes of the coupling of the inflaton to gauge fields (parameter ${\tilde f}^{-1}$) and of the slope of the inflaton potential in the intermediate region (parameter $v'$) affect the number of peaks obtained in the regime of validity of the homogeneous backreaction.~\footnote{The number of peaks can also be increased by increasing the width of the region of the potential with significant gauge field amplification, as shown in the example of Figure~\ref{fig:Results_Deltaphi_c5}. We do not elaborate on this possibility in this section. More in general, the other model parameters are kept fixed as in the `baseline' model, up to the rescaling~\eqref{kappa}.} To do so, we consider a two-dimensional grid with $14$ equally spaced values of the coupling and of the slope parameter. For each point in this grid we compute the background evolution and sourced modes under the assumption of homogeneous backreaction, and count the number of relative maxima in the SGWB spectrum having an amplitude $\Omega_{\rm GW} h^2 > 10^{-14}$. We stop the evolution (and, thus, the peak counting) whenever the ratio between the axion gradient and (zero mode) kinetic energy overcomes the $10 \%$ threshold. If this never occurs, the evolution is continued for the full $60$ e-folds of inflation, with the sequence of peaks ceasing when the inflaton exits from the intermediate region of the potential. Clearly, there is a degree of arbitrariness in the $10^{-14}$ and $10\%$ choices, but taking slightly different values would not alter the qualitative features of Figure~\ref{fig:peaks} and the present discussion. 

Regions in parameter space characterized by the same peak count are shown with the same color in Figure~\ref{fig:peaks}. Dashed lines in the figure are isocontours of the value of $\Omega_{\rm GW} h^2$ in the highest peak obtained in the evolution. Finally, the three marked points refer to the evolutions shown in Figure~\ref{fig:Results_57_27} (denoted here as ``3'') and in the top (``5T'') and bottom (``5B'') panel of Figure~\ref{fig:Results_f_c5}. 

Let us now discuss the results shown in the figure. The bottom-left portion of the figure corresponds to relatively small coupling and small slope. This is where backreaction is weakest, and no localized maximum appear in the evolution of the inflaton speed. The number of maxima increases when, starting from this corner, we increase either the coupling (namely, if we move towards the right portion of the figure) or the slope (namely, if we move towards the higher part of the figure). However, this increase does not continue indefinitely, as the height of the peaks also increases (as indicated by the dashed isocontours) as we move in these directions, so that the system reaches the threshold indicating the breakdown of the homogeneous approximation. This threshold is indicated with a (red) solid line in the figure. Not surprisingly, the largest number of peaks is obtained in the `intermediate' portion of the figure, where the gauge amplification is both sufficiently strong to lead to relevant backreaction, and sufficiently weak to preserve the homogeneous approximation.  

The five (black) dashed lines in the figure are contours of equal $\Omega_{\rm GW} h^2$ evaluated in the highest peak for the corresponding evolution. Naively, we expect that these lines give a slope that is a decreasing function of the coupling. This is because, generally speaking, the amount of sourced GW increases with both increasing coupling and slope. Therefore, if we fix the amount of production, an increase of the coupling must be accompanied by a decrease of the slope. This expected behaviour is confirmed in the region below the solid (red) line, for which the gradient energy never exceeds the $10\%$ fraction of the kinetic energy. On the contrary, the contours of equal $\Omega_{\rm GW} h^2$ are characterized by increasing coupling and slope in the portion on the top of the (red) solid line. As explained above, we only consider peaks generated before the breakdown of the homogeneous backreaction. This breakdown occurs progressively earlier as we move towards the top-right corner of the figure, so that increasing both parameters in this region does not results in an increase of the value of the highest peak produced before the breakdown of the approximation. We note, however, that the variation of the maximum height in this portion of the figure is much smaller than that in the portion below the solid (red) line. 

Finally, let us compare the results for the three marked points with those in the corresponding earlier figure. According to the contourplot, the evolution in Figure~\ref{fig:Results_57_27} should be characterized by $2$ peaks above the $10^{-14}$ threshold, with a maximum height slightly above $\Omega_{\rm GW} h^2 \simeq 6 \cdot 10^{-10}$, all generated in the regime of homogeneous backreaction. The height and the persistance of the approximation agree with what effectively observed in Figure~\ref{fig:Results_57_27}. We note, however, that the spectrum of the figure has actually a third peak, slightly above the threshold. This peak is slightly below threshold in the evolution used for the contourplot closest to the parameters of Figure~\ref{fig:Results_57_27}. Concerning instead Figure~\ref{fig:Results_f_c5}, according to the contourplot the evolution in the top panel should be characterized by either $2$ or $3$ peaks (we cannot determine the actual value between the two, due to the finite resolution of the grid used to generate the contourplot), with a top peak between $10^{-10}$ and $4 \cdot 10^{-10}$, and with a ratio $\rho_{\rm grad} / \rho_{\rm kin}$ that overcomes the $10\%$ threshold; instead, the evolution in the bottom panel should be characterized by either $2$ or $3$ peaks, with a top peak slightly below $10^{-10}$, and without ever exceeding the $10\%$ value. All this agrees with the results shown in the two panels of Figure~\ref{fig:Results_f_c5}.

\section{Conclusions}
\label{sec:conclusions}

Axion/natural inflation constitutes a well-motivated class of inflationary models, as the shift symmetry protects the potential from large radiative corrections~\cite{Freese:1990rb,Adams:1992bn}. However, the shift symmetry alone does not provide guidance regarding the form of the inflaton potential, which emerges once the symmetry is broken. The only viable means of determining it is through experimental observations. While CMB and Large-Scale Structure (LSS) measurements are in excellent agreement with the inflationary paradigm, they directly probe only about $\sim 7$ e-folds of inflation. A substantially broader range could, in principle, be accessed through measurements of the stochastic gravitational wave background (SGWB), an area that is currently experiencing rapid and significant progress. Unfortunately, vanilla slow-roll inflationary models predict a vacuum gravitational wave signal that is too weak to be detected in the near future. While the shift symmetry does not directly determine the inflaton potential, it does constrain the inflaton couplings to other fields. In particular, a pseudo-scalar interaction between the inflaton and a gauge field can generate detectable scalar and tensor perturbations across a wide range of scales.

The amplitude of these sourced signals is exponentially sensitive to the ratio $\xi \propto \frac{\dot{\phi}}{f}$ between the inflaton speed and the coupling to gauge fields. The former is sensitive to the “local” value of the potential, namely the value and shape it has in the neighborhood where the inflaton is situated while the modes probed by any given experiment are produced (which occurs around horizon exit). Therefore, when attempting to directly infer the local shape of the potential from a given observation, it is preferable to adopt an `agnostic' parametrization of the potential, to avoid constraints that the phenomenology of one part of the potential might impose on another in any specific model. This is the working procedure adopted in this work, which aims to study the possible SGWB generated in this model at scales probed by PTA, astrometry, and existing and planned interferometers. To this end, we considered an inflaton potential with a sufficiently small slope at CMB scales and in the final stages of inflation, allowing for a greater slope only at intermediate scales, where the potential is assumed to be linear. While we refrain from a full quantitative exploration of this potential (which clearly lacks a theoretical motivation), we are able to discuss how the phenomenology of the model is affected by the main qualitative features of this parametrization: the width and the position of the region where significant production occurs, the slope of the potential in this region, and the strength $f^{-1}$ of the pseudo-scalar interaction.

The lack of a motivated model, and in particular of the end-of-inflation/reheating stage (which we assume to occur via a modification of the potential beyond our scope), constitutes a first limitation of our work. A second, more relevant limitation is the approximate treatment of the backreaction of the produced gauge fields on the axion inflaton dynamics. While analytic studies of this mechanism date back more than $15$ years, only in the last $\sim 3$ years have full solutions of the system appeared in the literature using lattice simulations. These numerical integrations are extremely computationally expensive. Just to provide an example, the highly advanced solutions obtained in~\cite{Figueroa:2023oxc,Figueroa:2024rkr} for the regime of strong backreaction cover ``only'' about the last $\sim 7$ e-folds of inflation. To simulate the full duration of inflation, we resort to an approximate scheme in which the inflaton is taken to be homogeneous. Within this approximation, when backreaction becomes relevant, the inflaton speed exhibits an oscillatory behaviour with a period of about $4$–$5$ e-folds. This gives rise to a highly characteristic pattern of correlated peaks in the SGWB spectrum, which could be detected in the same or in different GW observatories.

The most recent lattice simulations challenge this result. For instance, a much more chaotic picture emerges from the examples of~\cite{Figueroa:2023oxc,Figueroa:2024rkr}, in which typically only one oscillation of $\dot{\phi}$ is observed, after which the inflaton spatial gradient energy takes over, signaling the breakdown of homogeneous backreaction and of the coherence needed for the pattern of oscillations. The situation is further complicated by the fact that the results presented in~\cite{Figueroa:2023oxc,Figueroa:2024rkr} are extremely sensitive to small variations of the model parameters, and that these simulations cover the end of inflation (where the oscillations of the inflaton speed obtained from the homogeneous approximation are much faster than those taking place earlier during inflation).

One factor contributing to this extreme sensitivity to the model parameters is the exponential dependence of the sourced fields on the $\frac{\dot{\phi}}{f}$ ratio mentioned above. This also implies that this general mechanism is highly tuned, and that an ${\rm O}(1)$ variation of the combination $\xi$ in the phenomenologically interesting range changes the sourced signals from negligible to incompatible with the present bounds. It is reasonable to expect that this narrow window is characterized by three intervals with increasing $\xi$: (1) one in which the production starts to be relevant for the perturbations, but still provides a negligible backreaction on the background dynamics; (2) one in which backreaction is well described by the homogeneous regime; and (3) one in which scalar gradients become relevant, leading the system into the regime of full backreaction. The presence of these three regimes can in fact be seen in Figure~\ref{fig:peaks}. 

Based on this reasoning, the present work aims to investigate whether a prolonged phase of several oscillations in regime (2) can indeed take place, and whether it is associated with a detectable phenomenology. Due to the extreme complexity of the system, we are unable to formulate mathematically rigorous criteria for the validity of the homogeneous backreaction regime, and we believe that this can ultimately be assessed only by comparison with the results of a full lattice simulation. In the absence of such a comparison, we consider the most informative simple procedure for assessing the validity of the approximation to be the comparison of the energy density in the inflaton spatial gradient, $\rho_{\rm grad}$ (computed in perturbation theory), with that of the kinetic energy of the zeroth mode, $\rho_{\rm kinetic}$.

Using this criterion, we find that the parameter choice employed in the example shown in Ref.~\cite{Garcia-Bellido:2023ser} leads to a breakdown of the homogeneous backreaction approximation already at the first peak, which most likely invalidates the phenomenological result presented there. However, by varying the model parameters, we present here several examples for which $\rho_{\rm grad}$ remains subdominant throughout inflation, and the SGWB spectrum presents multiple peaks above the planned sensitivity of future experiments. We provided examples in which $\rho_{\rm grad}$ is only slightly smaller than 
$\rho_{\rm kinetic}$. It is likely that, for these limiting cases, non-homogeneities in the inflaton, although subdominant in the energy density, can still play an important role in the evolution. We however also provided examples in which the gradient energy remains one or two orders of magnitude smaller than the kinetic energy. They are characterized by a slightly smaller SGWB, which remains however above the detection threshold.~\footnote{It should be stressed that the word `detectable'  in this work simply indicates a signal above the experimental sensitivity curves listed in Figure~\ref{fig:GW_57_495}. Determining whether or not the signal can be detected requires dedicated analyses, that also take into account the presence of other SGWBs, as for instance the signal reported in~\cite{NANOGrav:2023gor}.}

To conclude, it has long been understood that axion inflation coupled to gauge fields can possess a very rich phenomenology. Earlier analytical studies have unequivocally shown this, but they were based on several approximations and should not be regarded as precision computations. Very recently, sufficiently robust lattice simulations have emerged that can provide more accurate answers, although we are still in a position where improvement is needed to make precise predictions for the primordial perturbations in the case of full backreaction. Lattice methods remain in a unique position to treat the inhomogeneities of the inflaton that emerge after an initial, analytically treatable stage. This situation somewhat mirrors that of preheating~\cite{Kofman:1997yn}, in which the initial linear stage of negligible backreaction (that can be solved analytically) is followed by a regime in which backreaction can be reasonably treated in a mean-field approximation, and a subsequent phase in which, due to rescattering, the inflaton becomes fully inhomogeneous and the system needs to be solved on the lattice~\cite{Felder:2000hq}. While the onset of full inhomogeneities is unavoidable at reheating, we have shown examples in which the inflationary expansion keeps the gradient energy at a subdominant level, supporting the idea that these cases can be treated with sufficient accuracy within the framework of homogeneous backreaction. As we mentioned, it would be extremely interesting to check the validity of these results against lattice simulations.

\begin{acknowledgments}
We thank Angelo~Caravano, Valerie~Domcke, Daniel~G.~Figueroa, Alexandros~Papageorgiou, Kai~Schmitz, and Matias Zaldarriaga for insightful discussions. The \textit{Mathematica} code used in our simulations is an improvement of a code originally created by Alexandros~Papageorgiou and further developed by Federico~Greco. We thank them for their collaboration in related projects and for sharing their version of the code. We acknowledge support from Istituto Nazionale di Fisica Nucleare (INFN) through the Theoretical Astroparticle Physics (TAsP) project, and from the MIUR Progetti di Ricerca di Rilevante Interesse Nazionale (PRIN) Bando 2022 - grant 20228RMX4A. CloudVeneto is acknowledged for the use of computing and storage facilities. 
\end{acknowledgments} 

\appendix

\section{Approximate gauge mode functions and spectra of sourced perturbations for constant $\xi$ and $H$}
\label{app:constant-xi}

In this Appendix we test the validity of the expressions presented in Section~\ref{sec:perturbations} by evaluating them for constant $H$ and $\xi$ and by comparing them with the existing literature~\cite{Barnaby:2011vw}.~\footnote{For the tensor power spectrum, this check was also done in Appendix~B of~\cite{Garcia-Bellido:2023ser}, so we omit several intermediate steps in the present discussion. In the present work, we extend this check to include also the scalar power spectrum.} For large and constant $\xi$, and constant $H$, the amplified gauge field mode functions are well approximated by~\cite{Anber:2009ua} 
\begin{equation}
A_+ \left( \tau ,\, k \right) \simeq \frac{1}{\sqrt{2 k}} \, \left( \frac{-k \tau}{2 \xi} \right)^{1/4} \, {\rm e}^{\pi \xi - 2 \sqrt{-2 \xi k \tau}} \;\;,\;\; A_+' \left( \tau ,\, k \right) \simeq \sqrt{\frac{2 k \xi}{-\tau}} A_+ \left( \tau ,\, k \right) \;, 
\label{A-approx}
\end{equation} 
in the interval $\left( 8 \xi \right)^{-1} \lesssim k / \left( a H \right) \lesssim 2 \xi$ that accounts for most of the power in the produced gauge fluctuations. 

We evaluate the power spectra~\eqref{PS-sourced} of the sourced scalar and tensor modes by employing the relations (\ref{C-lambda}) and (\ref{C-phi}) for the correlators, with the late-time de Sitter (dS) Green function 
\begin{equation}
\lim_{- k \tau \to 0} \; \frac{{\tilde G}_k \left( \tau ,\, \tau' \right)}{a \left( \tau \right) \, a \left( \tau' \right) } = H^2 \, \frac{k \tau' \, \cos \left( k \tau' \right) - \sin \left( k \tau' \right)}{k^3} \;, 
\end{equation}
and with the approximate expressions~\eqref{A-approx} for the gauge field mode function and its derivative. The momentum integrands of these relations are symmetric functions of $p$ (the magnitude of the integration variable) and $q \equiv \left\vert \vec{k} - \vec{p} \right\vert$. For a generic function $f \left( p ,\, q \right)$, we can perform one trivial angular integration, and write
\begin{equation}
\int \frac{d^3 p}{\left( 2 \pi \right)^3} \, f \left( p ,\, q \right) = \frac{1}{4 \pi^2 k} \int_0^\infty d p \int_{\left\vert k - p \right\vert}^{k + p} d q \, p \, q \, f \left( p ,\, q \right) \;. 
\label{change1-d3p}
\end{equation}

Employing these relations, after some lengthy but straightforward algebra we reproduce the expressions (3.16) - (3.17) of~\cite{Barnaby:2011vw} for the scalar power spectrum. We also obtain a result that is a factor $\frac{1}{4}$ smaller than the combination of their eqs.~(3.39) and (3.40) (recalling that $\lambda = \pm 1$) for the tensor power spectrum. We verified that the origin of the discrepancy is a typo in the expression (3.40) of~\cite{Barnaby:2011vw}, in which, for example, their initial $\frac{1}{\xi}$ factor should be replaced by $\frac{1}{4 \, \xi}$. The numerical integration of our expression reproduces the numerical result (3.41) of~\cite{Barnaby:2011vw}, thus confirming that the missing $\frac{1}{4}$ in their (3.40) is a typo that does not impact their results.

This non-trivial check validates our expressions~\eqref{C-lambda} and~\eqref{C-phi}, that hold for arbitrary gauge field modes and Green function, namely for a background evolution beyond the dS + constant $\dot{\varphi}$ assumptions made in the analytic computations of~\cite{Barnaby:2011vw}.

\section{Expressions in code variables}
\label{app:code}

The code we use for the numerical evolutions presented in Section~\ref{sec:example} is an extension of the one developed for Ref.~\cite{Garcia-Bellido:2023ser}, so we employ the same code variables used in that work. For completeness, we report also in this appendix the full set of definitions and relevant equations, but we omit some intermediate equations given in~\cite{Garcia-Bellido:2023ser}. We also refer the reader to~\cite{Garcia-Bellido:2023ser} for the technical details of the code.~\footnote{We make two different choice of `internal' code parameters with respect to~\cite{Garcia-Bellido:2023ser}. Firstly, wile they use a grid of $400$ vector modes, equally log-spaced in wavenumber, we verified that using $300$ or $350$ modes already leads to converence of results. Secondly, we initialize gauge modes from the vacuum as in eq. (3.5) of~\cite{Garcia-Bellido:2023ser}, with a parameter $10$ rather than $10^{5/2}$ (in practice, we verified that a modes can be inizialized at a later moment with respect to what done in~\cite{Garcia-Bellido:2023ser}, with no significant loss of accuracy).} 

The initial moment of our evolutions is $60$ e-folds before the end of inflation. A suffix $0$ indicates quantities evaluated at this initial moment. The initial value of the scale factor are chosen, respectively, as $a_0 = 1$ and $N_0=0$. Since we assume a negligible backreaction from the gauge fields at the initial moment (to respect the CMB limits~\cite{Barnaby:2010vf} on the modes sourced at this time), inflation initially proceeds in the standard slow-roll fashion. Therefore we set the initial conformal time to $\tau_0 = - \frac{1}{a_0 \, H_0} = - \frac{1}{H_0}$, as conventionally done for dS spacetime. Therefore 
\begin{equation}
a = {\rm e}^N \;\; \rightarrow \;\; \tau = - \frac{1}{H_0} + \int_0^N \frac{d N'}{H \left( N' \right) \, {\rm e}^{N'}} \;. 
\end{equation}

\subsection{Definitions and background equations}
\label{app:code-background}

We use the number of e-folds as the ``time variable'' in the numerical integrations, and work in terms of the dimensionless quantities~\footnote{We correct a typo in the last rescaling of eq.~(A3) of~\cite{Garcia-Bellido:2023ser}.}
\begin{align}
& {\tilde k} \equiv \frac{M_p}{\sqrt{V_0}}\, k \;\;,\;\; 
{\tilde H} \equiv \frac{M_p}{\sqrt{V_0}} \, H \;\;,\;\; 
{\tilde \phi} \equiv \frac{\phi}{f} \;\;,\;\; 
{\tilde f} \equiv \frac{f}{M_p} \;\;,\;\; 
{\tilde V} \left( {\tilde \varphi} \right) \equiv \frac{V \left( \varphi \right)}{V_0} \;\;,\;\;
{\tilde V}_0 \equiv \frac{V_0}{M_p^4} \;\;, \nonumber\\
& {\bar A}_\lambda \equiv \sqrt{2 k} \, {\rm e}^{i k \left( \tau - \tau_0 \right)} \, A_\lambda = \sqrt{\frac{2 \, \sqrt{V_0} \, {\tilde k}}{M_p}} \, {\rm e}^{i {\tilde k} \int_0^N \frac{d N'}{{\tilde H} \left( N' \right) {\rm e}^{N'}}} \, A_\lambda \;, 
\label{rescaling}
\end{align}
where $V_0$ is the value of the inflaton potential at the initial moment $N=0$ and where the rescaling of the gauge mode eliminates the fast phase oscillation in the initial UV regime. We then obtain the following background equations~\cite{Garcia-Bellido:2023ser}
\begin{align}
& \frac{d^2 {\tilde \varphi}}{d N^2} = - \frac{d {\tilde \varphi}}{d N} + \frac{{\tilde f}^2}{6} \left( \frac{d {\tilde \varphi}}{d N} \right)^3 - \frac{{\tilde V'} \left( {\tilde \varphi} \right)}{{\tilde f^2} {\tilde H}^2} - \frac{2 {\tilde V} \left( {\tilde \varphi} \right)}{3 {\tilde H}^2} \frac{d {\tilde \varphi}}{d N} + \frac{\left\langle \vec{E} \cdot \vec{B} \right\rangle}{{\tilde f}^2 V_0 {\tilde H}^2} \;, \nonumber\\ 
& \frac{d {\tilde H}}{d N} =  - 2 {\tilde H} - \frac{{\tilde f}^2 {\tilde H}}{6} \left( \frac{d {\tilde \varphi}}{d N} \right)^2 + \frac{2 {\tilde V \left( {\tilde \varphi} \right)}}{3 {\tilde H}}  \;, \nonumber\\ 
& \frac{d^2 {\bar A}_\lambda}{d N^2} = \left( 1 + \frac{{\tilde f}^2}{6} \left( \frac{d {\tilde \varphi}}{d N} \right)^2 + \frac{2 i {\tilde k}}{{\tilde H} {\rm e}^N} - \frac{2 {\tilde V \left( {\tilde \varphi} \right)}}{3 {\tilde H}^2} \right) \frac{d {\bar A}_\lambda}{d N} + \frac{\lambda {\tilde k} {\bar A}_\lambda}{{\tilde H} e^N} \frac{d {\tilde \varphi}}{d N} \;. 
\end{align}

We integrate this system numerically, subject to the intial conditions at $N=0$ 
\begin{equation}
{\tilde \varphi} \left( 0 \right) = {\tilde \varphi}_0 \;\;,\;\;  \frac{d {\tilde \varphi}}{d N} \Big\vert_{N=0} = - \frac{1}{{\tilde f}^2} \frac{d {\tilde V} \left( {\tilde \varphi} \right)}{d {\tilde \varphi}} \Big\vert_{{\tilde \varphi} = {\tilde \varphi}_0} \;\;,\;\;  {\tilde H} \left( 0 \right) = \frac{1}{\sqrt{3}} \;\;,\;\; 
{\bar A}_\lambda \left( 0 \right) = 1 \;\;,\;\; 
\frac{d {\bar A}_\lambda}{d N} \Big\vert_{N=0} = 0 \;, 
\end{equation}
that assume that the inflaton potential dominates the energy density and that the vector mode is deep in the UV. 

In rescaled variables, the gauge field correlators read 
\begin{align}
\left\langle \vec{E} \cdot \vec{B} \right\rangle &= - \frac{V_0^2}{8 \pi^2 M_p^4} \, \frac{\tilde H}{{\rm e}^{3 N}} \int d {\tilde k} \, {\tilde k}^2 \sum_{\lambda = \pm} \lambda \frac{d \, \left\vert {\bar A}_\lambda \right\vert^2}{d N} \;, \nonumber\\
\left\langle \frac{\vec{E}^2+\vec{B}^2}{2} \right\rangle &= \frac{ V_0^2}{8 \pi^2  M_p^4\, {\rm e}^{4 N}} \int d {\tilde k} \, {\tilde k} \, \sum_{\lambda = \pm} \left[ {\rm e}^{2 N} {\tilde H}^2 \left\vert \frac{d {\bar A}_\lambda}{d N} - i \frac{{\tilde k} {\bar A}_\lambda}{{\tilde H} {\rm e}^N} \right\vert^2 + {\tilde k}^2 \, \left\vert {\bar A}_\lambda \right\vert^2 \right] \;. 
\end{align} 

\subsection{Green function}
\label{app:code-green}

Let us now turn our attention to the sourced perturbations. In this section we study how to formulate the Green function in code variables. In the next section we provide the expressions for the 2-point correlators of the perturbations, which are then employed in the computations of the energy densities, as reported in the next appendix. 

In code variables, the system~\eqref{system-X} rewrites
\begin{align}
& {\hat O}_{N,X} \; {\hat Q}_X \left( N ,\, \vec{k} \right) = \frac{\hat{\cal S}_X \left( N ,\, \vec{k} \right)}{H^2 \, {\rm e}^{2 N}} \;, \nonumber\\
& {\hat O}_{N,X} \equiv \frac{\partial^2}{\partial N^2} + \left[ 1 + \frac{1}{\tilde H} \, \frac{d {\tilde H}}{d N} \right] \frac{\partial}{\partial N} + \frac{{\tilde k}^2}{{\rm e}^{2 N} \, {\tilde H}^2} - 2 - \frac{1}{\tilde H} \, \frac{d {\tilde H}}{d N} + \delta_{X \phi} \, \frac{{\tilde V}'' \left( {\tilde \varphi} \right)}{{\tilde f}^2 \, {\tilde H}^2} \;, 
\label{system-X-N}
\end{align}
so that the sourced solution has the formal expression
\begin{equation}
{\hat Q}_{X,s} \left( N ,\, \vec{k} \right) = \int_{0}^N \, d N' \, \tilde{\cal G}_{X,k} \left( N ,\, N' \right) \, \frac{\hat{\cal S}_X \left( N' ,\, \vec{k} \right)}{\left( H \left( N' \right) \, {\rm e}^{N'} \right)^2} \;. 
\label{formal-sol-N}
\end{equation}

As this solution coincides with~\eqref{formal-sol}, rewritten in terms of $N$ instead of $\tau$, the two Green functions are related to each other by 
\begin{equation}
\tilde{\cal G}_{X,k} \left( N ,\, N' \right) = H \left( N' \right) \, {\rm e}^{N'} \, {\tilde G}_{X,k} \left( \tau \left( N \right) ,\, \tau' \left( N' \right) \right) \;, 
\label{gg}
\end{equation} 
showing that $\tilde{\cal G}_{X,k}$ is dimensionless. 

In terms of the function ${\cal F}_X \left( N ,\, k \right)$, satisfying
\begin{equation}
{\hat O}_{N,X} {\cal F}_X \left( N ,\, k \right) = 0 \;\;,\;\; \lim_{k \gg {\rm e}^N \, H} {\cal F}_X \left( N ,\, k \right) = \frac{{\rm e}^{-i k \tau \left( N \right)}}{\sqrt{2 k}} \;, 
\end{equation} 
the non distributional part of the Green function of
\begin{equation}
{\hat O}_{N,X} \left[ \tilde{\cal G}_{X,k} \left( N ,\, N' \right) \, \theta \left( N - N' \right) \right] = \delta \left( N - N' \right) \;, 
\end{equation}
is given by 
\begin{equation}
\tilde{\cal G}_{X,k} \left( N ,\, N' \right) = \frac{{\cal F}_X \left( N ,\, k \right) \, {\cal F}_X^* \left( N' ,\, k \right) - {\cal F}_X^* \left( N ,\, k \right) \, {\cal F}_X \left( N' ,\, k \right)}{\frac{\partial {\cal F}_X \left( N' ,\, k \right)}{\partial N'} \, {\cal F}_X^* \left( N' ,\, k\right) - \frac{\partial {\cal F}_X^* \left( N' ,\, k \right)}{\partial N'} \, {\cal F}_X  \left( N' ,\, k \right) } \;. 
\end{equation} 

As we did for the gauge field mode functions, it is numerically convenient to work in terms of the rescaled functions
\begin{equation}
\bar{\cal F}_X \equiv \sqrt{2 k} \, {\rm e}^{i k \left( \tau - \tau_0 \right)} \, {\cal F}_X = \sqrt{\frac{2 \, \sqrt{V_0} \, {\tilde k}}{M_p}} \, {\rm e}^{i {\tilde k} \int_0^N \frac{d N'}{{\tilde H} \left( N' \right) {\rm e}^{N'}}} \, {\cal F}_X \;, 
\end{equation}
that are determined by 
\begin{equation}
\left\{ \begin{array}{l}
\frac{\partial^2 \bar{\cal F}_X \left( N ,\, k \right)}{\partial N^2} + \left[ 1 + \frac{1}{\tilde H} \, \frac{d {\tilde H}}{d N} - \frac{2 i {\tilde k}}{{\tilde H} {\rm e}^N} \right] \frac{\partial \bar{\cal F}_X \left( N ,\, k \right)}{\partial N} + \left[ - 2 - \frac{1}{\tilde H} \, \frac{d {\tilde H}}{d N} + \delta_{X \phi} \, \frac{{\tilde V}'' \left( {\tilde \varphi} \right)}{{\tilde f}^2 \, {\tilde H}^2} \right] \bar{\cal F}_X \left( N ,\, k \right) = 0 \\ \\ 
\lim_{{\tilde k} \gg {\rm e}^N \, {\tilde H}} \bar{\cal F}_X \left( N ,\, k \right) = 1 \;\;,\;\; 
\lim_{{\tilde k} \gg {\rm e}^N \, {\tilde H}} \frac{\partial \bar{\cal F}_X \left( N ,\, k \right)}{\partial N} = 0  
\end{array} 
\right. \;, 
\label{F-code}
\end{equation} 
and in terms of which 
\begin{equation}
g_{X,{\tilde k}}^{(0)} \left( N ,\, N' \right) \equiv \tilde{\cal G}_{X,k} \left( N ,\, N' \right) = \frac{{\rm Im} \left( {\rm e}^{- i {\tilde k} \int_{N'}^N \frac{d n}{{\tilde H} \left( n \right) \, {\rm e}^{n}}} \, \bar{\cal F}_X \left( N ,\, k \right) \, \bar{\cal F}_X^* \left( N' ,\, k \right) \right)}{{\rm Im} \left( \frac{\partial \bar{\cal F}_X \left( N' ,\, k \right)}{\partial N'} \, \bar{\cal F}_X^* \left( N' ,\, k \right) \right) - \frac{\tilde k}{{\tilde H} \left( N' \right) \, {\rm e}^{N'}} \left\vert \bar{\cal F}_X \left( N' ,\, k \right) \right\vert^2 } \;. 
\label{Green-code}
\end{equation} 
For later convenience, we also give the explicit form of 
\begin{align}
g_{X,{\tilde k}}^{(1)} \left( N ,\, N' \right) &\equiv \frac{\partial \tilde{\cal G}_{X,k} \left( N ,\, N' \right)}{\partial N} - \tilde{\cal G}_{X,k} \left( N ,\, N' \right) \nonumber\\ 
&= \frac{{\rm Im} \left( {\rm e}^{- i {\tilde k} \int_{N'}^N \frac{d n}{{\tilde H} \left( n \right) \, {\rm e}^{n}}} \, \left[ \frac{\bar{\cal F}_X \left( N ,\, k \right)}{d N} - \left( 1 + \frac{i {\tilde k}}{{\tilde H} \left( N \right) {\rm e}^N} \right) \bar{\cal F}_X \left( N ,\, k \right) \right] \, \bar{\cal F}_X^* \left( N' ,\, k \right) \right)}{{\rm Im} \left( \frac{\partial \bar{\cal F}_X \left( N' ,\, k \right)}{\partial N'} \, \bar{\cal F}_X^* \left( N' ,\, k \right) \right) - \frac{\tilde k}{{\tilde H} \left( N' \right) \, {\rm e}^{N'}} \left\vert \bar{\cal F}_X \left( N' ,\, k \right) \right\vert^2 } \;. 
\label{Green-der-code}
\end{align}

The relations~\eqref{F-code} and~\eqref{Green-code} are written in terms of code variables only, and allow to numerically obtain the Green function that we employ in the expressions of the physical quantites reported in the next section.

\subsection{Correlators}
\label{app:code-rho}

We want to provide the explicit expressions for the correlators~\eqref{C-lambda} and~\eqref{C-phi}. 
Let us start from the Green function and its derivative. Using the expressions in the previous section, we write  
\begin{equation}
\frac{1}{a \left( \tau' \right)} \, \frac{\partial^m}{\partial \tau^m} \left( \frac{{\tilde G}_{X,k} \left( \tau ,\, \tau' \right)}{a \left( \tau \right)} \right) = \left( \frac{M_p}{\sqrt{V_0}} \right)^{1-m} \frac{\left( {\tilde H} \left( N \right) \, {\rm e}^N \right)^m \, g_{X,k}^{(m)} \left( N ,\, N' \right)}{{\tilde H} \left( N' \right) \, {\rm e}^{N+2 N'}} \;\;,\;\; m = 0 ,\, 1 \;. 
\end{equation}

Secondly, we perform one further change of variable in the momentum integral appearing in eqs.~\eqref{C-lambda} and~\eqref{C-phi}, extending the relation~\eqref{change1-d3p} to 
\begin{equation}
\int \frac{d^3 p}{\left( 2 \pi \right)^3} \, f \left( p ,\, q \right) = \frac{k^3}{2 \pi^2} \int_{\frac{1}{2}}^\infty d X \int_{-\frac{1}{2}}^{\frac{1}{2}} d Y \left( X^2 - Y^2 \right) \, f \left( k \left( X + Y \right) ,\, k \left( X - Y \right) \right) 
\;, 
\label{change2-d3p}
\end{equation}
resulting in constant extrema of integration (unlike the second expression in~\eqref{change1-d3p}). In these variables
\begin{align}
\left( 1 + \lambda \, {\hat k} \cdot {\hat p} \right)^2 \left( 1 + \lambda \, {\hat k} \cdot {\hat q} \right)^2 &= \frac{\left( 1 + 2 \lambda X \right)^4 \left( 1 - 4 Y^2 \right)^2}{16 \left( X^2 - Y^2 \right)^2} \;\;,\;\; \lambda = \pm \, 1 \;, \nonumber\\ 
\left( 1 - {\hat p} \cdot {\hat q} \right)^2 &= \frac{\left( 1 - 4 X^2 \right)^2}{4 \left( X^2 - Y^2 \right)^2} \;. 
\end{align} 

Inserting all this in eqs.~\eqref{C-lambda} and~\eqref{C-phi} we can rewrite the formal expressions of the correlators in terms of code variables only.  
\begin{align}
& \!\!\!\!\!\!\!\! \!\!\!\!\!\!\!\! \!\!\!\!\!\!\!\! \!\!\!\!\!\!\!\! {\cal C}_\lambda^{(m)} = \frac{{\tilde V}_0^2 \, {\rm e}^{-2N}}{2048 \pi^4} \, {\tilde k}^4 \left( \frac{{\tilde H} \left( N  \right) \, {\rm e}^N}{\tilde k} \right)^{2 m} \int_{\frac{1}{2}}^\infty d X \int_{-\frac{1}{2}}^\frac{1}{2} d Y \, \frac{\left( 1 - 4 Y^2 \right)^2 \left( 1 + 2 \lambda X \right)^4}{\left( X^2 - Y^2 \right)^2}   \Bigg\vert \int_0^N d N' {\rm e}^{-2 i {\tilde k} X \int_0^{N'} \; \frac{d n}{{\tilde H} \left( n \right) {\rm e}^n}} \; \frac{g_{\lambda,{\tilde k}}^{(m)} \left( N ,\, N' \right)}{{\rm e}^{N'}} \nonumber\\
& \quad\quad \times \Big\{ \frac{\partial {\bar A}_+ \left( N' ,\, {\tilde k} \left( X+Y \right) \right)}{\partial N'} \frac{\partial {\bar A}_+ \left( N' ,\, {\tilde k} \left( X-Y \right) \right)}{\partial N'} \nonumber\\
& \quad\quad \quad\quad - \frac{i {\tilde k}}{{\tilde H} \left( N' \right) \, {\rm e}^{N'}} \Big[ \left( X + Y \right) {\bar A}_+ \left( N' ,\, {\tilde k} \left( X+Y \right) \right) \frac{\partial {\bar A}_+ \left( N' ,\, {\tilde k} \left( X-Y \right) \right)}{\partial N'} \nonumber\\
& \quad\quad \quad\quad \quad\quad \quad\quad \quad\quad + \left( X - Y \right) {\bar A}_+ \left( N' ,\, {\tilde k} \left( X-Y \right) \right) \frac{\partial {\bar A}_+ \left( N' ,\, {\tilde k} \left( X+Y \right) \right)}{\partial N'} \Big] \Big\}  \Bigg\vert^2 \;, 
\label{C-lambda-app}
\end{align}
for the GW correlator, and 
\begin{align}
& \!\!\!\!\!\!\!\! \!\!\!\!\!\!\!\! \!\!\!\!\!\!\!\! \!\!\!\!\!\!\!\! {\cal C}_\phi^{(m)} =  \frac{{\tilde V}_0^2 \, {\rm e}^{-2N}}{512 \pi^4 {\tilde f}^2} \, {\tilde k}^6 \left( \frac{{\tilde H} \left( N  \right) \, {\rm e}^N}{\tilde k} \right)^{2 m} \int_{\frac{1}{2}}^\infty d X \int_{-\frac{1}{2}}^\frac{1}{2} d Y \, \frac{\left( 1 - 4 X^2 \right)^2}{ \left( X^2 - Y^2 \right)^2}  \Bigg\vert \int_0^N d N' \; {\rm e}^{-2 i {\tilde k} X \int_0^{N'} \frac{d n}{{\tilde H} \left( n \right) {\rm e}^n}} 
 \; \frac{g_{\phi,{\tilde k}}^{(m)} \left( N ,\, N' \right)}{{\tilde H} \left( N' \right) {\rm e}^{2 N'}} \nonumber\\
&\quad\quad \times \Bigg\{ \left( X + Y \right) {\bar A}_+ \left( N' ,\, {\tilde k} \left( X + Y \right) \right) \frac{\partial {\bar A}_+ \left( N' ,\, {\tilde k} \left( X-Y \right) \right)}{\partial N'} \nonumber\\
&\quad\quad \quad + \left( X - Y \right) {\bar A}_+ \left( N' ,\, {\tilde k} \left( X - Y \right) \right) \frac{\partial {\bar A}_+ \left( N' ,\, {\tilde k} \left( X+Y \right) \right)}{\partial N'} \nonumber\\
&\quad\quad \quad - \frac{2 i {\tilde k}}{{\tilde H \left( N' \right)} {\rm e}^{N'}} \left( X^2 - Y^2 \right) {\bar A}_+ \left( N' ,\, {\tilde k} \left( X + Y \right) \right) {\bar A}_+ \left( N' ,\, {\tilde k} \left( X - Y \right) \right) \Bigg\} \Bigg\vert^2 \;, 
\label{C-phi-app}
\end{align}
for the correlator of the inflaton perturbations. We recall that the rescaled Green functions and its derivative apperaing in these two expressions are given in eqs.~\eqref{Green-code} and~\eqref{Green-der-code}. 

The expression~\eqref{C-lambda-app} for $m=0$ corrects eq. (A.16) of~\cite{Garcia-Bellido:2023ser}, that erroneously omits the phase inside the third integral in the first line. We verified that this omission has only a marginal impact on their main result (Figure 4). The expression~\eqref{C-phi-app} is an original result of the present work.

\section{Energy densities}
\label{app:rho}

In this Appendix we provide the expressions for the energy densities that we evaluate and discuss in the main text. The zeroth mode inflaton and gauge field energy densities, are given, respectively, by
\begin{equation}
\rho_{\rm kinetic} \left( \tau \right) \equiv \frac{\varphi' \left( \tau \right)^2}{2 a \left( \tau \right)^2}  \;\;\;\; , \;\;\;\; \rho_{\rm potential} \left( \tau \right) \equiv V \left( \varphi \left( \tau \right) \right) \;, 
\label{rho-infla-bckg} 
\end{equation}
and by
\begin{equation}
\rho_{\rm gauge} \left( \tau \right) \equiv \left\langle \frac{\vec{E} \left( x \right)^2+\vec{B} \left( x \right)^2}{2} \right\rangle = \frac{1}{2 \, a \left( \tau \right)^4} \, \int d \ln k \; \frac{k^3}{2 \pi^2} \left[ \left\vert A_+' \left( \tau ,\, k \right) \right\vert^2 + k^2 \, \left\vert A_+ \left( \tau ,\, k \right) \right\vert^2 \right] \;. 
\label{rho-A} 
\end{equation} 
We are also interested in the GW energy density
\begin{align}
\rho_{\rm GW} \left( \tau \right) \equiv& \frac{M_p^2}{8 a\left( \tau \right)^2} \left\langle h_{ij}' \left( x \right) \, h_{ij}' \left( x \right) + \partial_k h_{ij} \left( x \right) \, \partial_k h_{ij} \left( x \right) \right\rangle \nonumber\\
=& \frac{M_p^2}{2 a \left( \tau \right)^2} \int d \ln k \; k^2 \; \sum_{\lambda=\pm} \left[ {\cal C}_\lambda^{(1)} \left( \tau ,\, k \right) + {\cal C}_\lambda^{(0)} \left( \tau ,\, k \right) \right] \;, 
\label{rho-GW}
\end{align} 
and in the inflaton gradient energy density 
\begin{align}
\rho_{\rm grad} \left( \tau \right) \equiv& \frac{1}{2 a\left( \tau \right)^2} \left\langle \partial_k \phi \left( x \right) \, \partial_k \phi \left( x \right) \right\rangle = \frac{M_p^2}{2 a \left( \tau \right)^2} \int d \ln k \; k^2 \; {\cal C}_\phi^{(0)} \left( \tau ,\, k \right) \;. 
\end{align} 

In code variables, these energy densities are expressed as 
\begin{align}
\rho_{\rm potential} \,/\, V_0 &= {\tilde V} \left( {\tilde \varphi} \right) \;,\nonumber\\
\rho_{\rm kinetic} \,/\, V_0 &= {\tilde f}^2 \, \frac{{\tilde H}^2}{2} \left( \frac{ d {\tilde \varphi}}{d N} \right)^2 \;,\nonumber\\
\rho_{\rm gauge} \,/\, V_0 &= \frac{{\tilde V}_0}{8 \pi^2 \, {\rm e}^{4 N}} \int d \ln {\tilde k} \, {\tilde k}^4 \left( \left\vert \frac{{\tilde H} \, {\rm e}^N}{\tilde k} \, \frac{\partial {\bar A}_+}{\partial N} - i \, {\bar A}_+ \right\vert^2 + \left\vert {\bar A}_+ \right\vert^2 \right) \;, \nonumber\\
\rho_{\rm GW} \,/\, V_0 &= \frac{1}{2 \, {\rm e}^{2 N}} \int d \ln {\tilde k} \; {\tilde k}^2 \; \sum_{\lambda=\pm} \left[ {\cal C}_\lambda^{(1)} \left( N ,\, {\tilde k} \right) + {\cal C}_\lambda^{(0)} \left( N ,\, {\tilde k} \right) \right] \;,  \nonumber\\
\rho_{\rm grad} \,/\, V_0 &= \frac{1}{2 \, {\rm e}^{2 N}} \int d \ln {\tilde k} \; {\tilde k}^2 \; {\cal C}_\phi^{(0)} \left( N ,\, {\tilde k} \right) \;. 
\label{rho-code}
\end{align}

The gauge energy density is regularized in the UV identically to the $\left\langle \vec{E} \cdot \vec{B} \right\rangle$ backreaction integral. The interested reader is referred to Section 3.1 of~\cite{Garcia-Bellido:2023ser}. The quantities $\rho_{\rm GW}$ and $\rho_{\rm grad}$ are regularized by the fact that we only include the (by far, dominant) contribution of the sourced modes up to a maximum wavenumber in which their power is no longer greater than that of the vacuum modes. We recall that, due to our choice of the potential, the sourced signals are significant only as long as the inflaton is in the intermediate portion of the potential. After that, the inflaton enters in a region with much milder slope, and the sourcing mechanisms shuts off. This implies that modes that were sub-horizon when the inflaton leaves the intermediate region are not sourced, effectively cutting-off the power in the UV. 

\begin{figure}[h!]
\centering
\includegraphics[width=0.75\linewidth]{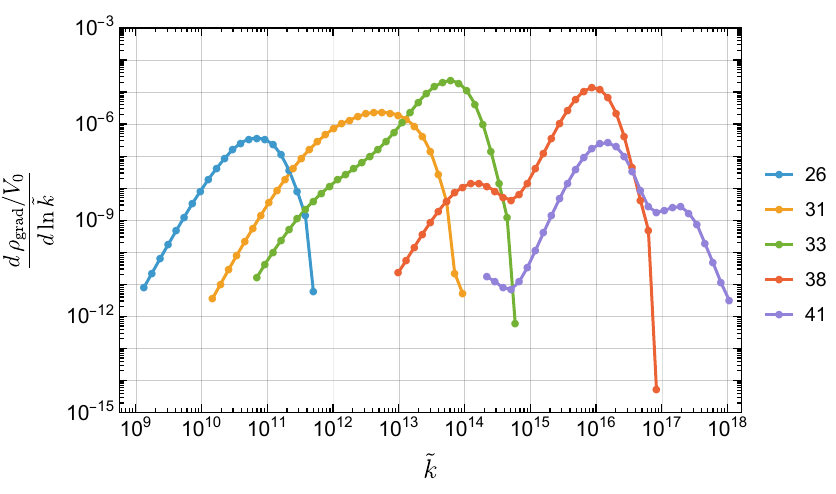}
\caption{Each curve in the figure shows the dominant portion of the integrand of the gradient inflaton energy density, evaluated at a given time during inflation. These results have been obtained from an evolution with ${\tilde f}^{-1}=80$ and $v' = -0.1157$, while the other parameters are chosen as in the `baseline' model (up to the rescaling~\eqref{kappa}). 
}
\label{fig:drhograd}
\end{figure}

This is visible in Figure~\ref{fig:drhograd}, where we show the integrand of the gradient energy density of the axion, resulting from a simulation with parameters provided in the caption of the figure. Each curve in the figure shows the integrand evaluated at a fixed moment during inflation. Each curves extends in the UV only up to the comoving momentum for which the sourced modes have been amplified (specifically we do not include UV modes for which the sourced power does not exceed that of the vacuum modes). We also do not include IR modes which contribute negligibly to the integral. 

The figure clearly shows the general tendency of the spectrum to ``move to the right'' as the number of e-folds increases. This can be easily understood. At any given time, the sourcing of GW and inflaton perturbations mostly takes place for modes of the order of the horizon. As $a H$ grows during inflation, this explains why modes of progressively greater comoving momentum are populated at increasing number of e-folds. Moreover, the production freezes out in the IR, and the physical gradient energy in this portion of the spectrum decreases due to redshift. Therefore, momenta that dominate the energy at any given time become subdominant at greater e-folds. 

If $\xi$ and $H$ were constant, we would expect all the spectra to have identical shape and amplitude. On the contrary, the spectra at the smallest and greatest $N$ shown in the figure are smaller than the other ones shown. We recall that in our model the gauge field amplification, and the consequent sourcing of GW and inflaton perturbations, is significant only as long as the inflaton is in the intermediate portion of the potential. As a reference, in the evolution under consideration the first and last peak on the parameter $\xi$ occur, respectively, at $N = 26$ and $N=41$, corresponding precisely to the first and last spectrum in the figure, and we find that the spectra at smaller and greater e-folds have a much smaller amplitude than those shown in the figure.

\section{Present frequency and fractional energy density of the SGWB}
\label{app:f-Omega}

In this Appendix we relate the comoving wavenumber $k$ and the primordial power spectrum $P_\lambda \left( k \right)$ to present frequency $f$ and fractional energy density of the SGWB, defined in eq.~(\ref{OmGW}).

We start by relating the parameter ${\tilde V}_0$, defined in eq.~\eqref{rescaling}, to the tensor-to-scalar ratio $r$ at CMB scales. Under the assumption that the sourced modes are highly subdominant to the vacuum ones at CMB scales and that the inflaton potential initially highly dominates the energy density of the universe, the amplitude of the promordial power spectrum is given by 
\begin{equation}
P_\zeta \simeq \frac{H_{\rm in}^2}{8 \pi^2 M_p^2 \epsilon} \simeq \frac{2 \, {\tilde V}_0}{3 \pi^2 \, r} \;, 
\end{equation}
where $H_{\rm in} \simeq \frac{\sqrt{V_0}}{\sqrt{3} M_p}$ is the Hubble rate at the initial times (when the CMB modes are produced) and where $\epsilon \equiv \frac{M_p^2}{2} \left( \frac{1}{V} \, \frac{d V}{d \varphi} \right)^2$ is a slow-roll parameter, related to the tensor-to-scalar ratio by the relation $r = 16 \, \epsilon$. 

We assume the ``Best Fit'' values in Table 1 of Ref.~\cite{Planck:2018vyg} for the cosmological parameter, giving $P_\zeta \simeq 2.1 \cdot 10^{-9}$. Therefore 
\begin{equation}
{\tilde V}_0^{1/4} \simeq 0.0055 \, r_{0.03}^{1/4} \;\;\;,\;\;\; r_{0.03} \equiv \frac{r}{0.03} \;,  
\label{V0t-r} 
\end{equation} 
where we normalized the tensor to scalar ratio to its current upper bound~\cite{Tristram:2021tvh,Galloni:2022mok}.  

We parametrize the energy density at any moment during inflation by its ratio to the initial inflaton potential as 
\begin{equation}
\rho^{1/4} \equiv q^{1/4} \, V_0^{1/4} \simeq q^{1/4} \, r_{0.03}^{1/4} \times 1.34 \cdot 10^{16} \, {\rm GeV} \;, 
\label{rho-q}
\end{equation}
where we stress that $q_{\rm in} = 1$ at the start of our simulation. 

We assume instantaneous reheating at the end of inflation, providing the reheating temperature
\begin{equation}
T_{\rm end} = \left( \frac{30 \, \rho_{\rm end}}{\pi^2 \, g_{*,{\rm end}}} \right)^{1/4} \simeq 5.50 \cdot 10^{15} \, {\rm GeV} \, q_{\rm end}^{1/4} \, r_{0.03}^{1/4} \;, 
\end{equation} 
where we have assumed the Standard Model degrees of freedom $g_{*,{\rm end}} = 106.75$ and where $q_{\rm end}$ is the parameter $q$ introduced in \eqref{rho-q} and evaluated at the end of inflation. This temperature can be related to the present CMB temperatue, $T_0 \simeq 2.35 \cdot 10^{-13} \, {\rm GeV}$, by conservation of entropy, $g_{*,s} \, a^3 \, T^3 = {\rm constant}$. From $g_{*,s,{\rm end}} = g_{*,{\rm end}} = 106.75$ and from $g_{*,s,0} \simeq 3.909$ (including the neutrinos contribution, since they decoupled while relativistic), we then have 
\begin{equation}
\frac{a_{\rm end}}{a_0} = \frac{g_{*,s,0}^{1/3}}{g_{*,s,{\rm end}}^{1/3}} \, \frac{T_0}{T_{\rm end}} \simeq \frac{1.42 \cdot 10^{-29}}{q_{\rm end}^{1/4} \, r_{0.03}^{1/4}} \;. 
\label{aend-a0}
\end{equation} 

A mode with comoving momentum $k \equiv \frac{\sqrt{V_0}}{M_p} \, {\tilde k}$ has the current frequency 
\begin{equation} 
f = \frac{k}{2 \pi a_0} \simeq 1.66 \cdot 10^{-16} \, {\rm GeV} \, \frac{r_{0.03}^{1/4}}{q_{\rm end}^{1/4}} \, \frac{\tilde k}{a_{\rm end}} \;,   
\end{equation} 
where the relations~\eqref{V0t-r} and~\eqref{aend-a0} have been used. In our convention $a_{\rm end} = {\rm e}^{60}$. Therefore, 
\begin{equation}
f \simeq 2.2 \cdot 10^{-18} \, {\rm Hz} \; {\tilde k} \; \left( \frac{r_{0.03}}{q_{\rm end}} \right)^{1/4} \;, 
\label{f-kt}
\end{equation} 
which provides the wanted relation between the rescaled comoving wavenumber ${\tilde k}$ labeling any mode in our notation to the present frequency $f$ of this mode. 

Let us now turn our attention to the parameter $\Omega_{\rm GW}$. From~\cite{Planck:2018vyg} we have the present values $\Omega_m = 0.316$ and $h = 0.67$ for, respectively, the fractional energy density in matter and rescaled Hubble rate. Including the contribution of neutrinos as if relativistic (giving $g_{*,0} \simeq 3.36$), the current fractional energy density of radiation is instead 
\begin{equation}
\Omega_{\rm rad} = \frac{\frac{\pi^2}{30} g_{*,0} T_0^4}{3 H_0^2 M_p^2} \simeq 9.3 \cdot 10^{-5} \;, 
\end{equation} 
so that matter-radiation equality occured at the redshift $z_{\rm eq} = \frac{\Omega_m}{\Omega_{\rm rad}} - 1 \simeq 3400$. Modes that re-entered the horizon at equality have therefore the present frequency
\begin{align}
f_{\rm eq,0} &= \frac{a_{\rm eq} \, H_{\rm eq}}{2 \pi a_0} \simeq \frac{1}{2 \pi z_{\rm eq}} \, \frac{1}{\sqrt{3} M_p} \, \sqrt{ \Omega_m \, \rho_{\rm crit,0} \, z_{\rm eq}^3 + \Omega_{\rm rad} \, \rho_{\rm crit,0} \, z_{\rm eq}^4} \nonumber\\
&\simeq \frac{H_0}{2 \pi z_{\rm eq}} \, \sqrt{\Omega_m \, z_{\rm eq}^3 + \Omega_{\rm rad} \, z_{\rm eq}^4} \simeq 1.6 \cdot 10^{-17} \, {\rm Hz} \;, 
\label{feq-0}
\end{align}
(where the two contributions are identical, by definition of equality), corresponding to the rescaled wavenumber 
\begin{equation} 
{\tilde k}_{\rm eq} \simeq 7.3 \, \left( \frac{q_{\rm end}}{r_{0.03}} \right)^{1/4} \;. 
\end{equation} 
Modes with ${\tilde k} > {\tilde k}_{\rm eq}$ (respectively,  ${\tilde k} < {\tilde k}_{\rm eq}$) have smaller (respectively, greater) wavelength, and therefore re-enter the horizon during the radiation-dominated (respectively, matter-dominated) era. 

To relate the present energy density in a SGWB mode to the one it had during inflation we need to study the evolution of the GW mode function. We refer the interested reader to Section 5.2 of~\cite{Caprini:2018mtu}, where the approximate relation
\begin{equation}
\Omega_{\rm GW} \left( f \right) \simeq \frac{\Omega_{\rm rad}}{24} \left[ 1 + \frac{9}{32} \left( \frac{f_{\rm eq,0}}{f} \right)^2  \right] \times \sum_{\lambda = \pm} P_\lambda \left( k = 2 \pi f a_0 \right) \;\;, 
\label{OmGW-Pla}
\end{equation} 
is shown to hold with very good accuracy.

\section{Inflaton potential and parameters adopted}
\label{app:parameters}

The potential employed in Section~\ref{sec:example} reads
\begin{equation}
{\tilde V} \left( {\tilde \phi} \right)= 
\left\{
\begin{array}{ll}
\left[ 1 - c_1 \left( {\tilde \phi} - {\tilde \phi}_0 \right)-c_2 \left( {\tilde \phi} - {\tilde \phi}_0 \right)^2-c_3 \left( {\tilde \phi} - {\tilde \phi}_0 \right)^3 - c_4 \left( {\tilde \phi} - {\tilde \phi}_0 \right)^4\right] 
& ,\; {\tilde \phi}_0 \leq {\tilde \phi} \leq {\tilde \phi}_1 \\
p_1 \, {\tilde \phi}^2 + p_2 \, {\tilde \phi} + p_3 
& ,\; {\tilde \phi}_1 \leq {\tilde \phi} \leq {\tilde \phi}_2 \\
c_5 \, {\tilde \phi} + p_4 
& ,\; \tilde{\phi}_2 \leq \tilde{\phi} \leq \tilde{\phi}_3 \\
p_5 \, {\tilde \phi}^2 + p_6 \, {\tilde \phi} + p_7 
& ,\; \tilde{\phi}_3 \leq \tilde{\phi} \leq \tilde{\phi}_4\\
c_6 \, {\tilde \phi} + p_8 
& ,\; \tilde{\phi}_4 \leq \tilde{\phi}\;, \\
\end{array} 
\right. 
\label{potential}
\end{equation}
where, as in eq.~\eqref{rescaling}, the inflaton $\phi$ and the parameters $\phi_i$ are given in units of the axion scale $f$, namely $\phi \equiv {\tilde \phi} \, f$ and $\phi_i \equiv {\tilde \phi}_i \, f$. The potential is given in units of $V_0$, namely $V \left( \phi \right) = {\tilde V} \left( {\tilde \phi} \right) \, V_0$. The parameter $\phi_0$ is the value of the inflaton at the beginning of our simulation, and therefore $V_0$ is the value of the inflaton potential at this initial time. We assume that the Planck pivot scale is generated at this value, so that $V_0$ can be derived by fitting the amplitude of the primordial scalar perturbations for CMB modes, as detailed in eq.~\eqref{V0t-r}.  

\begin{figure}[h!]
\centering
\hspace{-1cm}
\includegraphics[width=0.60\linewidth]{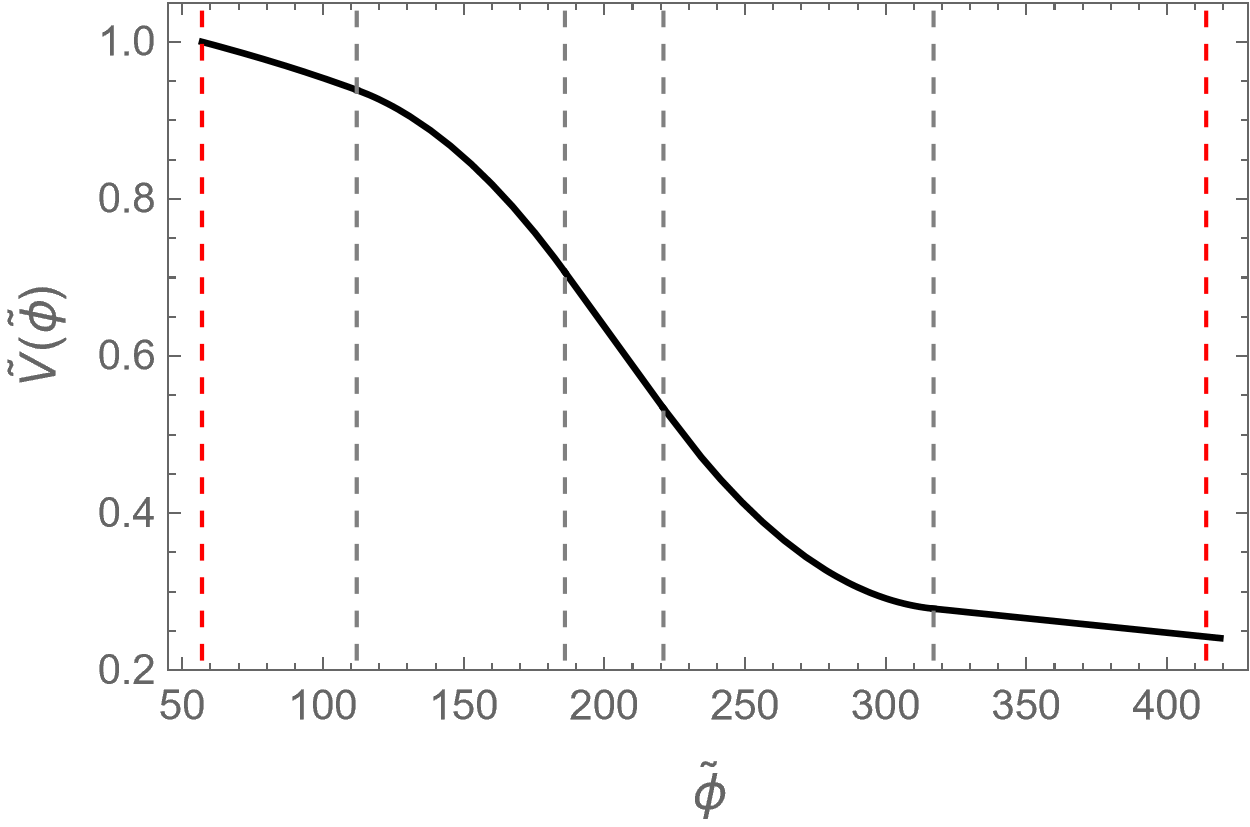}
\caption{The five‑segment potential defined in eq.~\eqref{potential}. Parameters values: $\tilde{\phi}_0=57,\,\tilde{\phi}_1=112,\,\tilde{\phi}_2=186,\,\tilde{\phi}_3=221,\,\tilde{\phi}_4=317$, $c_1=9.48\times10^{-4},\,c_2=2.39\times10^{-6},\,c_3=9.05\times10^{-9},\,c_4=3.97\times10^{-11}$, and line slopes $c_5=-4.95\times10^{-3},\,c_6=-3.68\times10^{-4}$. The outmost left and right (red) vertical dashed lines mark the field value at the beginning of the simulation and 60 e‑folds afterwards, where we set the end of inflation. The intermediate (black) dashed lines indicate the transition points between different segments of the potential.}
\label{fig:potential}
\end{figure}

As aleready remarked in the main text, the parameters $c_1$–$c_4$ that characterize the initial hill-top of the potential are chosen so that backreaction is negligible on CMB scales, and the predictions align with Planck results ($n_s=0.96$ and $r=0.023$ consistent at $2\sigma$ with present limits\cite{Planck:2018jri}). We note that the parameters $c_3$ and $c_4$ are not strictly necessary to reproduce the CMB phenomenology and, although we keep them to the values adopted in Ref.~\cite{Garcia-Bellido:2023ser} to directly compare with their results, none of our remarks would change if we set them to zero. The slopes $c_5$ and $c_6$ govern the axion velocity and the particle production at intermediate and late times, respectively. The remaining parameters $p_1$–$p_8$ are then determined by continuity of $V$ and $V'$ across the five segments. Therefore, the model is parametrized by $12$ parameters: $f$, $c_1 ,\, \dots ,\, c_6 ,\, {\tilde \phi}_0,\, {\tilde \phi}_1 ,\, \dots ,\, {\tilde \phi}_4$.

Figure~\ref{fig:potential} shows the potential~\eqref{potential} for the choice parameters employed in~\cite{Garcia-Bellido:2023ser} (provided in the caption), and in Subsection~\ref{subsec:baseline-parameters}. In Subsection~\ref{subsec:vary-parameters} we then study how a change in some of the model parameters impacts the phenomenology of the model. Table~\ref{tab:param-res} lists the modified parameters and some of the resulting phenomenological outcomes. At the technical level, we note that modifying the coupling strength $f$, while keeping the potential unchanged, in fact modifies all the rescaled dimensionless coefficients. This is due to the fact that the parametrization~\eqref{potential} is done in terms of the rescaled field ${\tilde \phi}$ and that the rescaling of the inflaton depends on $f$. Specifically, rescaling $f \to \kappa f$ at fixed potential results in 
\begin{align}
\tilde{f} \to \kappa\, \tilde{f} \;,\qquad
\tilde{\phi}_i \to \tilde{\phi}_i/\kappa \;,\qquad
c_1 &\to \kappa\, c_1 \;,\qquad
c_2 \to \kappa^2\, c_2 \;,\quad\;\;
c_3 \to \kappa^3\, c_3 \;,\qquad \nonumber\\
c_4 \to \kappa^4\, c_4 \;,\qquad
c_5 &\to \kappa\, c_5 \;,\qquad
c_6 \to \kappa\, c_6 \;. 
\label{kappa}
\end{align}

In the main text,  particular attention is paid to the dependence of the sourced signals on the coupling ${\tilde f}^{-1}$ and the slope of the potential $c_5$ in the intermediate region. For this discussion, it is convenient to parametrize the slope of the potential with a parameter that, unlike $c_5$, is not affected by the value of ${\tilde f}^{-1}$. Specifically, we introduce
\begin{equation}
v' \equiv \frac{M_p}{V_0} \, \frac{\partial V}{\partial \varphi} = \frac{c_5}{\tilde f} \;\;\;\;\;\;,\;\;\;\;\;\; \tilde{\phi}_2 \leq \tilde{\phi} \leq \tilde{\phi}_3 \;, 
\label{vprime}
\end{equation} 
which, for the `baseline' model, evaluates to $v' \simeq - 0.2822$. 

\begin{table}
\renewcommand{\arraystretch}{1.2}  
\begin{tabular}{|r||c|c|c|c|c|}
\hline
{\rm Figure} & {\rm Parameters changed} & ${\rm Max} \: \frac{\rho_{\rm grad}}{\rho_{\rm kin}}$ & $\Delta N_{\rm eff}$ & $\mu$ & $y$ \\
\hline \hline
\ref{fig:GW_57_495} & Baseline parameters & 17.68 & $2 \cdot 10^{-2}$ & $10^{-9}$ & $8 \cdot 10^{-10}$ \\ \hline
\ref{fig:Results_57_27} &  $v' = - 0.1516$ & 0.039 & $3 \cdot 10^{-4}$ & $9 \cdot 10^{-10}$ & $8 \cdot 10^{-10}$ \\ \hline
\ref{fig:Results_c495} & ${\tilde f}^{-1} = 22$ & 0.036 & $9 \cdot 10^{-4}$ & $5 \cdot 10^{-10}$ & $6 \cdot 10^{-10}$ \\ \hline
top \ref{fig:Results_f_c5} & ${\tilde f}^{-1} = 80 \;,\; v' = - 0.1237$ & 0.450 & $2 \cdot 10^{-4}$ & $10^{-9}$ & $8 \cdot 10^{-10}$ \\ \hline
bottom \ref{fig:Results_f_c5} & ${\tilde f}^{-1} = 75 \;,\; v' = - 0.1129$ & 0.016 & $5 \cdot 10^{-5}$ & $10^{-9}$ & $8 \cdot 10^{-10}$ \\ \hline
\ref{fig:Results_Deltaphi_c5} & $\tilde{\phi_3}=290 \;,\; v' = -0.1351$ & 0.0371 & $4 \cdot 10^{-4}$ & $9 \cdot 10^{-10}$ & $8 \cdot 10^{-10}$ \\ \hline
\end{tabular}
\caption{Parameters and results for the evolutions shown in the main text. For each evolution (each row): the second column lists the parameters changed in that evolution relative to those in the `baseline' model (which is the one employed in the first evolution in the table); the third column gives the maximum value attained by the ratio between the axion gradient and (zero mode) kinetic energy; the fourth column gives the present energy in the sourced SGWB (parametrized in terms of the effective number of extra relativistic degrees of freedom according to eq.~\eqref{Delta-N-eff}); the last two columns provide, respectively, the  CMB $\mu-$ and $y-$ distortions from the scalar perturbations evaluated with eq.~\eqref{mu-y}.
The amount of distortions (particularly the $y$ ones) is nearly identical for the various evolutions, as they are dominated by the vacuum scalar modes.}
\label{tab:param-res}
\end{table}

\bibliographystyle{apsrev}
\bibliography{paper-biblio}

\begin{thebibliography}{119}
\expandafter\ifx\csname natexlab\endcsname\relax\def\natexlab#1{#1}\fi
\expandafter\ifx\csname bibnamefont\endcsname\relax
  \def\bibnamefont#1{#1}\fi
\expandafter\ifx\csname bibfnamefont\endcsname\relax
  \def\bibfnamefont#1{#1}\fi
\expandafter\ifx\csname citenamefont\endcsname\relax
  \def\citenamefont#1{#1}\fi
\expandafter\ifx\csname url\endcsname\relax
  \def\url#1{\texttt{#1}}\fi
\expandafter\ifx\csname urlprefix\endcsname\relax\def\urlprefix{URL }\fi
\providecommand{\bibinfo}[2]{#2}
\providecommand{\eprint}[2][]{\url{#2}}

\bibitem[{\citenamefont{Akrami et~al.}(2020)}]{Planck:2018jri}
\bibinfo{author}{\bibfnamefont{Y.}~\bibnamefont{Akrami}} \bibnamefont{et~al.}
  (\bibinfo{collaboration}{Planck}), \bibinfo{journal}{Astron. Astrophys.}
  \textbf{\bibinfo{volume}{641}}, \bibinfo{pages}{A10} (\bibinfo{year}{2020}),
  \eprint{1807.06211}.

\bibitem[{\citenamefont{Ade et~al.}(2021)}]{BICEP:2021xfz}
\bibinfo{author}{\bibfnamefont{P.~A.~R.} \bibnamefont{Ade}}
  \bibnamefont{et~al.} (\bibinfo{collaboration}{BICEP, Keck}),
  \bibinfo{journal}{Phys. Rev. Lett.} \textbf{\bibinfo{volume}{127}},
  \bibinfo{pages}{151301} (\bibinfo{year}{2021}), \eprint{2110.00483}.

\bibitem[{\citenamefont{Guth}(1981)}]{Guth:1980zm}
\bibinfo{author}{\bibfnamefont{A.~H.} \bibnamefont{Guth}},
  \bibinfo{journal}{Phys. Rev. D} \textbf{\bibinfo{volume}{23}},
  \bibinfo{pages}{347} (\bibinfo{year}{1981}).

\bibitem[{\citenamefont{Linde}(1982)}]{Linde:1981mu}
\bibinfo{author}{\bibfnamefont{A.~D.} \bibnamefont{Linde}},
  \bibinfo{journal}{Phys. Lett. B} \textbf{\bibinfo{volume}{108}},
  \bibinfo{pages}{389} (\bibinfo{year}{1982}).

\bibitem[{\citenamefont{Albrecht and Steinhardt}(1982)}]{Albrecht:1982wi}
\bibinfo{author}{\bibfnamefont{A.}~\bibnamefont{Albrecht}} \bibnamefont{and}
  \bibinfo{author}{\bibfnamefont{P.~J.} \bibnamefont{Steinhardt}},
  \bibinfo{journal}{Phys. Rev. Lett.} \textbf{\bibinfo{volume}{48}},
  \bibinfo{pages}{1220} (\bibinfo{year}{1982}).

\bibitem[{\citenamefont{Lyth}(1997)}]{Lyth:1996im}
\bibinfo{author}{\bibfnamefont{D.~H.} \bibnamefont{Lyth}},
  \bibinfo{journal}{Phys. Rev. Lett.} \textbf{\bibinfo{volume}{78}},
  \bibinfo{pages}{1861} (\bibinfo{year}{1997}), \eprint{hep-ph/9606387}.

\bibitem[{\citenamefont{Freese et~al.}(1990)\citenamefont{Freese, Frieman, and
  Olinto}}]{Freese:1990rb}
\bibinfo{author}{\bibfnamefont{K.}~\bibnamefont{Freese}},
  \bibinfo{author}{\bibfnamefont{J.~A.} \bibnamefont{Frieman}},
  \bibnamefont{and} \bibinfo{author}{\bibfnamefont{A.~V.}
  \bibnamefont{Olinto}}, \bibinfo{journal}{Phys. Rev. Lett.}
  \textbf{\bibinfo{volume}{65}}, \bibinfo{pages}{3233} (\bibinfo{year}{1990}).

\bibitem[{\citenamefont{Adams et~al.}(1993)\citenamefont{Adams, Bond, Freese,
  Frieman, and Olinto}}]{Adams:1992bn}
\bibinfo{author}{\bibfnamefont{F.~C.} \bibnamefont{Adams}},
  \bibinfo{author}{\bibfnamefont{J.~R.} \bibnamefont{Bond}},
  \bibinfo{author}{\bibfnamefont{K.}~\bibnamefont{Freese}},
  \bibinfo{author}{\bibfnamefont{J.~A.} \bibnamefont{Frieman}},
  \bibnamefont{and} \bibinfo{author}{\bibfnamefont{A.~V.}
  \bibnamefont{Olinto}}, \bibinfo{journal}{Phys. Rev. D}
  \textbf{\bibinfo{volume}{47}}, \bibinfo{pages}{426} (\bibinfo{year}{1993}),
  \eprint{hep-ph/9207245}.

\bibitem[{\citenamefont{Pajer and Peloso}(2013)}]{Pajer:2013fsa}
\bibinfo{author}{\bibfnamefont{E.}~\bibnamefont{Pajer}} \bibnamefont{and}
  \bibinfo{author}{\bibfnamefont{M.}~\bibnamefont{Peloso}},
  \bibinfo{journal}{Class. Quant. Grav.} \textbf{\bibinfo{volume}{30}},
  \bibinfo{pages}{214002} (\bibinfo{year}{2013}), \eprint{1305.3557}.

\bibitem[{\citenamefont{Kim et~al.}(2005)\citenamefont{Kim, Nilles, and
  Peloso}}]{Kim:2004rp}
\bibinfo{author}{\bibfnamefont{J.~E.} \bibnamefont{Kim}},
  \bibinfo{author}{\bibfnamefont{H.~P.} \bibnamefont{Nilles}},
  \bibnamefont{and} \bibinfo{author}{\bibfnamefont{M.}~\bibnamefont{Peloso}},
  \bibinfo{journal}{JCAP} \textbf{\bibinfo{volume}{01}}, \bibinfo{pages}{005}
  (\bibinfo{year}{2005}), \eprint{hep-ph/0409138}.

\bibitem[{\citenamefont{Peloso and Unal}(2015)}]{Peloso:2015dsa}
\bibinfo{author}{\bibfnamefont{M.}~\bibnamefont{Peloso}} \bibnamefont{and}
  \bibinfo{author}{\bibfnamefont{C.}~\bibnamefont{Unal}},
  \bibinfo{journal}{JCAP} \textbf{\bibinfo{volume}{06}}, \bibinfo{pages}{040}
  (\bibinfo{year}{2015}), \eprint{1504.02784}.

\bibitem[{\citenamefont{Abazajian et~al.}(2022)}]{CMB-S4:2020lpa}
\bibinfo{author}{\bibfnamefont{K.}~\bibnamefont{Abazajian}}
  \bibnamefont{et~al.} (\bibinfo{collaboration}{CMB-S4}),
  \bibinfo{journal}{Astrophys. J.} \textbf{\bibinfo{volume}{926}},
  \bibinfo{pages}{54} (\bibinfo{year}{2022}), \eprint{2008.12619}.

\bibitem[{\citenamefont{Fuskeland et~al.}(2023)}]{LiteBIRD:2023iei}
\bibinfo{author}{\bibfnamefont{U.}~\bibnamefont{Fuskeland}}
  \bibnamefont{et~al.} (\bibinfo{collaboration}{LiteBIRD}),
  \bibinfo{journal}{Astron. Astrophys.} \textbf{\bibinfo{volume}{676}},
  \bibinfo{pages}{A42} (\bibinfo{year}{2023}), \eprint{2302.05228}.

\bibitem[{\citenamefont{Greco and Peloso}(2025)}]{Greco:2024ngr}
\bibinfo{author}{\bibfnamefont{F.}~\bibnamefont{Greco}} \bibnamefont{and}
  \bibinfo{author}{\bibfnamefont{M.}~\bibnamefont{Peloso}},
  \bibinfo{journal}{JCAP} \textbf{\bibinfo{volume}{01}}, \bibinfo{pages}{074}
  (\bibinfo{year}{2025}), \eprint{2409.01126}.

\bibitem[{\citenamefont{Turner and Widrow}(1988)}]{Turner:1987bw}
\bibinfo{author}{\bibfnamefont{M.~S.} \bibnamefont{Turner}} \bibnamefont{and}
  \bibinfo{author}{\bibfnamefont{L.~M.} \bibnamefont{Widrow}},
  \bibinfo{journal}{Phys. Rev. D} \textbf{\bibinfo{volume}{37}},
  \bibinfo{pages}{2743} (\bibinfo{year}{1988}).

\bibitem[{\citenamefont{Garretson et~al.}(1992)\citenamefont{Garretson, Field,
  and Carroll}}]{Garretson:1992vt}
\bibinfo{author}{\bibfnamefont{W.~D.} \bibnamefont{Garretson}},
  \bibinfo{author}{\bibfnamefont{G.~B.} \bibnamefont{Field}}, \bibnamefont{and}
  \bibinfo{author}{\bibfnamefont{S.~M.} \bibnamefont{Carroll}},
  \bibinfo{journal}{Phys. Rev. D} \textbf{\bibinfo{volume}{46}},
  \bibinfo{pages}{5346} (\bibinfo{year}{1992}), \eprint{hep-ph/9209238}.

\bibitem[{\citenamefont{Anber and Sorbo}(2006)}]{Anber:2006xt}
\bibinfo{author}{\bibfnamefont{M.~M.} \bibnamefont{Anber}} \bibnamefont{and}
  \bibinfo{author}{\bibfnamefont{L.}~\bibnamefont{Sorbo}},
  \bibinfo{journal}{JCAP} \textbf{\bibinfo{volume}{10}}, \bibinfo{pages}{018}
  (\bibinfo{year}{2006}), \eprint{astro-ph/0606534}.

\bibitem[{\citenamefont{Anber and Sorbo}(2010)}]{Anber:2009ua}
\bibinfo{author}{\bibfnamefont{M.~M.} \bibnamefont{Anber}} \bibnamefont{and}
  \bibinfo{author}{\bibfnamefont{L.}~\bibnamefont{Sorbo}},
  \bibinfo{journal}{Phys. Rev. D} \textbf{\bibinfo{volume}{81}},
  \bibinfo{pages}{043534} (\bibinfo{year}{2010}), \eprint{0908.4089}.

\bibitem[{\citenamefont{Adshead and Wyman}(2012)}]{Adshead:2012kp}
\bibinfo{author}{\bibfnamefont{P.}~\bibnamefont{Adshead}} \bibnamefont{and}
  \bibinfo{author}{\bibfnamefont{M.}~\bibnamefont{Wyman}},
  \bibinfo{journal}{Phys. Rev. Lett.} \textbf{\bibinfo{volume}{108}},
  \bibinfo{pages}{261302} (\bibinfo{year}{2012}), \eprint{1202.2366}.

\bibitem[{\citenamefont{Dimastrogiovanni
  et~al.}(2017)\citenamefont{Dimastrogiovanni, Fasiello, and
  Fujita}}]{Dimastrogiovanni:2016fuu}
\bibinfo{author}{\bibfnamefont{E.}~\bibnamefont{Dimastrogiovanni}},
  \bibinfo{author}{\bibfnamefont{M.}~\bibnamefont{Fasiello}}, \bibnamefont{and}
  \bibinfo{author}{\bibfnamefont{T.}~\bibnamefont{Fujita}},
  \bibinfo{journal}{JCAP} \textbf{\bibinfo{volume}{01}}, \bibinfo{pages}{019}
  (\bibinfo{year}{2017}), \eprint{1608.04216}.

\bibitem[{\citenamefont{Maleknejad and Komatsu}(2019)}]{Maleknejad:2018nxz}
\bibinfo{author}{\bibfnamefont{A.}~\bibnamefont{Maleknejad}} \bibnamefont{and}
  \bibinfo{author}{\bibfnamefont{E.}~\bibnamefont{Komatsu}},
  \bibinfo{journal}{JHEP} \textbf{\bibinfo{volume}{05}}, \bibinfo{pages}{174}
  (\bibinfo{year}{2019}), \eprint{1808.09076}.

\bibitem[{\citenamefont{Ishiwata et~al.}(2022)\citenamefont{Ishiwata, Komatsu,
  and Obata}}]{Ishiwata:2021yne}
\bibinfo{author}{\bibfnamefont{K.}~\bibnamefont{Ishiwata}},
  \bibinfo{author}{\bibfnamefont{E.}~\bibnamefont{Komatsu}}, \bibnamefont{and}
  \bibinfo{author}{\bibfnamefont{I.}~\bibnamefont{Obata}},
  \bibinfo{journal}{JCAP} \textbf{\bibinfo{volume}{03}}, \bibinfo{pages}{010}
  (\bibinfo{year}{2022}), \eprint{2111.14429}.

\bibitem[{\citenamefont{Iarygina et~al.}(2024)\citenamefont{Iarygina,
  Sfakianakis, Sharma, and Brandenburg}}]{Iarygina:2023mtj}
\bibinfo{author}{\bibfnamefont{O.}~\bibnamefont{Iarygina}},
  \bibinfo{author}{\bibfnamefont{E.~I.} \bibnamefont{Sfakianakis}},
  \bibinfo{author}{\bibfnamefont{R.}~\bibnamefont{Sharma}}, \bibnamefont{and}
  \bibinfo{author}{\bibfnamefont{A.}~\bibnamefont{Brandenburg}},
  \bibinfo{journal}{JCAP} \textbf{\bibinfo{volume}{04}}, \bibinfo{pages}{018}
  (\bibinfo{year}{2024}), \eprint{2311.07557}.

\bibitem[{\citenamefont{Dimastrogiovanni
  et~al.}(2025)\citenamefont{Dimastrogiovanni, Fasiello, Papageorgiou, and
  Gatica}}]{Dimastrogiovanni:2025snj}
\bibinfo{author}{\bibfnamefont{E.}~\bibnamefont{Dimastrogiovanni}},
  \bibinfo{author}{\bibfnamefont{M.}~\bibnamefont{Fasiello}},
  \bibinfo{author}{\bibfnamefont{A.}~\bibnamefont{Papageorgiou}},
  \bibnamefont{and} \bibinfo{author}{\bibfnamefont{C.~Z.} \bibnamefont{Gatica}}
  (\bibinfo{year}{2025}), \eprint{2504.17750}.

\bibitem[{\citenamefont{Barnaby and Peloso}(2011)}]{Barnaby:2010vf}
\bibinfo{author}{\bibfnamefont{N.}~\bibnamefont{Barnaby}} \bibnamefont{and}
  \bibinfo{author}{\bibfnamefont{M.}~\bibnamefont{Peloso}},
  \bibinfo{journal}{Phys. Rev. Lett.} \textbf{\bibinfo{volume}{106}},
  \bibinfo{pages}{181301} (\bibinfo{year}{2011}), \eprint{1011.1500}.

\bibitem[{\citenamefont{Ade et~al.}(2016)}]{Planck:2015zfm}
\bibinfo{author}{\bibfnamefont{P.~A.~R.} \bibnamefont{Ade}}
  \bibnamefont{et~al.} (\bibinfo{collaboration}{Planck}),
  \bibinfo{journal}{Astron. Astrophys.} \textbf{\bibinfo{volume}{594}},
  \bibinfo{pages}{A17} (\bibinfo{year}{2016}), \eprint{1502.01592}.

\bibitem[{\citenamefont{Jamieson et~al.}(2025)\citenamefont{Jamieson, Caravano,
  and Komatsu}}]{Jamieson:2025ngu}
\bibinfo{author}{\bibfnamefont{D.}~\bibnamefont{Jamieson}},
  \bibinfo{author}{\bibfnamefont{A.}~\bibnamefont{Caravano}}, \bibnamefont{and}
  \bibinfo{author}{\bibfnamefont{E.}~\bibnamefont{Komatsu}}
  (\bibinfo{year}{2025}), \eprint{2507.22285}.

\bibitem[{\citenamefont{Barnaby
  et~al.}(2012{\natexlab{a}})\citenamefont{Barnaby, Pajer, and
  Peloso}}]{Barnaby:2011qe}
\bibinfo{author}{\bibfnamefont{N.}~\bibnamefont{Barnaby}},
  \bibinfo{author}{\bibfnamefont{E.}~\bibnamefont{Pajer}}, \bibnamefont{and}
  \bibinfo{author}{\bibfnamefont{M.}~\bibnamefont{Peloso}},
  \bibinfo{journal}{Phys. Rev. D} \textbf{\bibinfo{volume}{85}},
  \bibinfo{pages}{023525} (\bibinfo{year}{2012}{\natexlab{a}}),
  \eprint{1110.3327}.

\bibitem[{\citenamefont{Cook and Sorbo}(2012)}]{Cook:2011hg}
\bibinfo{author}{\bibfnamefont{J.~L.} \bibnamefont{Cook}} \bibnamefont{and}
  \bibinfo{author}{\bibfnamefont{L.}~\bibnamefont{Sorbo}},
  \bibinfo{journal}{Phys. Rev. D} \textbf{\bibinfo{volume}{85}},
  \bibinfo{pages}{023534} (\bibinfo{year}{2012}), \bibinfo{note}{[Erratum:
  Phys.Rev.D 86, 069901 (2012)]}, \eprint{1109.0022}.

\bibitem[{\citenamefont{Domcke et~al.}(2016)\citenamefont{Domcke, Pieroni, and
  Bin{\'e}truy}}]{Domcke:2016bkh}
\bibinfo{author}{\bibfnamefont{V.}~\bibnamefont{Domcke}},
  \bibinfo{author}{\bibfnamefont{M.}~\bibnamefont{Pieroni}}, \bibnamefont{and}
  \bibinfo{author}{\bibfnamefont{P.}~\bibnamefont{Bin{\'e}truy}},
  \bibinfo{journal}{JCAP} \textbf{\bibinfo{volume}{06}}, \bibinfo{pages}{031}
  (\bibinfo{year}{2016}), \eprint{1603.01287}.

\bibitem[{\citenamefont{Sorbo}(2011)}]{Sorbo:2011rz}
\bibinfo{author}{\bibfnamefont{L.}~\bibnamefont{Sorbo}},
  \bibinfo{journal}{JCAP} \textbf{\bibinfo{volume}{06}}, \bibinfo{pages}{003}
  (\bibinfo{year}{2011}), \eprint{1101.1525}.

\bibitem[{\citenamefont{Gluscevic and Kamionkowski}(2010)}]{Gluscevic:2010vv}
\bibinfo{author}{\bibfnamefont{V.}~\bibnamefont{Gluscevic}} \bibnamefont{and}
  \bibinfo{author}{\bibfnamefont{M.}~\bibnamefont{Kamionkowski}},
  \bibinfo{journal}{Phys. Rev. D} \textbf{\bibinfo{volume}{81}},
  \bibinfo{pages}{123529} (\bibinfo{year}{2010}), \eprint{1002.1308}.

\bibitem[{\citenamefont{Cruz et~al.}(2024)\citenamefont{Cruz, Malhotra,
  Tasinato, and Zavala}}]{Cruz:2024esk}
\bibinfo{author}{\bibfnamefont{N.~M.~J.} \bibnamefont{Cruz}},
  \bibinfo{author}{\bibfnamefont{A.}~\bibnamefont{Malhotra}},
  \bibinfo{author}{\bibfnamefont{G.}~\bibnamefont{Tasinato}}, \bibnamefont{and}
  \bibinfo{author}{\bibfnamefont{I.}~\bibnamefont{Zavala}},
  \bibinfo{journal}{Phys. Rev. D} \textbf{\bibinfo{volume}{110}},
  \bibinfo{pages}{103505} (\bibinfo{year}{2024}), \eprint{2406.04957}.

\bibitem[{\citenamefont{Seto}(2006)}]{Seto:2006hf}
\bibinfo{author}{\bibfnamefont{N.}~\bibnamefont{Seto}}, \bibinfo{journal}{Phys.
  Rev. Lett.} \textbf{\bibinfo{volume}{97}}, \bibinfo{pages}{151101}
  (\bibinfo{year}{2006}), \eprint{astro-ph/0609504}.

\bibitem[{\citenamefont{Seto}(2007)}]{Seto:2006dz}
\bibinfo{author}{\bibfnamefont{N.}~\bibnamefont{Seto}}, \bibinfo{journal}{Phys.
  Rev. D} \textbf{\bibinfo{volume}{75}}, \bibinfo{pages}{061302}
  (\bibinfo{year}{2007}), \eprint{astro-ph/0609633}.

\bibitem[{\citenamefont{Domcke et~al.}(2020{\natexlab{a}})\citenamefont{Domcke,
  Garcia-Bellido, Peloso, Pieroni, Ricciardone, Sorbo, and
  Tasinato}}]{Domcke:2019zls}
\bibinfo{author}{\bibfnamefont{V.}~\bibnamefont{Domcke}},
  \bibinfo{author}{\bibfnamefont{J.}~\bibnamefont{Garcia-Bellido}},
  \bibinfo{author}{\bibfnamefont{M.}~\bibnamefont{Peloso}},
  \bibinfo{author}{\bibfnamefont{M.}~\bibnamefont{Pieroni}},
  \bibinfo{author}{\bibfnamefont{A.}~\bibnamefont{Ricciardone}},
  \bibinfo{author}{\bibfnamefont{L.}~\bibnamefont{Sorbo}}, \bibnamefont{and}
  \bibinfo{author}{\bibfnamefont{G.}~\bibnamefont{Tasinato}},
  \bibinfo{journal}{JCAP} \textbf{\bibinfo{volume}{05}}, \bibinfo{pages}{028}
  (\bibinfo{year}{2020}{\natexlab{a}}), \eprint{1910.08052}.

\bibitem[{\citenamefont{Orlando et~al.}(2021)\citenamefont{Orlando, Pieroni,
  and Ricciardone}}]{Orlando:2020oko}
\bibinfo{author}{\bibfnamefont{G.}~\bibnamefont{Orlando}},
  \bibinfo{author}{\bibfnamefont{M.}~\bibnamefont{Pieroni}}, \bibnamefont{and}
  \bibinfo{author}{\bibfnamefont{A.}~\bibnamefont{Ricciardone}},
  \bibinfo{journal}{JCAP} \textbf{\bibinfo{volume}{03}}, \bibinfo{pages}{069}
  (\bibinfo{year}{2021}), \eprint{2011.07059}.

\bibitem[{\citenamefont{Seto and Taruya}(2007)}]{Seto:2007tn}
\bibinfo{author}{\bibfnamefont{N.}~\bibnamefont{Seto}} \bibnamefont{and}
  \bibinfo{author}{\bibfnamefont{A.}~\bibnamefont{Taruya}},
  \bibinfo{journal}{Phys. Rev. Lett.} \textbf{\bibinfo{volume}{99}},
  \bibinfo{pages}{121101} (\bibinfo{year}{2007}), \eprint{0707.0535}.

\bibitem[{\citenamefont{Seto and Taruya}(2008)}]{Seto:2008sr}
\bibinfo{author}{\bibfnamefont{N.}~\bibnamefont{Seto}} \bibnamefont{and}
  \bibinfo{author}{\bibfnamefont{A.}~\bibnamefont{Taruya}},
  \bibinfo{journal}{Phys. Rev. D} \textbf{\bibinfo{volume}{77}},
  \bibinfo{pages}{103001} (\bibinfo{year}{2008}), \eprint{0801.4185}.

\bibitem[{\citenamefont{Crowder et~al.}(2013)\citenamefont{Crowder, Namba,
  Mandic, Mukohyama, and Peloso}}]{Crowder:2012ik}
\bibinfo{author}{\bibfnamefont{S.~G.} \bibnamefont{Crowder}},
  \bibinfo{author}{\bibfnamefont{R.}~\bibnamefont{Namba}},
  \bibinfo{author}{\bibfnamefont{V.}~\bibnamefont{Mandic}},
  \bibinfo{author}{\bibfnamefont{S.}~\bibnamefont{Mukohyama}},
  \bibnamefont{and} \bibinfo{author}{\bibfnamefont{M.}~\bibnamefont{Peloso}},
  \bibinfo{journal}{Phys. Lett. B} \textbf{\bibinfo{volume}{726}},
  \bibinfo{pages}{66} (\bibinfo{year}{2013}), \eprint{1212.4165}.

\bibitem[{\citenamefont{Mentasti et~al.}(2023)\citenamefont{Mentasti, Contaldi,
  and Peloso}}]{Mentasti:2023gmg}
\bibinfo{author}{\bibfnamefont{G.}~\bibnamefont{Mentasti}},
  \bibinfo{author}{\bibfnamefont{C.}~\bibnamefont{Contaldi}}, \bibnamefont{and}
  \bibinfo{author}{\bibfnamefont{M.}~\bibnamefont{Peloso}},
  \bibinfo{journal}{JCAP} \textbf{\bibinfo{volume}{08}}, \bibinfo{pages}{053}
  (\bibinfo{year}{2023}), \eprint{2304.06640}.

\bibitem[{\citenamefont{Abac et~al.}(2025)}]{Abac:2025saz}
\bibinfo{author}{\bibfnamefont{A.}~\bibnamefont{Abac}} \bibnamefont{et~al.}
  (\bibinfo{year}{2025}), \eprint{2503.12263}.

\bibitem[{\citenamefont{Barnaby
  et~al.}(2012{\natexlab{b}})\citenamefont{Barnaby, Moxon, Namba, Peloso, Shiu,
  and Zhou}}]{Barnaby:2012xt}
\bibinfo{author}{\bibfnamefont{N.}~\bibnamefont{Barnaby}},
  \bibinfo{author}{\bibfnamefont{J.}~\bibnamefont{Moxon}},
  \bibinfo{author}{\bibfnamefont{R.}~\bibnamefont{Namba}},
  \bibinfo{author}{\bibfnamefont{M.}~\bibnamefont{Peloso}},
  \bibinfo{author}{\bibfnamefont{G.}~\bibnamefont{Shiu}}, \bibnamefont{and}
  \bibinfo{author}{\bibfnamefont{P.}~\bibnamefont{Zhou}},
  \bibinfo{journal}{Phys. Rev. D} \textbf{\bibinfo{volume}{86}},
  \bibinfo{pages}{103508} (\bibinfo{year}{2012}{\natexlab{b}}),
  \eprint{1206.6117}.

\bibitem[{\citenamefont{Namba et~al.}(2016)\citenamefont{Namba, Peloso,
  Shiraishi, Sorbo, and Unal}}]{Namba:2015gja}
\bibinfo{author}{\bibfnamefont{R.}~\bibnamefont{Namba}},
  \bibinfo{author}{\bibfnamefont{M.}~\bibnamefont{Peloso}},
  \bibinfo{author}{\bibfnamefont{M.}~\bibnamefont{Shiraishi}},
  \bibinfo{author}{\bibfnamefont{L.}~\bibnamefont{Sorbo}}, \bibnamefont{and}
  \bibinfo{author}{\bibfnamefont{C.}~\bibnamefont{Unal}},
  \bibinfo{journal}{JCAP} \textbf{\bibinfo{volume}{01}}, \bibinfo{pages}{041}
  (\bibinfo{year}{2016}), \eprint{1509.07521}.

\bibitem[{\citenamefont{Linde et~al.}(2013)\citenamefont{Linde, Mooij, and
  Pajer}}]{Linde:2012bt}
\bibinfo{author}{\bibfnamefont{A.}~\bibnamefont{Linde}},
  \bibinfo{author}{\bibfnamefont{S.}~\bibnamefont{Mooij}}, \bibnamefont{and}
  \bibinfo{author}{\bibfnamefont{E.}~\bibnamefont{Pajer}},
  \bibinfo{journal}{Phys. Rev. D} \textbf{\bibinfo{volume}{87}},
  \bibinfo{pages}{103506} (\bibinfo{year}{2013}), \eprint{1212.1693}.

\bibitem[{\citenamefont{Bugaev and Klimai}(2014)}]{Bugaev:2013fya}
\bibinfo{author}{\bibfnamefont{E.}~\bibnamefont{Bugaev}} \bibnamefont{and}
  \bibinfo{author}{\bibfnamefont{P.}~\bibnamefont{Klimai}},
  \bibinfo{journal}{Phys. Rev. D} \textbf{\bibinfo{volume}{90}},
  \bibinfo{pages}{103501} (\bibinfo{year}{2014}), \eprint{1312.7435}.

\bibitem[{\citenamefont{Garcia-Bellido
  et~al.}(2016)\citenamefont{Garcia-Bellido, Peloso, and
  Unal}}]{Garcia-Bellido:2016dkw}
\bibinfo{author}{\bibfnamefont{J.}~\bibnamefont{Garcia-Bellido}},
  \bibinfo{author}{\bibfnamefont{M.}~\bibnamefont{Peloso}}, \bibnamefont{and}
  \bibinfo{author}{\bibfnamefont{C.}~\bibnamefont{Unal}},
  \bibinfo{journal}{JCAP} \textbf{\bibinfo{volume}{12}}, \bibinfo{pages}{031}
  (\bibinfo{year}{2016}), \eprint{1610.03763}.

\bibitem[{\citenamefont{Garcia-Bellido
  et~al.}(2017)\citenamefont{Garcia-Bellido, Peloso, and
  Unal}}]{Garcia-Bellido:2017aan}
\bibinfo{author}{\bibfnamefont{J.}~\bibnamefont{Garcia-Bellido}},
  \bibinfo{author}{\bibfnamefont{M.}~\bibnamefont{Peloso}}, \bibnamefont{and}
  \bibinfo{author}{\bibfnamefont{C.}~\bibnamefont{Unal}},
  \bibinfo{journal}{JCAP} \textbf{\bibinfo{volume}{09}}, \bibinfo{pages}{013}
  (\bibinfo{year}{2017}), \eprint{1707.02441}.

\bibitem[{\citenamefont{{\"O}zsoy and Tasinato}(2023)}]{Ozsoy:2023ryl}
\bibinfo{author}{\bibfnamefont{O.}~\bibnamefont{{\"O}zsoy}} \bibnamefont{and}
  \bibinfo{author}{\bibfnamefont{G.}~\bibnamefont{Tasinato}},
  \bibinfo{journal}{Universe} \textbf{\bibinfo{volume}{9}},
  \bibinfo{pages}{203} (\bibinfo{year}{2023}), \eprint{2301.03600}.

\bibitem[{\citenamefont{Caravano et~al.}(2023)\citenamefont{Caravano, Komatsu,
  Lozanov, and Weller}}]{Caravano:2022epk}
\bibinfo{author}{\bibfnamefont{A.}~\bibnamefont{Caravano}},
  \bibinfo{author}{\bibfnamefont{E.}~\bibnamefont{Komatsu}},
  \bibinfo{author}{\bibfnamefont{K.~D.} \bibnamefont{Lozanov}},
  \bibnamefont{and} \bibinfo{author}{\bibfnamefont{J.}~\bibnamefont{Weller}},
  \bibinfo{journal}{Phys. Rev. D} \textbf{\bibinfo{volume}{108}},
  \bibinfo{pages}{043504} (\bibinfo{year}{2023}), \eprint{2204.12874}.

\bibitem[{\citenamefont{Creminelli et~al.}(2023)\citenamefont{Creminelli,
  Kumar, Salehian, and Santoni}}]{Creminelli:2023aly}
\bibinfo{author}{\bibfnamefont{P.}~\bibnamefont{Creminelli}},
  \bibinfo{author}{\bibfnamefont{S.}~\bibnamefont{Kumar}},
  \bibinfo{author}{\bibfnamefont{B.}~\bibnamefont{Salehian}}, \bibnamefont{and}
  \bibinfo{author}{\bibfnamefont{L.}~\bibnamefont{Santoni}},
  \bibinfo{journal}{JCAP} \textbf{\bibinfo{volume}{08}}, \bibinfo{pages}{076}
  (\bibinfo{year}{2023}), \eprint{2305.07695}.

\bibitem[{\citenamefont{Figueroa et~al.}(2023)\citenamefont{Figueroa,
  Lizarraga, Urio, and Urrestilla}}]{Figueroa:2023oxc}
\bibinfo{author}{\bibfnamefont{D.~G.} \bibnamefont{Figueroa}},
  \bibinfo{author}{\bibfnamefont{J.}~\bibnamefont{Lizarraga}},
  \bibinfo{author}{\bibfnamefont{A.}~\bibnamefont{Urio}}, \bibnamefont{and}
  \bibinfo{author}{\bibfnamefont{J.}~\bibnamefont{Urrestilla}},
  \bibinfo{journal}{Phys. Rev. Lett.} \textbf{\bibinfo{volume}{131}},
  \bibinfo{pages}{151003} (\bibinfo{year}{2023}), \eprint{2303.17436}.

\bibitem[{\citenamefont{Cheng et~al.}(2016)\citenamefont{Cheng, Lee, and
  Ng}}]{Cheng:2015oqa}
\bibinfo{author}{\bibfnamefont{S.-L.} \bibnamefont{Cheng}},
  \bibinfo{author}{\bibfnamefont{W.}~\bibnamefont{Lee}}, \bibnamefont{and}
  \bibinfo{author}{\bibfnamefont{K.-W.} \bibnamefont{Ng}},
  \bibinfo{journal}{Phys. Rev. D} \textbf{\bibinfo{volume}{93}},
  \bibinfo{pages}{063510} (\bibinfo{year}{2016}), \eprint{1508.00251}.

\bibitem[{\citenamefont{Notari and Tywoniuk}(2016)}]{Notari:2016npn}
\bibinfo{author}{\bibfnamefont{A.}~\bibnamefont{Notari}} \bibnamefont{and}
  \bibinfo{author}{\bibfnamefont{K.}~\bibnamefont{Tywoniuk}},
  \bibinfo{journal}{JCAP} \textbf{\bibinfo{volume}{12}}, \bibinfo{pages}{038}
  (\bibinfo{year}{2016}), \eprint{1608.06223}.

\bibitem[{\citenamefont{Dall'Agata et~al.}(2020)\citenamefont{Dall'Agata,
  Gonz{\'a}lez-Mart{\'\i}n, Papageorgiou, and Peloso}}]{DallAgata:2019yrr}
\bibinfo{author}{\bibfnamefont{G.}~\bibnamefont{Dall'Agata}},
  \bibinfo{author}{\bibfnamefont{S.}~\bibnamefont{Gonz{\'a}lez-Mart{\'\i}n}},
  \bibinfo{author}{\bibfnamefont{A.}~\bibnamefont{Papageorgiou}},
  \bibnamefont{and} \bibinfo{author}{\bibfnamefont{M.}~\bibnamefont{Peloso}},
  \bibinfo{journal}{JCAP} \textbf{\bibinfo{volume}{08}}, \bibinfo{pages}{032}
  (\bibinfo{year}{2020}), \eprint{1912.09950}.

\bibitem[{\citenamefont{Domcke et~al.}(2020{\natexlab{b}})\citenamefont{Domcke,
  Guidetti, Welling, and Westphal}}]{Domcke:2020zez}
\bibinfo{author}{\bibfnamefont{V.}~\bibnamefont{Domcke}},
  \bibinfo{author}{\bibfnamefont{V.}~\bibnamefont{Guidetti}},
  \bibinfo{author}{\bibfnamefont{Y.}~\bibnamefont{Welling}}, \bibnamefont{and}
  \bibinfo{author}{\bibfnamefont{A.}~\bibnamefont{Westphal}},
  \bibinfo{journal}{JCAP} \textbf{\bibinfo{volume}{09}}, \bibinfo{pages}{009}
  (\bibinfo{year}{2020}{\natexlab{b}}), \eprint{2002.02952}.

\bibitem[{\citenamefont{Peloso and Sorbo}(2023)}]{Peloso:2022ovc}
\bibinfo{author}{\bibfnamefont{M.}~\bibnamefont{Peloso}} \bibnamefont{and}
  \bibinfo{author}{\bibfnamefont{L.}~\bibnamefont{Sorbo}},
  \bibinfo{journal}{JCAP} \textbf{\bibinfo{volume}{01}}, \bibinfo{pages}{038}
  (\bibinfo{year}{2023}), \eprint{2209.08131}.

\bibitem[{\citenamefont{Gorbar et~al.}(2021)\citenamefont{Gorbar, Schmitz,
  Sobol, and Vilchinskii}}]{Gorbar:2021rlt}
\bibinfo{author}{\bibfnamefont{E.~V.} \bibnamefont{Gorbar}},
  \bibinfo{author}{\bibfnamefont{K.}~\bibnamefont{Schmitz}},
  \bibinfo{author}{\bibfnamefont{O.~O.} \bibnamefont{Sobol}}, \bibnamefont{and}
  \bibinfo{author}{\bibfnamefont{S.~I.} \bibnamefont{Vilchinskii}},
  \bibinfo{journal}{Phys. Rev. D} \textbf{\bibinfo{volume}{104}},
  \bibinfo{pages}{123504} (\bibinfo{year}{2021}), \eprint{2109.01651}.

\bibitem[{\citenamefont{Durrer et~al.}(2023)\citenamefont{Durrer, Sobol, and
  Vilchinskii}}]{Durrer:2023rhc}
\bibinfo{author}{\bibfnamefont{R.}~\bibnamefont{Durrer}},
  \bibinfo{author}{\bibfnamefont{O.}~\bibnamefont{Sobol}}, \bibnamefont{and}
  \bibinfo{author}{\bibfnamefont{S.}~\bibnamefont{Vilchinskii}},
  \bibinfo{journal}{Phys. Rev. D} \textbf{\bibinfo{volume}{108}},
  \bibinfo{pages}{043540} (\bibinfo{year}{2023}), \eprint{2303.04583}.

\bibitem[{\citenamefont{von Eckardstein et~al.}(2023)\citenamefont{von
  Eckardstein, Peloso, Schmitz, Sobol, and Sorbo}}]{vonEckardstein:2023gwk}
\bibinfo{author}{\bibfnamefont{R.}~\bibnamefont{von Eckardstein}},
  \bibinfo{author}{\bibfnamefont{M.}~\bibnamefont{Peloso}},
  \bibinfo{author}{\bibfnamefont{K.}~\bibnamefont{Schmitz}},
  \bibinfo{author}{\bibfnamefont{O.}~\bibnamefont{Sobol}}, \bibnamefont{and}
  \bibinfo{author}{\bibfnamefont{L.}~\bibnamefont{Sorbo}},
  \bibinfo{journal}{JHEP} \textbf{\bibinfo{volume}{11}}, \bibinfo{pages}{183}
  (\bibinfo{year}{2023}), \eprint{2309.04254}.

\bibitem[{\citenamefont{Sobol et~al.}(2019)\citenamefont{Sobol, Gorbar, and
  Vilchinskii}}]{Sobol:2019xls}
\bibinfo{author}{\bibfnamefont{O.~O.} \bibnamefont{Sobol}},
  \bibinfo{author}{\bibfnamefont{E.~V.} \bibnamefont{Gorbar}},
  \bibnamefont{and} \bibinfo{author}{\bibfnamefont{S.~I.}
  \bibnamefont{Vilchinskii}}, \bibinfo{journal}{Phys. Rev. D}
  \textbf{\bibinfo{volume}{100}}, \bibinfo{pages}{063523}
  (\bibinfo{year}{2019}), \eprint{1907.10443}.

\bibitem[{\citenamefont{Domcke et~al.}(2024)\citenamefont{Domcke, Ema, and
  Sandner}}]{Domcke:2023tnn}
\bibinfo{author}{\bibfnamefont{V.}~\bibnamefont{Domcke}},
  \bibinfo{author}{\bibfnamefont{Y.}~\bibnamefont{Ema}}, \bibnamefont{and}
  \bibinfo{author}{\bibfnamefont{S.}~\bibnamefont{Sandner}},
  \bibinfo{journal}{JCAP} \textbf{\bibinfo{volume}{03}}, \bibinfo{pages}{019}
  (\bibinfo{year}{2024}), \eprint{2310.09186}.

\bibitem[{\citenamefont{Durrer et~al.}(2024)\citenamefont{Durrer, von
  Eckardstein, Garg, Schmitz, Sobol, and Vilchinskii}}]{Durrer:2024ibi}
\bibinfo{author}{\bibfnamefont{R.}~\bibnamefont{Durrer}},
  \bibinfo{author}{\bibfnamefont{R.}~\bibnamefont{von Eckardstein}},
  \bibinfo{author}{\bibfnamefont{D.}~\bibnamefont{Garg}},
  \bibinfo{author}{\bibfnamefont{K.}~\bibnamefont{Schmitz}},
  \bibinfo{author}{\bibfnamefont{O.}~\bibnamefont{Sobol}}, \bibnamefont{and}
  \bibinfo{author}{\bibfnamefont{S.}~\bibnamefont{Vilchinskii}},
  \bibinfo{journal}{Phys. Rev. D} \textbf{\bibinfo{volume}{110}},
  \bibinfo{pages}{043533} (\bibinfo{year}{2024}), \eprint{2404.19694}.

\bibitem[{\citenamefont{Garcia-Bellido
  et~al.}(2024)\citenamefont{Garcia-Bellido, Papageorgiou, Peloso, and
  Sorbo}}]{Garcia-Bellido:2023ser}
\bibinfo{author}{\bibfnamefont{J.}~\bibnamefont{Garcia-Bellido}},
  \bibinfo{author}{\bibfnamefont{A.}~\bibnamefont{Papageorgiou}},
  \bibinfo{author}{\bibfnamefont{M.}~\bibnamefont{Peloso}}, \bibnamefont{and}
  \bibinfo{author}{\bibfnamefont{L.}~\bibnamefont{Sorbo}},
  \bibinfo{journal}{JCAP} \textbf{\bibinfo{volume}{01}}, \bibinfo{pages}{034}
  (\bibinfo{year}{2024}), \eprint{2303.13425}.

\bibitem[{\citenamefont{{\"O}zsoy et~al.}(2024)\citenamefont{{\"O}zsoy,
  Papageorgiou, and Fasiello}}]{Ozsoy:2024apn}
\bibinfo{author}{\bibfnamefont{O.}~\bibnamefont{{\"O}zsoy}},
  \bibinfo{author}{\bibfnamefont{A.}~\bibnamefont{Papageorgiou}},
  \bibnamefont{and} \bibinfo{author}{\bibfnamefont{M.}~\bibnamefont{Fasiello}},
  \bibinfo{journal}{JCAP} \textbf{\bibinfo{volume}{12}}, \bibinfo{pages}{008}
  (\bibinfo{year}{2024}), \eprint{2405.14963}.

\bibitem[{\citenamefont{von Eckardstein et~al.}(2025)\citenamefont{von
  Eckardstein, Schmitz, and Sobol}}]{vonEckardstein:2025oic}
\bibinfo{author}{\bibfnamefont{R.}~\bibnamefont{von Eckardstein}},
  \bibinfo{author}{\bibfnamefont{K.}~\bibnamefont{Schmitz}}, \bibnamefont{and}
  \bibinfo{author}{\bibfnamefont{O.}~\bibnamefont{Sobol}}
  (\bibinfo{year}{2025}), \eprint{2508.00798}.

\bibitem[{\citenamefont{Caprini et~al.}(2019)\citenamefont{Caprini, Figueroa,
  Flauger, Nardini, Peloso, Pieroni, Ricciardone, and
  Tasinato}}]{Caprini:2019pxz}
\bibinfo{author}{\bibfnamefont{C.}~\bibnamefont{Caprini}},
  \bibinfo{author}{\bibfnamefont{D.~G.} \bibnamefont{Figueroa}},
  \bibinfo{author}{\bibfnamefont{R.}~\bibnamefont{Flauger}},
  \bibinfo{author}{\bibfnamefont{G.}~\bibnamefont{Nardini}},
  \bibinfo{author}{\bibfnamefont{M.}~\bibnamefont{Peloso}},
  \bibinfo{author}{\bibfnamefont{M.}~\bibnamefont{Pieroni}},
  \bibinfo{author}{\bibfnamefont{A.}~\bibnamefont{Ricciardone}},
  \bibnamefont{and} \bibinfo{author}{\bibfnamefont{G.}~\bibnamefont{Tasinato}},
  \bibinfo{journal}{JCAP} \textbf{\bibinfo{volume}{11}}, \bibinfo{pages}{017}
  (\bibinfo{year}{2019}), \eprint{1906.09244}.

\bibitem[{\citenamefont{Auclair
  et~al.}(2023)}]{LISACosmologyWorkingGroup:2022jok}
\bibinfo{author}{\bibfnamefont{P.}~\bibnamefont{Auclair}} \bibnamefont{et~al.}
  (\bibinfo{collaboration}{LISA Cosmology Working Group}),
  \bibinfo{journal}{Living Rev. Rel.} \textbf{\bibinfo{volume}{26}},
  \bibinfo{pages}{5} (\bibinfo{year}{2023}), \eprint{2204.05434}.

\bibitem[{\citenamefont{Braglia
  et~al.}(2024)}]{LISACosmologyWorkingGroup:2024hsc}
\bibinfo{author}{\bibfnamefont{M.}~\bibnamefont{Braglia}} \bibnamefont{et~al.}
  (\bibinfo{collaboration}{LISA Cosmology Working Group}),
  \bibinfo{journal}{JCAP} \textbf{\bibinfo{volume}{11}}, \bibinfo{pages}{032}
  (\bibinfo{year}{2024}), \eprint{2407.04356}.

\bibitem[{\citenamefont{Thorne et~al.}(2018)\citenamefont{Thorne, Fujita,
  Hazumi, Katayama, Komatsu, and Shiraishi}}]{Thorne:2017jft}
\bibinfo{author}{\bibfnamefont{B.}~\bibnamefont{Thorne}},
  \bibinfo{author}{\bibfnamefont{T.}~\bibnamefont{Fujita}},
  \bibinfo{author}{\bibfnamefont{M.}~\bibnamefont{Hazumi}},
  \bibinfo{author}{\bibfnamefont{N.}~\bibnamefont{Katayama}},
  \bibinfo{author}{\bibfnamefont{E.}~\bibnamefont{Komatsu}}, \bibnamefont{and}
  \bibinfo{author}{\bibfnamefont{M.}~\bibnamefont{Shiraishi}},
  \bibinfo{journal}{Phys. Rev. D} \textbf{\bibinfo{volume}{97}},
  \bibinfo{pages}{043506} (\bibinfo{year}{2018}), \eprint{1707.03240}.

\bibitem[{\citenamefont{Talebian et~al.}(2023)\citenamefont{Talebian,
  Hosseini~Mansoori, and Firouzjahi}}]{Talebian:2022cwk}
\bibinfo{author}{\bibfnamefont{A.}~\bibnamefont{Talebian}},
  \bibinfo{author}{\bibfnamefont{S.~A.} \bibnamefont{Hosseini~Mansoori}},
  \bibnamefont{and}
  \bibinfo{author}{\bibfnamefont{H.}~\bibnamefont{Firouzjahi}},
  \bibinfo{journal}{Astrophys. J.} \textbf{\bibinfo{volume}{948}},
  \bibinfo{pages}{48} (\bibinfo{year}{2023}), \eprint{2210.13822}.

\bibitem[{\citenamefont{Dimastrogiovanni
  et~al.}(2024)\citenamefont{Dimastrogiovanni, Fasiello, Leedom, Putti, and
  Westphal}}]{Dimastrogiovanni:2023juq}
\bibinfo{author}{\bibfnamefont{E.}~\bibnamefont{Dimastrogiovanni}},
  \bibinfo{author}{\bibfnamefont{M.}~\bibnamefont{Fasiello}},
  \bibinfo{author}{\bibfnamefont{J.~M.} \bibnamefont{Leedom}},
  \bibinfo{author}{\bibfnamefont{M.}~\bibnamefont{Putti}}, \bibnamefont{and}
  \bibinfo{author}{\bibfnamefont{A.}~\bibnamefont{Westphal}},
  \bibinfo{journal}{JHEP} \textbf{\bibinfo{volume}{08}}, \bibinfo{pages}{072}
  (\bibinfo{year}{2024}), \eprint{2312.13431}.

\bibitem[{\citenamefont{Putti et~al.}(2024)\citenamefont{Putti, Bartolo,
  Bhattacharya, and Peloso}}]{Putti:2024uyr}
\bibinfo{author}{\bibfnamefont{M.}~\bibnamefont{Putti}},
  \bibinfo{author}{\bibfnamefont{N.}~\bibnamefont{Bartolo}},
  \bibinfo{author}{\bibfnamefont{S.}~\bibnamefont{Bhattacharya}},
  \bibnamefont{and} \bibinfo{author}{\bibfnamefont{M.}~\bibnamefont{Peloso}},
  \bibinfo{journal}{JCAP} \textbf{\bibinfo{volume}{08}}, \bibinfo{pages}{016}
  (\bibinfo{year}{2024}), \eprint{2403.08594}.

\bibitem[{\citenamefont{Caravano and Peloso}(2025)}]{Caravano:2024xsb}
\bibinfo{author}{\bibfnamefont{A.}~\bibnamefont{Caravano}} \bibnamefont{and}
  \bibinfo{author}{\bibfnamefont{M.}~\bibnamefont{Peloso}},
  \bibinfo{journal}{JCAP} \textbf{\bibinfo{volume}{01}}, \bibinfo{pages}{104}
  (\bibinfo{year}{2025}), \eprint{2407.13405}.

\bibitem[{\citenamefont{He et~al.}(2025)\citenamefont{He, Zhang, Fu, and
  Guo}}]{He:2025ieo}
\bibinfo{author}{\bibfnamefont{J.-F.} \bibnamefont{He}},
  \bibinfo{author}{\bibfnamefont{K.-G.} \bibnamefont{Zhang}},
  \bibinfo{author}{\bibfnamefont{C.}~\bibnamefont{Fu}}, \bibnamefont{and}
  \bibinfo{author}{\bibfnamefont{Z.-K.} \bibnamefont{Guo}},
  \bibinfo{journal}{Phys. Rev. D} \textbf{\bibinfo{volume}{111}},
  \bibinfo{pages}{103525} (\bibinfo{year}{2025}), \eprint{2502.13158}.

\bibitem[{\citenamefont{Zhang et~al.}(2025)\citenamefont{Zhang, He, Fu, and
  Guo}}]{Zhang:2025cyd}
\bibinfo{author}{\bibfnamefont{K.-G.} \bibnamefont{Zhang}},
  \bibinfo{author}{\bibfnamefont{J.-F.} \bibnamefont{He}},
  \bibinfo{author}{\bibfnamefont{C.}~\bibnamefont{Fu}}, \bibnamefont{and}
  \bibinfo{author}{\bibfnamefont{Z.-K.} \bibnamefont{Guo}}
  (\bibinfo{year}{2025}), \eprint{2507.02611}.

\bibitem[{\citenamefont{Adshead et~al.}(2018)\citenamefont{Adshead, Giblin, and
  Weiner}}]{Adshead:2018doq}
\bibinfo{author}{\bibfnamefont{P.}~\bibnamefont{Adshead}},
  \bibinfo{author}{\bibfnamefont{J.~T.} \bibnamefont{Giblin}},
  \bibnamefont{and} \bibinfo{author}{\bibfnamefont{Z.~J.}
  \bibnamefont{Weiner}}, \bibinfo{journal}{Phys. Rev. D}
  \textbf{\bibinfo{volume}{98}}, \bibinfo{pages}{4} (\bibinfo{year}{2018}),
  \eprint{1805.04550}.

\bibitem[{\citenamefont{Adshead
  et~al.}(2020{\natexlab{a}})\citenamefont{Adshead, Giblin, Pieroni, and
  Weiner}}]{Adshead:2019lbr}
\bibinfo{author}{\bibfnamefont{P.}~\bibnamefont{Adshead}},
  \bibinfo{author}{\bibfnamefont{J.~T.} \bibnamefont{Giblin}},
  \bibinfo{author}{\bibfnamefont{M.}~\bibnamefont{Pieroni}}, \bibnamefont{and}
  \bibinfo{author}{\bibfnamefont{Z.~J.} \bibnamefont{Weiner}},
  \bibinfo{journal}{Phys. Rev. D} \textbf{\bibinfo{volume}{101}},
  \bibinfo{pages}{8} (\bibinfo{year}{2020}{\natexlab{a}}), \eprint{1909.12842}.

\bibitem[{\citenamefont{Adshead
  et~al.}(2020{\natexlab{b}})\citenamefont{Adshead, Giblin, Pieroni, and
  Weiner}}]{Adshead:2019igv}
\bibinfo{author}{\bibfnamefont{P.}~\bibnamefont{Adshead}},
  \bibinfo{author}{\bibfnamefont{J.~T.} \bibnamefont{Giblin}},
  \bibinfo{author}{\bibfnamefont{M.}~\bibnamefont{Pieroni}}, \bibnamefont{and}
  \bibinfo{author}{\bibfnamefont{Z.~J.} \bibnamefont{Weiner}},
  \bibinfo{journal}{Phys. Rev. Lett.} \textbf{\bibinfo{volume}{124}},
  \bibinfo{pages}{17} (\bibinfo{year}{2020}{\natexlab{b}}),
  \eprint{1909.12843}.

\bibitem[{\citenamefont{Bastero-Gil and Manso}(2023)}]{Bastero-Gil:2022fme}
\bibinfo{author}{\bibfnamefont{M.}~\bibnamefont{Bastero-Gil}} \bibnamefont{and}
  \bibinfo{author}{\bibfnamefont{A.~T.} \bibnamefont{Manso}},
  \bibinfo{journal}{JCAP} \textbf{\bibinfo{volume}{08}}, \bibinfo{pages}{001}
  (\bibinfo{year}{2023}), \eprint{2209.15572}.

\bibitem[{\citenamefont{Michelotti et~al.}(2025)\citenamefont{Michelotti,
  Gonzalez~Quaglia, Dimastrogiovanni, Fasiello, and
  Roest}}]{Michelotti:2024bbc}
\bibinfo{author}{\bibfnamefont{M.}~\bibnamefont{Michelotti}},
  \bibinfo{author}{\bibfnamefont{R.}~\bibnamefont{Gonzalez~Quaglia}},
  \bibinfo{author}{\bibfnamefont{E.}~\bibnamefont{Dimastrogiovanni}},
  \bibinfo{author}{\bibfnamefont{M.}~\bibnamefont{Fasiello}}, \bibnamefont{and}
  \bibinfo{author}{\bibfnamefont{D.}~\bibnamefont{Roest}},
  \bibinfo{journal}{JCAP} \textbf{\bibinfo{volume}{04}}, \bibinfo{pages}{058}
  (\bibinfo{year}{2025}), \eprint{2411.19892}.

\bibitem[{\citenamefont{Kume et~al.}(2025)\citenamefont{Kume, Peloso, and
  Bartolo}}]{Kume:2025lvz}
\bibinfo{author}{\bibfnamefont{J.}~\bibnamefont{Kume}},
  \bibinfo{author}{\bibfnamefont{M.}~\bibnamefont{Peloso}}, \bibnamefont{and}
  \bibinfo{author}{\bibfnamefont{N.}~\bibnamefont{Bartolo}},
  \bibinfo{journal}{JCAP} \textbf{\bibinfo{volume}{08}}, \bibinfo{pages}{044}
  (\bibinfo{year}{2025}), \eprint{2501.02890}.

\bibitem[{\citenamefont{Galanti et~al.}(2024)\citenamefont{Galanti, Conzinu,
  Marozzi, and Santos~da Costa}}]{Galanti:2024jhw}
\bibinfo{author}{\bibfnamefont{D.~C.} \bibnamefont{Galanti}},
  \bibinfo{author}{\bibfnamefont{P.}~\bibnamefont{Conzinu}},
  \bibinfo{author}{\bibfnamefont{G.}~\bibnamefont{Marozzi}}, \bibnamefont{and}
  \bibinfo{author}{\bibfnamefont{S.}~\bibnamefont{Santos~da Costa}},
  \bibinfo{journal}{Phys. Rev. D} \textbf{\bibinfo{volume}{110}},
  \bibinfo{pages}{123510} (\bibinfo{year}{2024}), \eprint{2406.19960}.

\bibitem[{\citenamefont{Caravano et~al.}(2022)\citenamefont{Caravano, Komatsu,
  Lozanov, and Weller}}]{Caravano:2021bfn}
\bibinfo{author}{\bibfnamefont{A.}~\bibnamefont{Caravano}},
  \bibinfo{author}{\bibfnamefont{E.}~\bibnamefont{Komatsu}},
  \bibinfo{author}{\bibfnamefont{K.~D.} \bibnamefont{Lozanov}},
  \bibnamefont{and} \bibinfo{author}{\bibfnamefont{J.}~\bibnamefont{Weller}},
  \bibinfo{journal}{Phys. Rev. D} \textbf{\bibinfo{volume}{105}},
  \bibinfo{pages}{123530} (\bibinfo{year}{2022}), \eprint{2110.10695}.

\bibitem[{\citenamefont{Sharma et~al.}(2025)\citenamefont{Sharma, Brandenburg,
  Subramanian, and Vikman}}]{Sharma:2024nfu}
\bibinfo{author}{\bibfnamefont{R.}~\bibnamefont{Sharma}},
  \bibinfo{author}{\bibfnamefont{A.}~\bibnamefont{Brandenburg}},
  \bibinfo{author}{\bibfnamefont{K.}~\bibnamefont{Subramanian}},
  \bibnamefont{and} \bibinfo{author}{\bibfnamefont{A.}~\bibnamefont{Vikman}},
  \bibinfo{journal}{JCAP} \textbf{\bibinfo{volume}{05}}, \bibinfo{pages}{079}
  (\bibinfo{year}{2025}), \eprint{2411.04854}.

\bibitem[{\citenamefont{Figueroa et~al.}(2025)\citenamefont{Figueroa,
  Lizarraga, Loayza, Urio, and Urrestilla}}]{Figueroa:2024rkr}
\bibinfo{author}{\bibfnamefont{D.~G.} \bibnamefont{Figueroa}},
  \bibinfo{author}{\bibfnamefont{J.}~\bibnamefont{Lizarraga}},
  \bibinfo{author}{\bibfnamefont{N.}~\bibnamefont{Loayza}},
  \bibinfo{author}{\bibfnamefont{A.}~\bibnamefont{Urio}}, \bibnamefont{and}
  \bibinfo{author}{\bibfnamefont{J.}~\bibnamefont{Urrestilla}},
  \bibinfo{journal}{Phys. Rev. D} \textbf{\bibinfo{volume}{111}},
  \bibinfo{pages}{063545} (\bibinfo{year}{2025}), \eprint{2411.16368}.

\bibitem[{\citenamefont{Lizarraga et~al.}(2025)\citenamefont{Lizarraga,
  L{\'o}pez-Mediavilla, and Urio}}]{Lizarraga:2025aiw}
\bibinfo{author}{\bibfnamefont{J.}~\bibnamefont{Lizarraga}},
  \bibinfo{author}{\bibfnamefont{C.}~\bibnamefont{L{\'o}pez-Mediavilla}},
  \bibnamefont{and} \bibinfo{author}{\bibfnamefont{A.}~\bibnamefont{Urio}}
  (\bibinfo{year}{2025}), \eprint{2505.19950}.

\bibitem[{\citenamefont{Adshead et~al.}(2015)\citenamefont{Adshead, Giblin,
  Scully, and Sfakianakis}}]{Adshead:2015pva}
\bibinfo{author}{\bibfnamefont{P.}~\bibnamefont{Adshead}},
  \bibinfo{author}{\bibfnamefont{J.~T.} \bibnamefont{Giblin}},
  \bibinfo{author}{\bibfnamefont{T.~R.} \bibnamefont{Scully}},
  \bibnamefont{and} \bibinfo{author}{\bibfnamefont{E.~I.}
  \bibnamefont{Sfakianakis}}, \bibinfo{journal}{JCAP}
  \textbf{\bibinfo{volume}{12}}, \bibinfo{pages}{034} (\bibinfo{year}{2015}),
  \eprint{1502.06506}.

\bibitem[{\citenamefont{Cuissa and Figueroa}(2019)}]{Cuissa:2018oiw}
\bibinfo{author}{\bibfnamefont{J.~R.~C.} \bibnamefont{Cuissa}}
  \bibnamefont{and} \bibinfo{author}{\bibfnamefont{D.~G.}
  \bibnamefont{Figueroa}}, \bibinfo{journal}{JCAP}
  \textbf{\bibinfo{volume}{06}}, \bibinfo{pages}{002} (\bibinfo{year}{2019}),
  \eprint{1812.03132}.

\bibitem[{\citenamefont{Adshead et~al.}(2024)\citenamefont{Adshead, Giblin,
  Grutkoski, and Weiner}}]{Adshead:2023mvt}
\bibinfo{author}{\bibfnamefont{P.}~\bibnamefont{Adshead}},
  \bibinfo{author}{\bibfnamefont{J.~T.} \bibnamefont{Giblin}},
  \bibinfo{author}{\bibfnamefont{R.}~\bibnamefont{Grutkoski}},
  \bibnamefont{and} \bibinfo{author}{\bibfnamefont{Z.~J.}
  \bibnamefont{Weiner}}, \bibinfo{journal}{JCAP} \textbf{\bibinfo{volume}{03}},
  \bibinfo{pages}{017} (\bibinfo{year}{2024}), \eprint{2311.01504}.

\bibitem[{\citenamefont{Barnaby et~al.}(2011)\citenamefont{Barnaby, Namba, and
  Peloso}}]{Barnaby:2011vw}
\bibinfo{author}{\bibfnamefont{N.}~\bibnamefont{Barnaby}},
  \bibinfo{author}{\bibfnamefont{R.}~\bibnamefont{Namba}}, \bibnamefont{and}
  \bibinfo{author}{\bibfnamefont{M.}~\bibnamefont{Peloso}},
  \bibinfo{journal}{JCAP} \textbf{\bibinfo{volume}{04}}, \bibinfo{pages}{009}
  (\bibinfo{year}{2011}), \eprint{1102.4333}.

\bibitem[{\citenamefont{Schmitz}(2021)}]{Schmitz:2020syl}
\bibinfo{author}{\bibfnamefont{K.}~\bibnamefont{Schmitz}},
  \bibinfo{journal}{JHEP} \textbf{\bibinfo{volume}{01}}, \bibinfo{pages}{097}
  (\bibinfo{year}{2021}), \eprint{2002.04615}.

\bibitem[{\citenamefont{Garcia-Bellido
  et~al.}(2021)\citenamefont{Garcia-Bellido, Murayama, and
  White}}]{Garcia-Bellido:2021zgu}
\bibinfo{author}{\bibfnamefont{J.}~\bibnamefont{Garcia-Bellido}},
  \bibinfo{author}{\bibfnamefont{H.}~\bibnamefont{Murayama}}, \bibnamefont{and}
  \bibinfo{author}{\bibfnamefont{G.}~\bibnamefont{White}},
  \bibinfo{journal}{JCAP} \textbf{\bibinfo{volume}{12}}, \bibinfo{pages}{023}
  (\bibinfo{year}{2021}), \eprint{2104.04778}.

\bibitem[{\citenamefont{Caprini and Figueroa}(2018)}]{Caprini:2018mtu}
\bibinfo{author}{\bibfnamefont{C.}~\bibnamefont{Caprini}} \bibnamefont{and}
  \bibinfo{author}{\bibfnamefont{D.~G.} \bibnamefont{Figueroa}},
  \bibinfo{journal}{Class. Quant. Grav.} \textbf{\bibinfo{volume}{35}},
  \bibinfo{pages}{163001} (\bibinfo{year}{2018}), \eprint{1801.04268}.

\bibitem[{\citenamefont{Aghanim et~al.}(2020)}]{Planck:2018vyg}
\bibinfo{author}{\bibfnamefont{N.}~\bibnamefont{Aghanim}} \bibnamefont{et~al.}
  (\bibinfo{collaboration}{Planck}), \bibinfo{journal}{Astron. Astrophys.}
  \textbf{\bibinfo{volume}{641}}, \bibinfo{pages}{A6} (\bibinfo{year}{2020}),
  \bibinfo{note}{[Erratum: Astron.Astrophys. 652, C4 (2021)]},
  \eprint{1807.06209}.

\bibitem[{\citenamefont{Tristram et~al.}(2022)}]{Tristram:2021tvh}
\bibinfo{author}{\bibfnamefont{M.}~\bibnamefont{Tristram}}
  \bibnamefont{et~al.}, \bibinfo{journal}{Phys. Rev. D}
  \textbf{\bibinfo{volume}{105}}, \bibinfo{pages}{083524}
  (\bibinfo{year}{2022}), \eprint{2112.07961}.

\bibitem[{\citenamefont{Galloni et~al.}(2023)\citenamefont{Galloni, Bartolo,
  Matarrese, Migliaccio, Ricciardone, and Vittorio}}]{Galloni:2022mok}
\bibinfo{author}{\bibfnamefont{G.}~\bibnamefont{Galloni}},
  \bibinfo{author}{\bibfnamefont{N.}~\bibnamefont{Bartolo}},
  \bibinfo{author}{\bibfnamefont{S.}~\bibnamefont{Matarrese}},
  \bibinfo{author}{\bibfnamefont{M.}~\bibnamefont{Migliaccio}},
  \bibinfo{author}{\bibfnamefont{A.}~\bibnamefont{Ricciardone}},
  \bibnamefont{and} \bibinfo{author}{\bibfnamefont{N.}~\bibnamefont{Vittorio}},
  \bibinfo{journal}{JCAP} \textbf{\bibinfo{volume}{04}}, \bibinfo{pages}{062}
  (\bibinfo{year}{2023}), \eprint{2208.00188}.

\bibitem[{\citenamefont{Silk}(1968)}]{Silk:1967kq}
\bibinfo{author}{\bibfnamefont{J.}~\bibnamefont{Silk}},
  \bibinfo{journal}{Astrophys. J.} \textbf{\bibinfo{volume}{151}},
  \bibinfo{pages}{459} (\bibinfo{year}{1968}).

\bibitem[{\citenamefont{Sunyaev and Zeldovich}(1970)}]{Sunyaev:1970plh}
\bibinfo{author}{\bibfnamefont{R.~A.} \bibnamefont{Sunyaev}} \bibnamefont{and}
  \bibinfo{author}{\bibfnamefont{Y.~B.} \bibnamefont{Zeldovich}},
  \bibinfo{journal}{Astrophys. Space Sci.} \textbf{\bibinfo{volume}{9}},
  \bibinfo{pages}{368} (\bibinfo{year}{1970}).

\bibitem[{\citenamefont{Daly}(1991)}]{Daly:1991uob}
\bibinfo{author}{\bibfnamefont{R.~A.} \bibnamefont{Daly}},
  \bibinfo{journal}{Astrophys. J.} \textbf{\bibinfo{volume}{371}}
  (\bibinfo{year}{1991}).

\bibitem[{\citenamefont{{Barrow} and {Coles}}(1991)}]{1991MNRAS.248...52B}
\bibinfo{author}{\bibfnamefont{J.~D.} \bibnamefont{{Barrow}}} \bibnamefont{and}
  \bibinfo{author}{\bibfnamefont{P.}~\bibnamefont{{Coles}}},
  \textbf{\bibinfo{volume}{248}}, \bibinfo{pages}{52} (\bibinfo{year}{1991}).

\bibitem[{\citenamefont{Hu et~al.}(1994)\citenamefont{Hu, Scott, and
  Silk}}]{Hu:1994bz}
\bibinfo{author}{\bibfnamefont{W.}~\bibnamefont{Hu}},
  \bibinfo{author}{\bibfnamefont{D.}~\bibnamefont{Scott}}, \bibnamefont{and}
  \bibinfo{author}{\bibfnamefont{J.}~\bibnamefont{Silk}},
  \bibinfo{journal}{Astrophys. J. Lett.} \textbf{\bibinfo{volume}{430}},
  \bibinfo{pages}{L5} (\bibinfo{year}{1994}), \eprint{astro-ph/9402045}.

\bibitem[{\citenamefont{Chluba et~al.}(2015{\natexlab{a}})\citenamefont{Chluba,
  Dai, Grin, Amin, and Kamionkowski}}]{Chluba:2014qia}
\bibinfo{author}{\bibfnamefont{J.}~\bibnamefont{Chluba}},
  \bibinfo{author}{\bibfnamefont{L.}~\bibnamefont{Dai}},
  \bibinfo{author}{\bibfnamefont{D.}~\bibnamefont{Grin}},
  \bibinfo{author}{\bibfnamefont{M.}~\bibnamefont{Amin}}, \bibnamefont{and}
  \bibinfo{author}{\bibfnamefont{M.}~\bibnamefont{Kamionkowski}},
  \bibinfo{journal}{Mon. Not. Roy. Astron. Soc.}
  \textbf{\bibinfo{volume}{446}}, \bibinfo{pages}{2871}
  (\bibinfo{year}{2015}{\natexlab{a}}), \eprint{1407.3653}.

\bibitem[{\citenamefont{Chluba et~al.}(2012{\natexlab{a}})\citenamefont{Chluba,
  Khatri, and Sunyaev}}]{Chluba:2012gq}
\bibinfo{author}{\bibfnamefont{J.}~\bibnamefont{Chluba}},
  \bibinfo{author}{\bibfnamefont{R.}~\bibnamefont{Khatri}}, \bibnamefont{and}
  \bibinfo{author}{\bibfnamefont{R.~A.} \bibnamefont{Sunyaev}},
  \bibinfo{journal}{Mon. Not. Roy. Astron. Soc.}
  \textbf{\bibinfo{volume}{425}}, \bibinfo{pages}{1129}
  (\bibinfo{year}{2012}{\natexlab{a}}), \eprint{1202.0057}.

\bibitem[{\citenamefont{Chluba et~al.}(2012{\natexlab{b}})\citenamefont{Chluba,
  Erickcek, and Ben-Dayan}}]{Chluba:2012we}
\bibinfo{author}{\bibfnamefont{J.}~\bibnamefont{Chluba}},
  \bibinfo{author}{\bibfnamefont{A.~L.} \bibnamefont{Erickcek}},
  \bibnamefont{and}
  \bibinfo{author}{\bibfnamefont{I.}~\bibnamefont{Ben-Dayan}},
  \bibinfo{journal}{Astrophys. J.} \textbf{\bibinfo{volume}{758}},
  \bibinfo{pages}{76} (\bibinfo{year}{2012}{\natexlab{b}}), \eprint{1203.2681}.

\bibitem[{\citenamefont{Chluba et~al.}(2015{\natexlab{b}})\citenamefont{Chluba,
  Hamann, and Patil}}]{Chluba:2015bqa}
\bibinfo{author}{\bibfnamefont{J.}~\bibnamefont{Chluba}},
  \bibinfo{author}{\bibfnamefont{J.}~\bibnamefont{Hamann}}, \bibnamefont{and}
  \bibinfo{author}{\bibfnamefont{S.~P.} \bibnamefont{Patil}},
  \bibinfo{journal}{Int. J. Mod. Phys. D} \textbf{\bibinfo{volume}{24}},
  \bibinfo{pages}{1530023} (\bibinfo{year}{2015}{\natexlab{b}}),
  \eprint{1505.01834}.

\bibitem[{\citenamefont{Lucca et~al.}(2020)\citenamefont{Lucca, Sch{\"o}neberg,
  Hooper, Lesgourgues, and Chluba}}]{Lucca:2019rxf}
\bibinfo{author}{\bibfnamefont{M.}~\bibnamefont{Lucca}},
  \bibinfo{author}{\bibfnamefont{N.}~\bibnamefont{Sch{\"o}neberg}},
  \bibinfo{author}{\bibfnamefont{D.~C.} \bibnamefont{Hooper}},
  \bibinfo{author}{\bibfnamefont{J.}~\bibnamefont{Lesgourgues}},
  \bibnamefont{and} \bibinfo{author}{\bibfnamefont{J.}~\bibnamefont{Chluba}},
  \bibinfo{journal}{JCAP} \textbf{\bibinfo{volume}{02}}, \bibinfo{pages}{026}
  (\bibinfo{year}{2020}), \eprint{1910.04619}.

\bibitem[{\citenamefont{Fixsen et~al.}(1996)\citenamefont{Fixsen, Cheng, Gales,
  Mather, Shafer, and Wright}}]{Fixsen:1996nj}
\bibinfo{author}{\bibfnamefont{D.~J.} \bibnamefont{Fixsen}},
  \bibinfo{author}{\bibfnamefont{E.~S.} \bibnamefont{Cheng}},
  \bibinfo{author}{\bibfnamefont{J.~M.} \bibnamefont{Gales}},
  \bibinfo{author}{\bibfnamefont{J.~C.} \bibnamefont{Mather}},
  \bibinfo{author}{\bibfnamefont{R.~A.} \bibnamefont{Shafer}},
  \bibnamefont{and} \bibinfo{author}{\bibfnamefont{E.~L.}
  \bibnamefont{Wright}}, \bibinfo{journal}{Astrophys. J.}
  \textbf{\bibinfo{volume}{473}}, \bibinfo{pages}{576} (\bibinfo{year}{1996}),
  \eprint{astro-ph/9605054}.

\bibitem[{\citenamefont{Bianchini and Fabbian}(2022)}]{Bianchini:2022dqh}
\bibinfo{author}{\bibfnamefont{F.}~\bibnamefont{Bianchini}} \bibnamefont{and}
  \bibinfo{author}{\bibfnamefont{G.}~\bibnamefont{Fabbian}},
  \bibinfo{journal}{Phys. Rev. D} \textbf{\bibinfo{volume}{106}},
  \bibinfo{pages}{063527} (\bibinfo{year}{2022}), \eprint{2206.02762}.

\bibitem[{\citenamefont{Kogut et~al.}(2010)}]{Kogut:2010xfw}
\bibinfo{author}{\bibfnamefont{A.~J.} \bibnamefont{Kogut}}
  \bibnamefont{et~al.}, \bibinfo{journal}{Proc. SPIE Int. Soc. Opt. Eng.}
  \textbf{\bibinfo{volume}{7731}}, \bibinfo{pages}{77311S}
  (\bibinfo{year}{2010}).

\bibitem[{\citenamefont{Andr{\'e} et~al.}(2014)}]{PRISM:2013fvg}
\bibinfo{author}{\bibfnamefont{P.}~\bibnamefont{Andr{\'e}}}
  \bibnamefont{et~al.} (\bibinfo{collaboration}{PRISM}),
  \bibinfo{journal}{JCAP} \textbf{\bibinfo{volume}{02}}, \bibinfo{pages}{006}
  (\bibinfo{year}{2014}), \eprint{1310.1554}.

\bibitem[{\citenamefont{Masi et~al.}(2021)}]{Masi:2021azs}
\bibinfo{author}{\bibfnamefont{S.}~\bibnamefont{Masi}} \bibnamefont{et~al.}, in
  \emph{\bibinfo{booktitle}{{16th Marcel Grossmann Meeting on~Recent
  Developments in Theoretical and Experimental General Relativity, Astrophysics
  and Relativistic Field Theories}}} (\bibinfo{year}{2021}),
  \eprint{2110.12254}.

\bibitem[{\citenamefont{Maffei et~al.}(2021)}]{Maffei:2021xur}
\bibinfo{author}{\bibfnamefont{B.}~\bibnamefont{Maffei}} \bibnamefont{et~al.},
  in \emph{\bibinfo{booktitle}{{16th Marcel Grossmann Meeting on~Recent
  Developments in Theoretical and Experimental General Relativity, Astrophysics
  and Relativistic Field Theories}}} (\bibinfo{year}{2021}),
  \eprint{2111.00246}.

\bibitem[{\citenamefont{Chluba et~al.}(2019)}]{Chluba:2019kpb}
\bibinfo{author}{\bibfnamefont{J.}~\bibnamefont{Chluba}} \bibnamefont{et~al.},
  \bibinfo{journal}{Bull. Am. Astron. Soc.} \textbf{\bibinfo{volume}{51}},
  \bibinfo{pages}{184} (\bibinfo{year}{2019}), \eprint{1903.04218}.

\bibitem[{\citenamefont{Chluba et~al.}(2021)}]{Chluba:2019nxa}
\bibinfo{author}{\bibfnamefont{J.}~\bibnamefont{Chluba}} \bibnamefont{et~al.},
  \bibinfo{journal}{Exper. Astron.} \textbf{\bibinfo{volume}{51}},
  \bibinfo{pages}{1515} (\bibinfo{year}{2021}), \eprint{1909.01593}.

\bibitem[{\citenamefont{Abbott et~al.}(2021)}]{KAGRA:2021kbb}
\bibinfo{author}{\bibfnamefont{R.}~\bibnamefont{Abbott}} \bibnamefont{et~al.}
  (\bibinfo{collaboration}{KAGRA, Virgo, LIGO Scientific}),
  \bibinfo{journal}{Phys. Rev. D} \textbf{\bibinfo{volume}{104}},
  \bibinfo{pages}{022004} (\bibinfo{year}{2021}), \eprint{2101.12130}.

\bibitem[{\citenamefont{Agazie et~al.}(2023)}]{NANOGrav:2023gor}
\bibinfo{author}{\bibfnamefont{G.}~\bibnamefont{Agazie}} \bibnamefont{et~al.}
  (\bibinfo{collaboration}{NANOGrav}), \bibinfo{journal}{Astrophys. J. Lett.}
  \textbf{\bibinfo{volume}{951}}, \bibinfo{pages}{L8} (\bibinfo{year}{2023}),
  \eprint{2306.16213}.

\bibitem[{\citenamefont{Kofman et~al.}(1997)\citenamefont{Kofman, Linde, and
  Starobinsky}}]{Kofman:1997yn}
\bibinfo{author}{\bibfnamefont{L.}~\bibnamefont{Kofman}},
  \bibinfo{author}{\bibfnamefont{A.~D.} \bibnamefont{Linde}}, \bibnamefont{and}
  \bibinfo{author}{\bibfnamefont{A.~A.} \bibnamefont{Starobinsky}},
  \bibinfo{journal}{Phys. Rev. D} \textbf{\bibinfo{volume}{56}},
  \bibinfo{pages}{3258} (\bibinfo{year}{1997}), \eprint{hep-ph/9704452}.

\bibitem[{\citenamefont{Felder and Tkachev}(2008)}]{Felder:2000hq}
\bibinfo{author}{\bibfnamefont{G.~N.} \bibnamefont{Felder}} \bibnamefont{and}
  \bibinfo{author}{\bibfnamefont{I.}~\bibnamefont{Tkachev}},
  \bibinfo{journal}{Comput. Phys. Commun.} \textbf{\bibinfo{volume}{178}},
  \bibinfo{pages}{929} (\bibinfo{year}{2008}), \eprint{hep-ph/0011159}.

\end{thebibliography}

\end{document}